\documentclass[twocolumn,twocolappendix,times]{aastex63}

\usepackage{CJK}    
\usepackage{multirow}   
\usepackage{amsmath}

\newcommand{\hst}{{\it HST\/}}       
\newcommand{\herschel}{{\it Herschel\/}}
\newcommand{\chandra}{{\it Chandra\/}}

\newcommand{\jwst}{{\it JWST\/}}
\newcommand{\spitzer}{{\it Spitzer\/}}

\newcommand{\akari}{{\it AKARI\/}}

\newcommand{\xray}{\hbox{X-ray}}  

\newcommand{\lx}{L_{\rm X}}
\newcommand{\nh}{N_{\rm H}}
\newcommand{\ld}{L_{\rm disk}}
\newcommand{\lsix}{L_{\rm 6\mu m}}

\newcommand{\fracA}{{\rm frac_{AGN}}}
\newcommand{\fracM}{{\rm frac^{\it K}_{MIR, AGN}}}

\newcommand{\smad}{\sigma_{\rm MAD}}

\newcommand{\xcig}{\hbox{\sc x-cigale}}
\newcommand{\mirisim}{\hbox{\sc mirisim}}

\shorttitle{JWST/MIRI Simulated Imaging}
\shortauthors{Yang et al.}

\begin{document}
\begin{CJK*}{UTF8}{gbsn}
\title{JWST/MIRI Simulated Imaging:  Insights into Obscured Star-Formation and AGN for Distant Galaxies in Deep Surveys}

\correspondingauthor{G. Yang}
\email{gyang206265@gmail.com}

\author[0000-0001-8835-7722]{G. Yang (杨光)}
\affiliation{Department of Physics and Astronomy, Texas A\&M
  University, College Station, TX, 77843-4242 USA}
\affiliation{George P.\ and Cynthia Woods Mitchell Institute for
 Fundamental Physics and Astronomy, Texas A\&M University, College
 Station, TX, 77843-4242 USA}

\author[0000-0001-7503-8482]{C. Papovich}
\affiliation{Department of Physics and Astronomy, Texas A\&M
  University, College Station, TX, 77843-4242 USA}
\affiliation{George P.\ and Cynthia Woods Mitchell Institute for
 Fundamental Physics and Astronomy, Texas A\&M University, College
 Station, TX, 77843-4242 USA}

\author{M. B. Bagley}
\affiliation{Department of Astronomy, University of Texas at Austin, Austin, TX 78712, USA}

\author{V. Buat}
\affiliation{Aix Marseille Univ, CNRS, CNES, LAM, Marseille, France}

\author{D. Burgarella}
\affiliation{Aix Marseille Univ, CNRS, CNES, LAM, Marseille, France}

\author{M. Dickinson}
\affiliation{National Optical Astronomy Observatory, 950 North Cherry Avenue, Tucson, AZ 85719, USA}

\author{D. Elbaz}
\affiliation{AIM, CEA, CNRS, Universit\'e Paris-Saclay, Universit\'e Paris
Diderot, Sorbonne Paris Cit\'e, F-91191 Gif-sur-Yvette, France}

\author{S. L. Finkelstein}
\affiliation{Department of Astronomy, University of Texas at Austin, Austin, TX 78712, USA}

\author{A. Fontana}
\affiliation{INAF – Osservatorio Astronomico di Roma, via Frascati 33, 00078, Monteporzio Catone, Italy}

\author{N. A. Grogin}
\affiliation{Space Telescope Science Institute, 3700 San Martin Drive, Baltimore, MD 21218, USA}

\author{I. Jung}
\affiliation{Department of Physics, The Catholic University of America, Washington, DC, 20064 USA}
\affiliation{Astrophysics Science Division, Goddard Space Flight Center, Greenbelt, MD, 20771 USA}

\author{J. S. Kartaltepe}
\affiliation{School of Physics and Astronomy, Rochester Institute of Technology, 84 Lomb Memorial Drive, Rochester NY 14623, USA}

\author{A. Kirkpatrick}
\affiliation{Department of Physics \& Astronomy, University of Kansas, Lawrence, KS 66045, USA}

\author{P. G. P\'erez-Gonz\'alez}
\affiliation{Centro de Astrobiolog\'ia, (CAB, CSIC-INTA), Carretera de Ajalvir, km 4, E-28850 Torrej\'on de Ardoz, Madrid, Spain}

\author{N. Pirzkal}
\affiliation{Space Telescope Science Institute, 3700 San Martin Drive, Baltimore, MD 21218, USA}

\author{L. Y. A. Yung}
\affiliation{Department of Physics and Astronomy, Rutgers University, 136 Frelinghuysen Road, Piscataway, NJ 08854, USA}
\affiliation{Center for Computational Astrophysics, Flatiron Institute, 162 5th Ave, New York, NY 10010, USA}
\affiliation{Astrophysics Science Division, Goddard Space Flight Center, Greenbelt, MD, 20771 USA}



\begin{abstract}
The \jwst\ MIRI instrument will revolutionize extragalactic astronomy 
{with unprecedented sensitivity and angular resolution in mid-IR}.
Here, we assess the potential of MIRI photometry to constrain
galaxy properties in the \textit{Cosmic Evolution Early Release Science} 
(CEERS) survey. 
We {derive estimated MIRI fluxes} from the spectral energy
distributions (SEDs) of real sources that fall in a planned MIRI pointing.
{We also obtain MIRI fluxes for hypothetical AGN-galaxy mixed models
varying the AGN fractional contribution to the total 
IR luminosity (${\fracA}$).}
{Based on these model fluxes}, we simulate CEERS imaging 
(3.6-hour exposure) in 6 bands from F770W to F2100W using {\sc mirisim}, and 
reduce these data using \textsc{jwst pipeline}.
We perform PSF-matched photometry with {\sc tphot}, and 
fit the source SEDs with \xcig, simultaneously modeling photometric
redshift and other physical properties.
Adding the MIRI data, the accuracy of both redshift and $\fracA$ is 
generally improved by factors of $\gtrsim 2$ for all sources at $z\lesssim 3$.
Notably, for pure-galaxy inputs ($\fracA=0$), the accuracy of $\fracA$ is 
improved by $\sim 100$ times thanks to MIRI.
The simulated CEERS MIRI data are slightly more sensitive to AGN 
detections than the deepest \xray\ survey, based on the empirical $\lx$-$\lsix$ 
relation.
Like \xray\ observations, MIRI can also be used to constrain the AGN
accretion power (accuracy $\approx 0.3$~dex).
Our work demonstrates that MIRI will be able to place strong constraints
on the mid-IR luminosities from star formation and AGN,
and thereby facilitate studies of the galaxy/AGN co-evolution.
\vspace{1.5 cm}
\end{abstract}



\section{Introduction} \label{sec:intro}
Mid-infrared (mid-IR) wavelengths provide extremely valuable information for 
extragalactic sources.
Galaxies often contain a large amount of polycyclic aromatic hydrocarbon (PAH) 
molecules (see \citealt{tielens08} for a review).
When heated by starlight, these PAH molecules produce strong emission features 
mainly at $\approx 3\text{--}18\ \mu$m.
In star-forming galaxies, PAHs can emit as much as $\sim 20\%$ of the total IR 
luminosity and the $7.7\ \mu$m PAH feature can contribute as much as 50\% of 
the total PAH emission \citep[e.g.,][]{shipley16}.
The PAH emission features not only can reveal important information about, e.g., 
metallicity and ionization parameter \citep[e.g.,][]{draine07, shivaei17},
but also are potential redshift indicators for high-redshift objects 
\citep[e.g.,][]{chary07}.

Active galactic nuclei (AGNs) are often surrounded by large amounts of dust 
\citep[e.g.,][]{antonucci93, urry95, netzer15}. 
The dust absorbs UV/optical radiation from the central engine and reaches 
temperatures of above a few hundred kelvin, much higher than the typical 
temperatures of interstellar dust heated by starlight (a few ten kelvin).
This AGN-heated hot dust reemits the absorbed energy mainly at mid-IR wavelengths. 
Therefore, AGNs can be identified based on the hot dust emission 
\citep[e.g.,][]{stern12b, assef13, kirkpatrick15}.
The mid-IR selection of AGNs has significant advantages over other methods in 
terms of dust obscuration (e.g., \citealt{hickox18}; {\citealt{alberts20}}).
The UV/optical selection is often affected by dust obscuration and biased to 
bright type~1 AGNs. 
{\xray\ data are currently the most robust method of AGN selection 
\citep[e.g.,][]{brandt15, xue17}.
Deep \xray\ surveys can sample low-luminosity AGNs below 
${\lx \sim 10^{43}}$~erg~s$^{-1}$ up to ${z\sim 4}$ 
\citep{xue16,luo17,vito18}.
At such a low ${\lx}$ level, most (${\approx 90\%}$) of the 
\xray\ selected AGNs are type~2, which are potentially missed by UV/optical 
selections \citep[e.g.,][]{merloni14}. 
However, \xray\ selection could miss a large population of extremely obscured
``Compton-thick'' AGNs (neutral-hydrogen column density, 
${\nh > 10^{24}}$~cm$^{-2}$; e.g., \citealt{brandt15}; \citealt{hickox18}).  These obscured AGN can be identified through warm-dust emission from the AGN in the mid-IR \citep[e.g.,][]{alexander08, del_moro16}, and therefore it is useful to explore the utility of mid-IR observations to identify these objects.
}

Past and current facilities have not been sufficiently sensitive to capture 
the mid-IR fluxes from most sources in the distant universe.
Until now, the most sensitive mid-IR facility has been \spitzer/IRAC$+$MIPS, 
covering wavelengths 3.6--24~\micron. 
However, the IRAC+MIPS coverage has a large uncovered gap between the longest 
IRAC band ($8\ \mu$m) and the shortest MIPS band ($24\ \mu$m), 
leaving a large room for model degeneracy at mid-IR wavelengths.
For example, observations have found some $z\sim 2$ galaxies with elevated 
24~$\mu$m emission compared to that expected from star formation 
\citep[mid-IR-excess galaxies; e.g.,][]{daddi07b, papovich07}.
The mid-IR excess emission may be interpreted as either PAH emission or 
AGN-heated dust radiation, and it is challenging to decompose these two 
components with broad-band imaging alone that lacks contiguous mid-IR 
wavelength coverage \citep[e.g.][]{daddi07, azadi18}.
Furthermore, the \spitzer\ imaging has a large point spread function (PSF, 
FWHM~$\approx 6\arcsec$ at $24\mu$m), causing source-confusion issues in 
crowded deep fields \citep[e.g.,][]{dole04, ashby18}.
\akari/IRC provided contiguous wavelength imaging over \hbox{2--26~$\mu$m}. 
However, its sensitivity is relatively low compared to \spitzer, and many 
\spitzer\ faint sources are undetectable by \akari\ 
\citep[e.g.,][]{papovich04, clements11}.
Due to the lack of sensitive \hbox{8--24 $\mu$m} imaging, many of the IR AGN 
selection techniques are forced to use $\lesssim 8 \mu$m colors that are also 
biased toward type~1 AGNs, {similar to UV/optical selections} (e.g., 
\citealt{donley12}; \citealt{li20}).

The upcoming \textit{James Webb Space Telescope} (\jwst) mission will 
revolutionize the field of infrared astronomy.  
The onboard Mid-Infrared Instrument (MIRI), equipped with both an imaging camera 
and a spectrograph, covers a wavelength range from 
$\approx 5$ to 28$~\mu$m.
The sensitivity of MIRI is generally $\approx 1\text{--}2$ orders of magnitude
higher than \spitzer\ \citep[e.g.,][]{glasse15}, able to capture the bulk of 
cosmic mid-IR emission \citep[e.g.,][]{bonato17, rieke19}.
The MIRI imager has a PSF FWHM of $\simeq$ $0\farcs2$--$0\farcs8$.  
The sub-arcsecond angular resolution provided by MIRI reduces greater source 
blending and source confusion.
The detector's field of view (FOV) is $1.3\arcmin \times 1.7 \arcmin$, 
sufficient for extragalactic surveys.
MIRI has a total of 9 broad-band filters with contiguous wavelength coverage, and able to well characterize the mid-IR spectral shape.
%

Aside from observational facilities, reliable techniques are also essential for 
the estimation of source properties.
Past methods have relied on color-selection (and color-color selection) to 
classify IR sources \citep[e.g.,][]{stern05, donley12, messias12, kirkpatrick17}. 
The color-color methods provide a mechanism to classify galaxies as AGN versus 
non-AGN, and to identify candidates for ``composite sources'' 
\citep[e.g,.][]{kirkpatrick17}.  
These methods can be be straightforwardly applied to a large sample of sources.
Indeed, previous work has shown that MIRI mid-IR color-color diagrams can 
reliably identify AGNs in galaxies down to low AGN luminosities (Eddington ratios 
of $\sim 0.01$, e.g., \citealt{kirkpatrick17}). 
However, color-color methods have disadvantages. 
They often require knowledge of the redshift of the source \textit{a priori} to 
interpret (for example, the redshift can determine which colors best separate 
warm dust heated by AGN from emission from PAHs). 
In addition they are typically binary (either a source is an AGN or it is not), or trinary (including composite sources). 
Lastly, typical color-selection methods use only a portion of the available 
information (e.g., a color-color diagram only uses flux densities measured at 
$\leq 4$ passbands).  
Historically, these methods have been very successful.  However, as the amount of 
available data and passbands increases (as will be the case in the \jwst/MIRI era)
it is prudent to explore methods that can provide more detailed constraints using the full dataset,  and thereby improve our understanding of 
the co-evolution of AGN and star formation in distant galaxies. 
%

An alternative type of quantitative method to characterize the IR emission 
from sources is spectral energy distribution (SED) fitting. 
{SED fitting has been successfully applied to AGN 
identification/characterization in previous works, based on the fact that 
AGNs and star-forming galaxies typically have distinctive SED shapes 
\citep[e.g.,][]{alonso_herrero06, caputi13, chang17}.}
SED-fitting methods typically fit the photometric data with numerous model 
templates.  
They are more computationally intensive than color-color methods, but they have 
the advantage that they can simultaneously take into account all photometric 
data, and fit for multiple model parameters, including fitting for a photometric 
redshift \citep[e.g.,][]{malek14}.  
{Simulations indicate that MIRI data, when included in the SED fitting, 
can significantly improve the accuracies of fitting results
\citep[e.g.,][]{bisigello16, bisigello17, kauffmann20}.}


Here, we employ one such SED-fitting method, {\sc x-cigale}, which is an efficient 
{\sc python} code that can model multiwavelength galaxy/AGN 
SEDs from \xray\ to radio \citep{yang20}.
Compared to its original version, {\sc cigale} \citep{boquien19}, 
\xcig\ implements many AGN-related improvements.
For example, \xcig\ allows a polar-dust component that has been 
recently observed with high-resolution imaging of local AGNs 
\citep[e.g.,][]{asmus19, stalevski19}.
Thanks to its optimized parallel algorithm, \xcig\ is able to fit thousands of SEDs 
using millions of models within a few hours on a typical multi-core desktop/laptop.
\xcig\ adopts physical models following the law of energy conservation.
Aside from the traditional least-$\chi^2$ analysis, \xcig\ also allows simultaneous 
analysis for source properties such as redshift and $\fracA$ (where $\fracA$ is the 
fraction of total dust IR luminosity attributed to the AGN) in a Bayesian style. 
The Bayesian analysis considers the full probably density functions (PDFs). 
In contrast, the least-$\chi^2$ analysis only considers the best-fit SED model and 
the non-negligible probability of other models may be neglected.
Therefore, the Bayesian results provide estimates of the PDFs for model parameters 
that are more informative than minimum-$\chi^2$ results 
\citep[e.g.,][]{pirzkal12, boquien19}.  
%

In this paper, we investigate the potential ability of deep imaging with \jwst/MIRI 
to constrain the properties of distant galaxies.  
We use as our fiducial example the MIRI observations expected as part of the Cosmic 
Evolution Early Release Science (CEERS) survey, including the existing multiwavelength 
photometry expected with that dataset.
We simulate a realistic set of galaxy flux densities in the MIRI bands, using predictions 
from real galaxies observed in deep \hst\ imaging from CANDELS 
\citep{grogin11,koekemoer11,stefanon17}.  
We then generate simulated raw MIRI images that account for realistic estimates of 
the noise.  
We then reduce MIRI raw imaging data to obtain MIRI mosaics that are matched to the 
existing \hst/CANDELS imaging.  
We measure the mid-IR fluxes in the MIRI bands. 
We then perform SED fitting with \xcig\ on the fluxes of MIRI and currently existing
bands in UV-to-IR wavelengths.
The redshift and other source properties are fit simultaneously in this 
SED-fitting process, where results are then derived by marginalizing over the other parameters.
Finally, we evaluate the results by comparing the SED-fitting source properties 
with the model input ones.

This paper is structured as follows.
In \S\ref{sec:sample}, we describe the CEERS survey, the existing CANDELS catalog, 
the construction of our model MIRI flux densities, and how we generated the simulated 
MIRI (raw) data.
In \S\ref{sec:data_process}, we discuss the reduction of the simulated MIRI raw data, 
method for measuring matched-aperture flux densities, and we analyze the SEDs (using 
SED fitting from \xcig). 
We then assess the results, focusing on the ability to constrain the photometric 
redshifts and $\fracA$ of distant galaxies by including the MIRI data.
In \S\ref{sec:discuss}, we discuss our results and future prospects using \jwst/MIRI 
for this application.
We summarize our work in \S\ref{sec:summary}.

Throughout this paper, we assume the same cosmology as \xcig, i.e., a flat 
$\Lambda$CDM cosmology with $H_0=70.4$~km~s$^{-1}$~Mpc$^{-1}$ and 
$\Omega_M=0.272$ (WMAP7; \citealt{komatsu11}).
We adopt a Chabrier initial mass function (IMF; \hbox{\citealt{chabrier03}}) for 
relevant quantities (stellar masses, etc.).
Quoted uncertainties are at the $1\sigma$\ (68\%) confidence level.
All magnitudes are in AB units \citep{oke83}, 
where $m_\mathrm{AB} = -48.6 - 2.5 \log(f_\nu)$ for $f_\nu$ in units of 
erg~s$^{-1}$~cm$^{-2}$ Hz$^{-1}$.   

\section{Sample and data}
\label{sec:sample}

\subsection{The CEERS survey}
\label{sec:ceers}
The Cosmic Evolution Early Release Science (CEERS) Survey is an approved 
\jwst\ program \citep{finkelstein17}, covering $\approx 100$~arcmin$^2$ 
in the Extended Groth Strip field (EGS).
CEERS consists of 63~hours of NIRCam
(\hbox{1--5 $\mu$m}) and MIRI (\hbox{5--21 $\mu$m}) imaging, NIRSpec $R\sim100$
and $R\sim1000$ spectroscopy, and NIRCam/grism $R\sim1500$ spectroscopy.
CEERS will be one of the first public \jwst\ surveys with data publicly 
available 5 months after acquisition in Cycle~1.  
%

\begin{figure*}
    \centering
	\includegraphics[width=2\columnwidth]{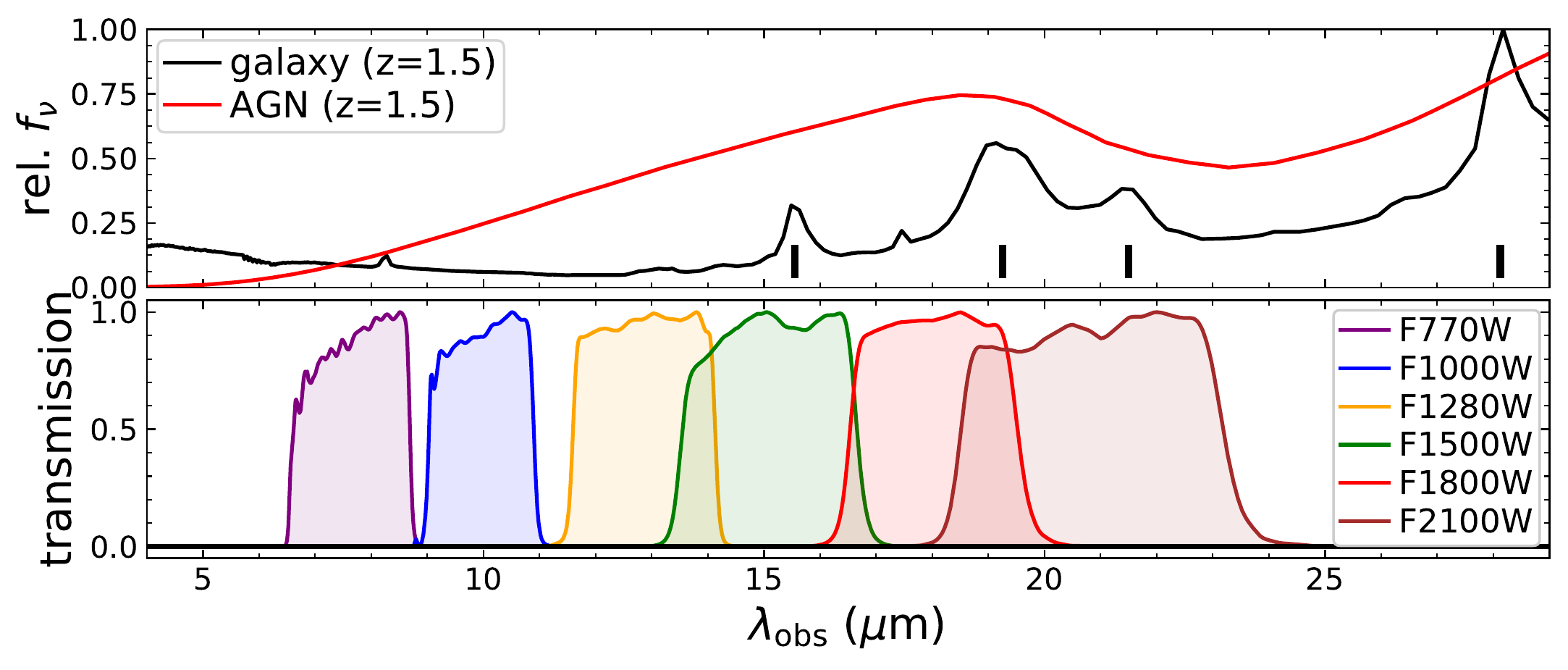}
    \caption{\textit{Top}: A star-forming galaxy (black) and AGN (red) SEDs at $z=1.5$ 
    generated by \xcig. 
    The prominent PAH emission features are marked by the vertical lines (at rest-frame 6.2, 7.7, 8.6, and 11.2~\micron, see \citealt{tielens08}).
    \textit{Bottom}: The transmission curves of the MIRI filters 
    for the CEERS/MIRI2 pointing (adopted in this work).   
    By covering the PAH features and AGN hot-dust emission, the MIRI data 
    can potentially improve the constraints on photo-$z$ and AGN component.
    }
    \label{fig:filters}
\end{figure*}

CEERS has four MIRI pointings, labeled as \hbox{MIRI~1--4} fields, 
respectively.   
In this paper, we focus on the MIRI2 pointing that has the widest wavelength 
coverage of six bands: F770W, F1000W, F1280W, F1500W, F1800W, and F2100W.
MIRI2 is centered at ($214^\circ.95330$, $52^\circ.95101$).\footnote{The coordinate 
is based on a spring launch configuration, which might differ from the 
future true pointing. 
However, the sample in this work is broadly representative of  
an arbitrary extragalactic pointing, and the conclusions should not depend on
the exact pointing coordinates.} 
For the five bands from F770W to F1800W, the planned exposure time is 
1665~seconds/band (3 dithers/band); for F2100W, the exposure time is 4662~seconds
(6 dithers).
The designed $5\sigma$ limiting magnitudes in the proposal are 25.5 (F770W), 
24.8 (F1000W), 24.3 (F1280W), 23.8 (F1500W), 22.9 (F1800W), and 
{22.8} (F2100W).
Fig.~\ref{fig:filters} displays the transmission curves of the available 
MIRI bands in the MIRI2 pointing.

With the goal of making our simulation as realistic as possible, we use all 
available data for the galaxies in the CEERS/MIRI2 field.
Our work is based on the F160W ($H$-band) selected CANDELS/EGS 
catalog \citep{stefanon17}.
In this catalog, there are 463 sources down to $H_{160}\approx 26.6$ 
within the MIRI2 pointing. 
Only 8 of these sources have secure spectroscopic redshift (spec-$z$) 
measurements, while the rest have photometric redshifts (photo-$z$) from 
\citet{stefanon17}.  
These CANDELS/EGS sources are mostly in the redshift range of \hbox{$z=0$--3}.
The CEERS/MIRI2 region is covered by 15 broad bands from CFHT/MegaCam $u^*$ to 
\spitzer/MIPS 24~$\mu$m \citep{dickinson06, stefanon17}.
Although this region is also covered by \herschel\ 
\citep{lutz11, oliver12}, most of our sources are beyond the sensitivity
of \herschel\ (only two sources in the current MIRI2 field are detected ($>5\sigma$) by PACS 
and none by SPIRE). 
Therefore, we do not use \herschel\ data in this work due to the 
low detection rate.
Based on these photometric and redshift data, we obtain the model 
input MIRI fluxes in \S\ref{sec:mod_flux}.

\subsection{The model MIRI fluxes}
\label{sec:mod_flux}
We have incomplete knowledge as to which galaxies in the CANDELS catalog 
(within MIRI2 pointing) host AGNs, because the currently existing photometric 
data provide very weak constraints on AGN emission (see \S\ref{sec:xcig_res_gal}).
However, considering the relatively small area of MIRI FOV (2.2~arcmin$^2$), we 
estimate that only $\sim 10$ AGNs (out of 463 sources; \S\ref{sec:ceers}) actually 
exist within the MIRI2 pointing, based on the AGN number density derived from the 
deepest \xray\ survey, 7~Ms \chandra\ Deep Field-South \cite[CDF-S;][]{luo17}.
Actually, none of the MIRI2 galaxies have been detected by the existing 800~ks 
\xray\ data in the EGS field (\hbox{2--7 keV}; \citealt{nandra15}).
Therefore, in \S\ref{sec:pure_gal}, we first model the SEDs of {real} 
existing photometry with pure galaxy models (i.e., $\fracA=0$), and integrate 
the best-fit SEDs with the MIRI filter transmission curves to obtain the 
bandpass-averaged flux densities as inputs to the MIRI simulations. 
To test the cases when AGN is present, we also add a {hypothetical} AGN 
component to the best-fit galaxy SEDs and rederive the MIRI flux densities to 
test for this possibility in additional sets of simulations in \S\ref{sec:mod_agn}.
These MIRI fluxes are used as the model input in our simulations of MIRI imaging 
data \S\ref{sec:mirisim}.

\subsubsection{{Pure galaxy models based on empirical SEDs}}
\label{sec:pure_gal}
We obtain model MIRI fluxes based on the SED fitting of real observations. 
We fit the currently existing broad-band photometric data (from \citealt{stefanon17}) 
that exist for the galaxies in the nominal MIRI2 pointing (\S\ref{sec:ceers}) 
with \xcig\ \citep{boquien19, yang20}.
We do not use the \xray\ module of \xcig\ due to the lack of \xray\ detections 
(\S\ref{sec:mod_flux}).
In the \xcig\ analysis, we fix the redshift at the spec-$z$ value (when available) 
or otherwise the photo-$z$ from the CANDELS/EGS catalog \citep{stefanon17}.
The fitting parameters are similar to those used by \cite{yang20} and are summarized 
in Table~\ref{tab:xcig_par}.
As in Table~\ref{tab:xcig_par}, we adopt the model of \cite{dale14} for 
galactic dust emission.
In this model, a single parameter of radiation $\alpha$ 
slope\footnote{$d M_{\rm d} \propto U^{-\alpha} d U$, where $M_{\rm d}$ and $U$ 
represent dust mass and radiation strength, respectively.} 
controls the IR SED shape (e.g., the MIR/FIR ratio and PAH emission strength).  
The model is derived from observations of local galaxies, and admittedly it may deviate 
from the true IR SEDs of distant sources.  
However, for this exercise it is sufficient as our goal is to recover the mid-IR emission. 
In the future, new models dedicated for distant galaxies can be obtained using empirical 
measurements from the MIRI data themselves (especially if additional spectroscopy from 
the medium-resolution spectrometer, MRS, becomes available).  
We will implement these new models to \xcig\ to facilitate the modeling of distant 
galaxies with real MIRI data.

\begin{figure}
    \centering
	\includegraphics[width=\columnwidth]{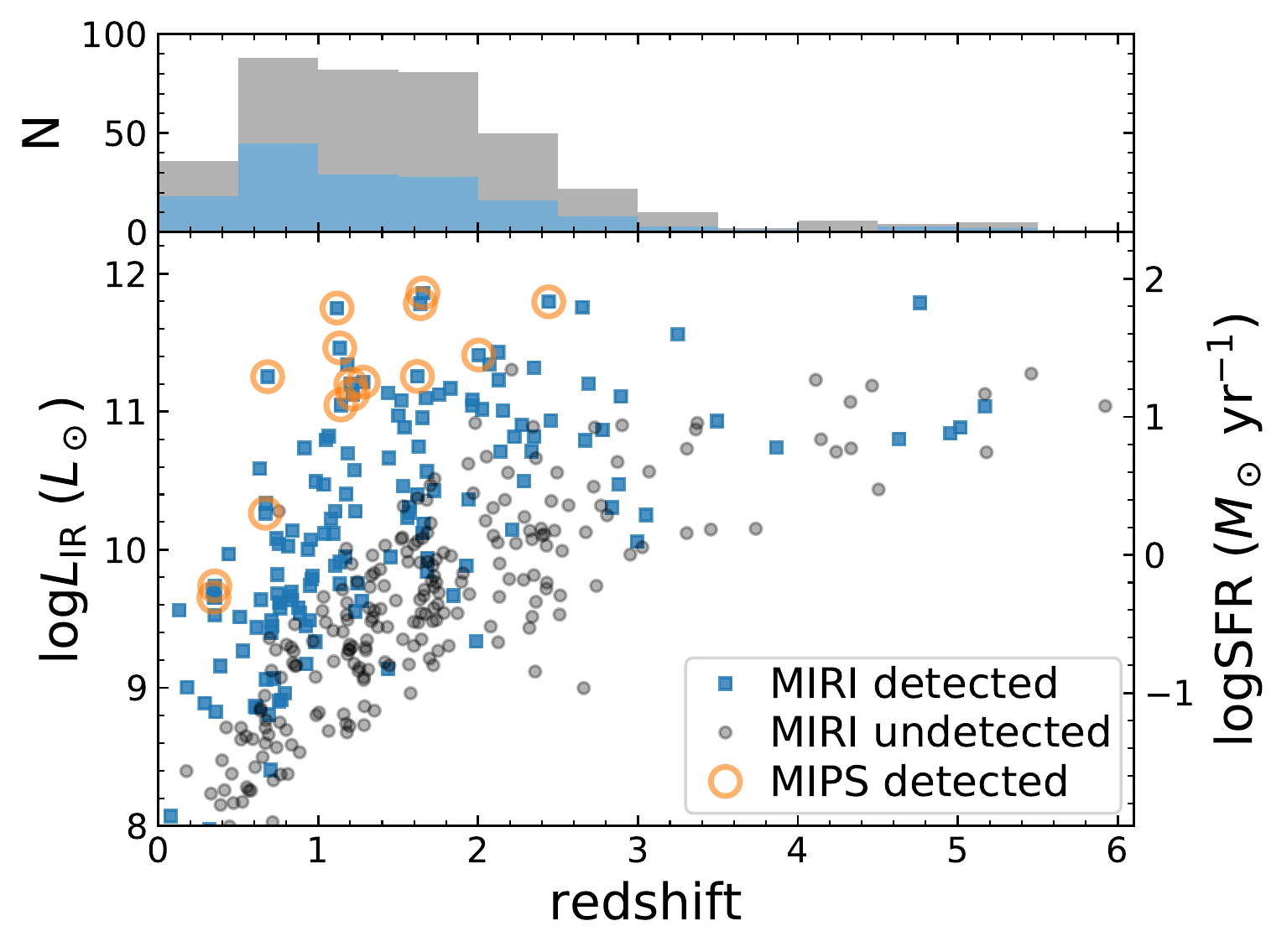}
    \caption{Total dust IR luminosity vs.\ redshift for our simulated sources in 
    the CEERS/MIRI2 pointing. 
    The redshift distribution is displayed in the top panel. 
    The IR luminosities are obtained from \xcig\ SED fitting with pure-galaxy 
    templates (\S\ref{sec:pure_gal}).
    The SFR is labeled on the right, which is converted from $L_{\rm IR}$ 
    using \citet{kennicutt98} modified for our adopted Chabrier IMF \citep{salim07}.
    The sources detected ($>5\sigma$) in at least two MIRI bands are 
    highlighted as blue squares. 
    As expected, the MIRI detected sources tend to be more IR luminous than  
    the undetected ones.
    The orange circles mark the sources detected ($>5\sigma$) by MIPS $24\mu$m.  The CEERS/MIRI data will probe galaxies with IR luminosities an order of magnitude lower than the current MIPS data (at fixed redshift). 
    }
    \label{fig:Lir_vs_z}
\end{figure}

After running \xcig, we obtain the model MIRI fluxes by integrating the 
best-fit SEDs with the transmission curves of MIRI filters (Fig.~\ref{fig:filters}).
Fig.~\ref{fig:Lir_vs_z} shows the best-fit dust IR luminosity as a function of 
redshift.  Because we do not have FIR data to directly constrain galaxy cold-dust emission, 
the IR luminosity largely comes from the modelling of UV/optical stellar 
extinction, but this is satisfactory for our purposes here where we study the 
measured mid-IR emission from MIRI based on the simulations.  
\xcig\ follows energy conservation, and thus the extincted UV/optical luminosity
equals the dust re-emitted IR luminosity.

\begin{table}
\centering
\caption{\xcig\ parameters for the fitting of existing data}
\label{tab:xcig_par}
\begin{tabular}{lll} \hline\hline
Module & Parameter & Values \\
\hline
    \multirow{2}{*}{\shortstack[l]{Star formation history\\
                                  $\mathrm{SFR}\propto t \exp(-t/\tau)$ }}
    & $\tau$ (Gyr) & 0.1, 0.5, 1, 5 \\
    & $t$ (Gyr) & 0.5, 1, 3, 5, 7 \\
\hline
\multirow{2}{*}{\shortstack[l]{Simple stellar population\\ \cite{bruzual03}}}
    & IMF & \cite{chabrier03} \\
    &     & \\
\hline
    \multirow{3}{*}{\shortstack[l]{Stellar extinction\\ \cite{calzetti00} \\ \cite{leitherer02} }}
    & \multirow{3}{*}{$E(B-V)$} 
        & \multirow{3}{*}{\shortstack[l]{0.1, 0.2, 0.3, 0.4, \\ 0.5, 0.7, 0.9}} \\
    \\
    & 
    \\
\hline
\multirow{2}{*}{\shortstack[l]{Galactic dust emission\\ \cite{dale14}}}
    & Radiation      & 1.0, 1.5, 2.0, 2.5 \\
    & $\alpha$ slope & \\
\hline
    \multirow{2}{*}{Redshifting} & $z$ & Fixed at $z$ of \\ 
                                 &     & \cite{stefanon17} \\
\hline
\end{tabular}
\begin{flushleft}
{\sc Note.} --- For parameters not listed here, we use \xcig\ default values.
\end{flushleft}
\end{table}

\subsubsection{{Hypothetical AGN-galaxy mixed models}}
\label{sec:mod_agn}
In \S\ref{sec:pure_gal}, we derive the model-input MIRI fluxes from 
pure-galaxy templates {based on the real existing photometry}.
Now, we add a {hypothetical} AGN component to the SEDs to 
explore the cases when an AGN is present.
{The purpose of using realistic galaxy templates is to test, 
if a galaxy in our field hosts an AGN, how well we can 
detect/characterize the AGN component.  
As we argue below (\S~4.2) MIRI observations may be sensitive to the 
presence of emission from obscured AGN even in galaxies that are not 
detected in the X-ray data.  
For this reason, it is useful to test the ability to recover the 
AGN emission in these ``real'' objects like those that will be seen in 
CEERS.}
 
{To generate AGN-galaxy mixed SEDs}, we again employ \xcig, 
which not only fit the photometric data but also generate model SEDs with 
given physical parameters.
For each source, we adopt the best-fit parameters in \S\ref{sec:pure_gal} for 
the galaxy component. 
Then, we add an AGN component with the SKIRTOR model in \xcig\ 
\citep{stalevski12, stalevski16}.
SKIRTOR is a modern clumpy torus model based on Monte Carlo simulations.
\cite{yang20} introduced a polar-dust component to this model when 
implementing it to \xcig.

For the AGN model parameters, we set the torus angle (between horizontal 
plane and torus edge) as 40$^\circ$ (the \xcig\ default value), which is 
favored by the observations of \cite{stalevski16}.
We randomly choose a viewing angle (between AGN axis and line of sight)
among 60$^\circ$, 70$^\circ$, 80$^\circ$, and 90$^\circ$.
We do not include smaller face-on viewing angles that correspond to type~1 
(broad-line) AGNs.
These AGNs are typically luminous as the UV/optical emission from the central 
engine is directly visible.
Therefore, their properties such as redshift and AGN luminosity can be well 
measured with UV/optical spectroscopy and photometry.
{Also, most AGNs selected in deep \xray\ surveys are type~2 
(see \S\ref{sec:intro}).}
In this paper, we focus on type~2 AGNs for which the UV/optical emission
is entirely obscured and therefore our choice of the larger viewing 
angles is appropriate. 

For other AGN torus parameters such as the 9.7~$\mu$m optical depth ($\tau_{9.7}$) 
and radial profile index $p$ (density $\propto r^{-p}$), we randomly 
assign values to each source from all allowed values in the SKIRTOR model
(e.g., $\tau_{9.7}=3, 5, 7, 9, 11$ and $p=0, 0.5, 1, 1.5$).
For the AGN polar-dust $E(B-V)$, temperature, and emissivity, we randomly 
assign values for each source from ranges of \hbox{0--0.3} (SMC extinction 
law), \hbox{100--300 K}, and \hbox{1--2}. 
This random choice of model parameters is to represent the diversity of torus 
physical properties based on current constraints \citep{stalevski12, stalevski16}. 
{There are a total of nine parameters describing the hypothetical 
AGN component.
We note that four of these parameters are free when we perform the SED
fitting while the other five parameters are fixed to canonical values (see \S\ref{sec:sed_fit}).  
We further note that the values of these fixed parameters do not impact 
the conclusions here \citep[see also][]{yang19}.}

\begin{figure}
    \centering
	\includegraphics[width=\columnwidth]{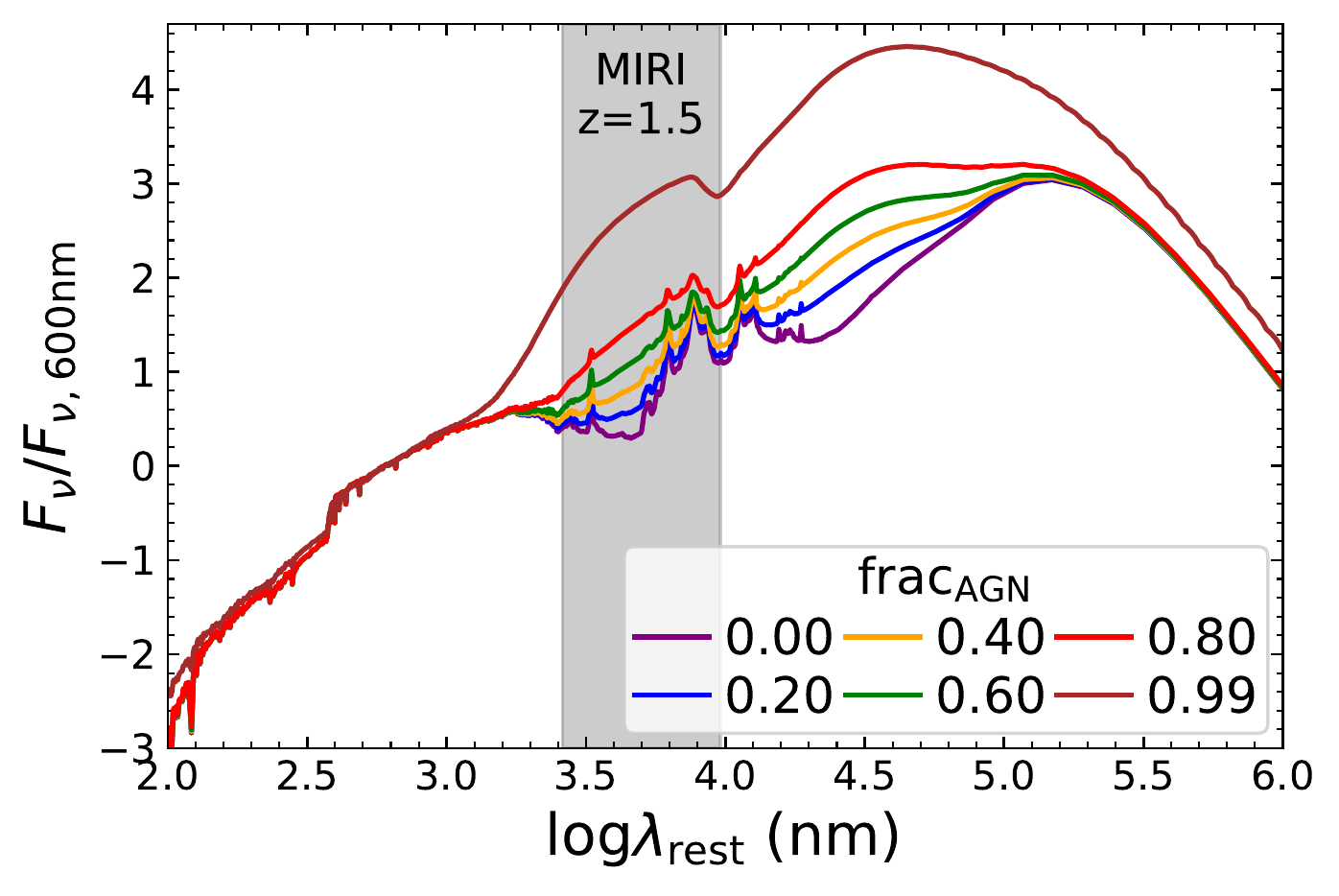}
    \caption{A set of \xcig\ SED templates with different $\fracA$.
    The $F_\nu$ is normalized at 600~nm.
    The grey shaded region indicates the MIRI coverage 
    (F770W to F2100W) at $z=1.5$.
    The type~2 AGN component mainly affects the SED of mid-IR wavelengths, 
    which are covered by MIRI.
    MIRI also covers the galactic PAH emission features.
    }
    \label{fig:sed_templates}
\end{figure}

After determining the AGN parameters above, we generate the total (galaxy $+$ 
AGN) SEDs with a given $\fracA$, and convolve the SEDs with MIRI filter curves
to obtain model-input MIRI fluxes.
We adopt five $\fracA$ values of 0.2, 0.4, 0.6, 0.8, and 0.99, for different 
AGN strengths.\footnote{The pure-galaxy models in \S\ref{sec:pure_gal} can be 
considered as $\fracA=0$.}
For each $\fracA$, we perform a set of MIRI simulations in \S\ref{sec:mirisim}. 
Fig.~\ref{fig:sed_templates} shows a set of SED models with different $\fracA$
values (as a reminder, $\fracA$ is the fraction of the IR luminosity 
[3--1000~\micron] from the AGN component). 
As $\fracA$ increases, the shape of mid-IR SEDs changes significantly, because
AGN hot-dust emission mostly concentrates on mid-IR wavelengths.
After inserting the {hypothetical} AGN component, the flux densities 
at other bands may exceed the observed fluxes (this is especially for the IRAC 
and MIPS bands), and we deal with this issue 
in \S\ref{sec:sed_fit} by generating mock fluxes for these bands from the new SED.

\subsection{The simulations of imaging data}
\label{sec:mirisim}
We use \mirisim\ (version 2.2.0; {\citealt{klaassen20}}) to simulate 
the images of MIRI bands.
\mirisim, developed by the MIRI European Consortium, is a dedicated simulation 
package for the \jwst/MIRI instrument.
\mirisim\ accepts source positions, fluxes, and morphological shapes as input. 
The MIRI images generated by \mirisim\ are preliminary (raw) ``level 0'' data products, 
and need to be processed through the {\sc jwst calibration 
pipeline}\footnote{https://github.com/spacetelescope/jwst} (\textsc{pipeline} 
hereafter) for further reduction (\S\ref{sec:pipeline}).  
In this way, \mirisim\ provides data products equivalent in format to what we expect 
\jwst\ to deliver. 

We set the background noise as ``low'' in \mirisim, as the EGS region has weak
zodiacal background \citep{finkelstein17}.
We set the background spatial gradient to 5\%/arcmin (i.e., the background changes
by 5\% per arcmin) with a position angle (PA) of $45^\circ$.
We adopt a dither pattern identical to the one in the CEERS Astronomer's Proposal 
Tool (APT) file, optimized for parallel observations between Near Infrared Camera 
(NIRCam) and MIRI \citep{finkelstein17}.
The exposure duration for each dither is the same as that in the CEERS APT configuration 
(using the same Groups/integration, integrations/exposure, and exposures/dither 
as expected for CEERS; \S\ref{sec:ceers}).
The detector readout mode is set to ``fast'', the same as proposed for CEERS.
The input fluxes are the model fluxes obtained in \S\ref{sec:mod_flux}.

Besides a point-like source profile, \mirisim\ also allows galaxy morphologies 
with S\'ersic profiles. 
Based on the \hst\ F160W imaging data, \cite{van_der_wel12} performed S\'ersic
fitting \citep{sersic68} for $H_{160} \leq 24.5$ galaxies in the CANDELS/EGS field.
We adopt their results for sources with good fitting quality (flag $\leq 1$). 
For the rest of the sources (flag~$>1$ or $H_{160} >24.5$), we adopt a point-like profile
as \mirisim\ input to test results for unresolved sources. 
The adopted galaxy profiles have diameters ($2\times R_e$) of 
$0.27\arcsec$--$0.85\arcsec$ (20\%--80\% percentile), while the MIRI PSFs have 
FWHMs ranging from $0.24\arcsec$ (F770W) to $0.67\arcsec$ (F2100W).  
Therefore, MIRI can spatially resolve a non-negligible fraction of the CANDELS 
galaxies even at $\gtrsim 20 \mu$m.
In this work, we assume that the AGN and galaxy emissions have the same 
morphology, but in reality the former is likely more compact than the latter.   
Therefore, a source could be more compact in MIRI than in F160W due to the 
presence of an AGN.
Such a morphological difference between MIRI and F160W could serve as an AGN 
indicator. 
We leave the investigation of this potential technique of AGN identification 
to future works.

\begin{figure*}
    \centering
	\includegraphics[width=2\columnwidth]{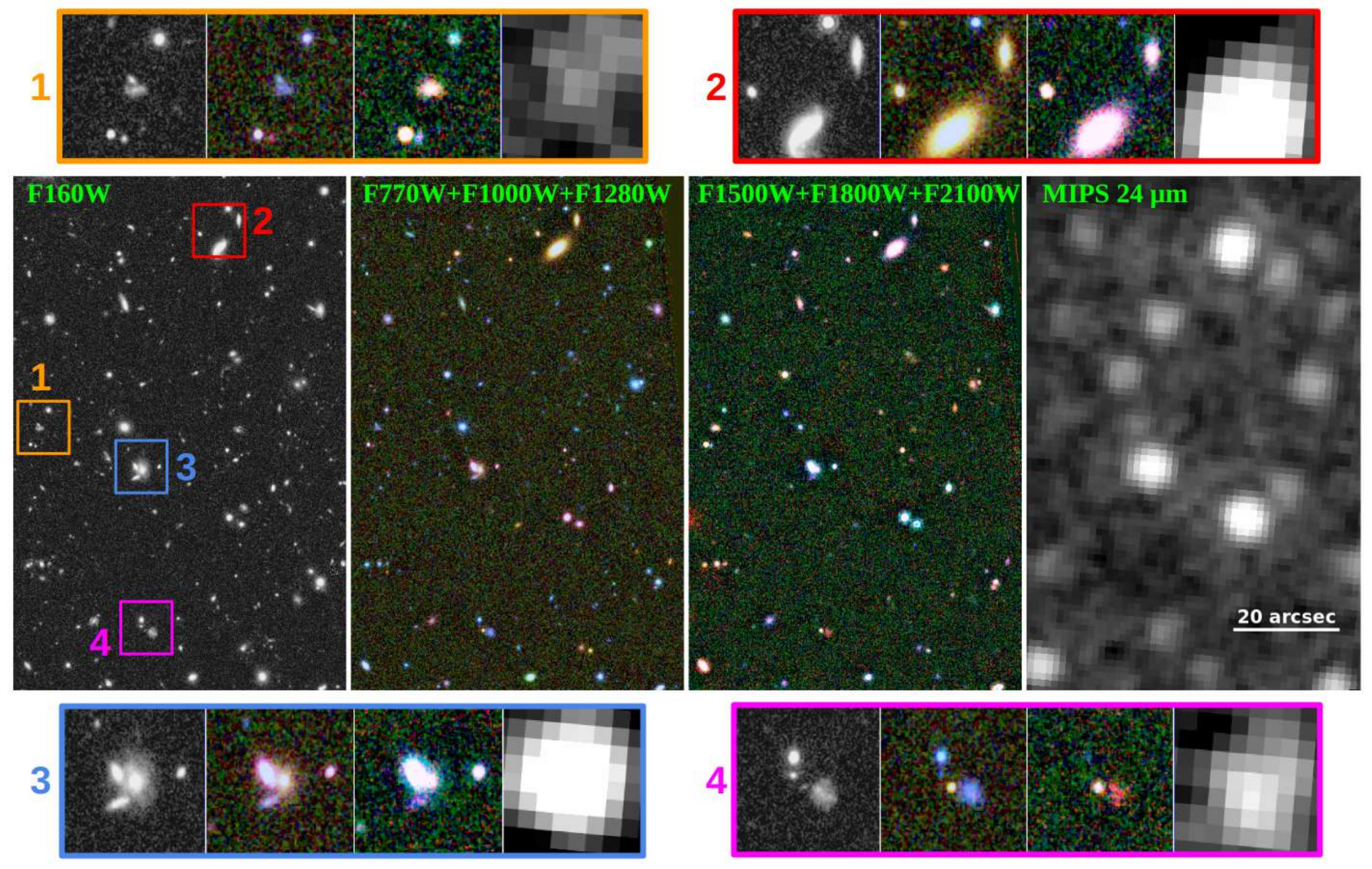}
    \caption{The images of \hst/WFC3~F160W, 
    \jwst/MIRI~F770W$+$F1000W$+$F1280W (Blue-Green-Red false color), 
    \jwst/MIRI~F1500W$+$F1800W$+$F2100W (Blue-Green-Red false color), and 
    \spitzer/MIPS~24$\mu$m.
    The F160W and MIPS images are real; the MIRI images are simulated
    The four zoom-in boxes have a size of $10\arcsec\times 10\arcsec$, 
    and highlight various examples of galaxies with different MIRI colors 
    and levels of crowding.
    The MIRI cutouts are from the simulation set of pure-galaxy models
    (\S\ref{sec:pure_gal}).
    Different MIRI colors reveal different observed-frame wavelengths 
    of the PAH features due to redshifting 
    (e.g., Figs.~\ref{fig:filters} and \ref{fig:sed_templates}).
    Note that the angular resolution of MIRI is much higher than that
    of MIPS and enables detections of galaxies previously blended by MIPS. 
    } 
    \label{fig:cutout}
\end{figure*}

For each model-input $\fracA$ (\S\ref{sec:mod_flux}), we perform an independent 
set of \mirisim\ simulations for all the six MIRI bands (from F770W to F2100W).
Fig.~\ref{fig:cutout} displays the simulated images of the pure-galaxy 
case in false (RGB) color (after the data reduction in \S\ref{sec:pipeline}).  
It is clear that the colors and morphologies from MIRI will provide an 
unprecedented view of the IR emission from distant galaxies compared to 
previous missions (e.g., MIPS 24~\micron).  

Since our simulations are based on the CANDELS F160W sources, we assume that all
MIRI sources are detected in F160W. 
We acknowledge that a new type of optically-faint MIR-bright sources could actually 
exist, and such sources may be detectable by MIRI but not by F160W.
The real MIRI images after launch will provide a great chance to explore this
unknown regime of distant galaxies.

In addition to the real CANDELS sources, we also simulated a set of bright
point sources in Appendix~\ref{app:cor} using the CEERS MIRI exposure configuration. 
These point sources are used for the purposes of data validation, photometric tests, and 
corrections (see Appendix~\ref{app:cor}).
For the photometry extraction purpose (see \S\ref{sec:tphot_prep} for details), 
we also need to simulate \hst\ F160W images using the identical galaxy morphologies 
(i.e., pure S\'ersic models) that are equivalent to  those in the simulated MIRI images.
We describe the simulated F160W images in  Appendix~\ref{app:f160w}.    

\section{Data processing and analyses}
\label{sec:data_process}

\subsection{The reduction of MIRI imaging data}
\label{sec:pipeline}
We run the \jwst\ \textsc{pipeline} (version 0.15.2) to process the \mirisim\ output of the 
level~0 data from \mirisim\ (\S\ref{sec:mirisim}).
The level~0 data for each dither is in the format of a FITS file.
There are three stages of data processing.
Stage~1 performs detector-level corrections for individual dithers, producing
count-rate maps from the ``up-the-ramp'' readouts.
Stage~2 applies physical corrections and calibrations to individual dithers, 
producing flux maps with image header containing standard World Coordinate System 
(WCS) information. 
Stage~3 combines individual dithers to a single image.

\begin{figure}
    \centering
	\includegraphics[width=\columnwidth]{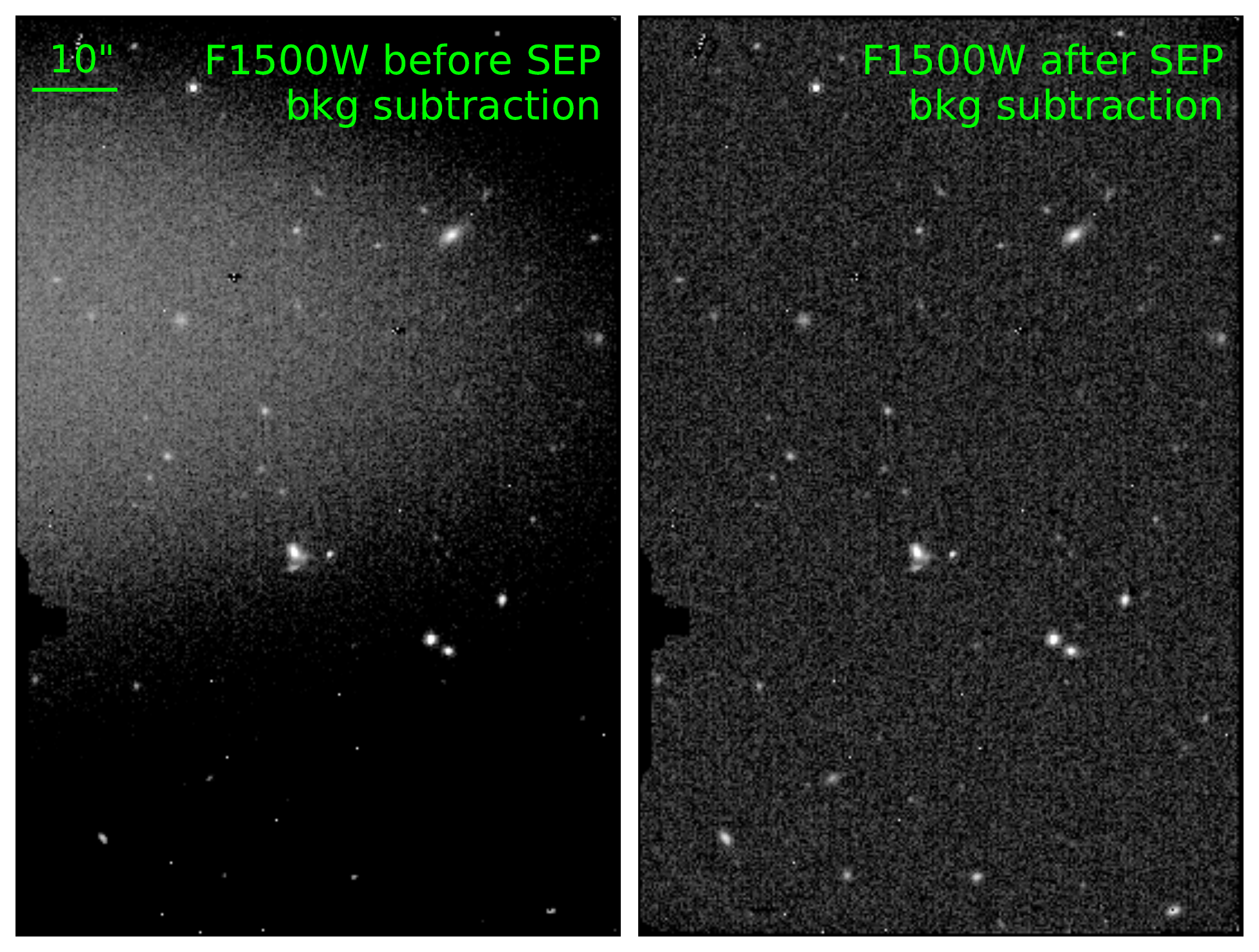}
    \caption{The F1500W (dither 1) images before (left) and after (right) 
    our background-subtraction process with {\sc sep}.
    The two images share the same arcsinh color scale.
    The background subtraction step largely removes the spatial-dependent background. 
    }
    \label{fig:bkg_maps}
\end{figure}

Stage~1 involves the rejection of cosmic rays, which are deposited onto the 
imaging data during the exposure. 
While this process successfully rejects most ($>99\%$) of the cosmic rays, it 
still misses up to $\approx 20$ cosmic rays per dither after visual 
inspection.
To clean up these outliers, we run 
{\sc astro-scrappy}\footnote{https://github.com/astropy/astroscrappy} which
implements the ``Laplacian Edge'' algorithm of cosmic-ray detection (see \citealt{van_dokkum01}), 
and mask the pixels polluted by the detected cosmic rays. 
Stage~3 involves background subtraction before merging the dithers.   
However, the current version of \textsc{pipeline} only subtracts a single 
global background value, although the actual background may have spatial 
dependence due to instrumental thermal emission and/or scattered light. 
Therefore, before merging the dithers, we perform an additional background
subtraction with {\sc sep} (version 1.0.3; \citealt{barbary16}), which 
realizes the core algorithms of {\sc source extractor} \citep{bertin96} in 
{\sc python}. 
We adopt a cell size of 32 pixels (pixel size $= 0.11\arcsec$) and a 
median-filtering size of 3 cells. 
Fig.~\ref{fig:bkg_maps} demonstrates the effect of our background subtraction.   
After this background subtraction, we find there is a small fraction of 
pixels ($\lesssim 0.1\%$) that have extremely negative values 
($<-10\times$ uncertainties) in the F1800W and F2100W images. 
The status of these pixels are not marked as abnormal 
in the associated quality map.  
Therefore, we manually mask these pixels by setting their values to zero and 
their status as abnormal.

Stage~3 provides the a final image that merges all the individual images for that band. 
We find the merged images also show traces of a residual background, 
especially for the red bands (F1800W and F2100W) where the background is stronger.
Therefore, we repeat the background subtraction step again with the same {\sc sep}
parameters as above.
This process effectively removes the residual background for each band.  
The merged images have a native pixel size of $0.11\arcsec$. 
We find that the background-subtraction procedure above does not bias the measured 
flux densities (see Appendix~\ref{app:cor}), 
so we conclude this background is mostly additive and removable by this procedure.

However, for photometry we make an additional set of MIRI mosaics at the pixel 
scale of the \hst\ F160W image.  
The simulated \hst\ F160W image has a pixel size of $0.06\arcsec$.
Our process of photometry extraction (\S\ref{sec:photo}) requires that 
MIRI and F160W images have pixel scales that are integer multiples of this pixel scale. 
Because the current version of \textsc{pipeline} does not allow a user-defined pixel 
size for the final image, we reprojected the MIRI images to the frame of the 
simulated F160W image for our photometric step (utilizing the {\sc reproject\_exact} 
function of {\sc astropy} \citep{astropy}).
The algorithm of {\sc reproject\_exact} is designed to conserve fluxes, 
so the final image is oversampled to the F160W pixel scale ($0\farcs06$/pixel where 
the fluxes are conserved. 
%
%

\subsection{The extraction of MIRI photometry}
\label{sec:photo}
We utilize {\sc tphot} {\citep[version~2.1;][]{merlin15, merlin16}} 
to measure MIRI fluxes.
{\sc tphot} performs PSF-matched photometric measurements based on a 
``template-fitting'' technique \citep{laidler07}. 
First, for each source from a given ``high-resolution'' image, it crops a template 
(i.e., a cutout containing this source) and normalize the total flux to unity.
{\sc tphot} then convolves these templates with a given PSF kernel. 
This kernel matches the PSF of the high-resolution image to that of the 
``low-resolution'' image (i.e., the image for photometric measurements).
Next, {\sc tphot} makes a model low-resolution image by convolving these templates 
with the kernel and normalizing it to have unity flux.
Finally, the code scales the templates for sources near the source of interest 
simultaneously so that the model image is an optimal match to the values in the 
input low-resolution image (through $\chi^2$ minimization).
The scaling factor for each template is the output best-fit flux for the 
corresponding source.

{We opt to use ``forced photometry'' models (specifically {\sc tphot} \citealt{merlin15, merlin16}) instead of methods that measure direct aperture photometry (e.g., {\sc source extractor} 
\citealt{bertin96}) for MIRI photometry extraction, as the former ({\sc tphot}) accounts for 
different PSF sizes in different bands without degrading the image quality and can provide flux densities measured in matched apertures for all objects detected in a catalog.
%
%
It is also possible to perform PSF-matched photometry under the ``dual image'' mode of 
{\sc source extractor}. 
However, this typically requires to degrade all images to a dataset's largest PSF 
(F2100W in our case) so that all PSFs across different bands become similar 
\citep[e.g.,][]{yang14}. 
In our case, this approach means a huge loss for the imaging quality of the blue MIRI
bands, because, e.g., the F770W PSF is $\approx 3$ times sharper than the F2100W PSF. 
Therefore, we consider that our {\sc tphot} approach is superior to this one.}

{
We note that it is possible to run {\sc source extractor} on each MIRI image 
independently, and then match the detected single-band MIRI sources with the CANDELS/EGS
catalog.  However, another advantage to using forced-photometry methods, such as {\sc TPHOT} is that they are able to measure flux densities (or upper limits) for all sources in a catalog.  This provides useful constraints on the emission (in the case here on the mid-IR emission) which can place limits on the amount of warm dust from AGN or molecular emission from star-formation on all galaxies in our catalog. 
We present a detailed comparison between the photometry of {\sc tphot} and 
{\sc source extractor} in Appendix~\ref{app:tphot_vs_se}.
The results show {\sc tphot} performs better than {\sc source extractor}, justifying our
choice of the former.
}

\subsubsection{TPHOT preparation}
\label{sec:tphot_prep}
We adopt the simulated \hst\ F160W map (Appendix~\ref{app:f160w}) as the input 
high-resolution image and the simulated MIRI maps as the low-resolution 
images.  
We adopt this step because we expect (nearly) all galaxies detected in the red-MIRI bands to be detected in the deep CANDELS \hst\ F160W imaging.
The full width at half maximum (FWHM) of F160W PSF is $0.20\arcsec$, smaller 
than those of MIRI bands (ranging from F770W $0.24\arcsec$ to F2100W 
$\rm 0.67\arcsec$).
As a reminder, for the test here we use the \textit{simulated} F160W image (using identical morphological models as used in the MIRI simulations). 
The reason to use the simulated F160W image rather than the observed one is that, 
some sources in the observed F160W image have morphologies with structures that are not represented by the simple S\'ersic profiles used to create the MIRI images. This is especially true for some large (angular-size) galaxies that show asymmetric features (tails, star-forming knots, disturbed morphologies, see Figure~\ref{fig:f160w}, for example). 
When processing the real observed MIRI images in the future, the observed F160W
image rather than the simulated one should be used, as we expect the morphologies between the real F160W and actual MIRI imaging will be more consistent than the models used here. 

Our {\sc tphot} run also needs the kernels that transform (through convolution) the F160W PSF 
(Appendix~\ref{app:f160w}) to match the MIRI PSFs.
We obtain the MIRI PSF for each band using {\sc webbpsf}.
{\sc webbpsf} can only generate PSFs with pixel sizes of 0.11$\arcsec/n$, where
$n$ is a user-defined integer.
We first generate a PSF with 0.055$\arcsec$/pixel.
We then re-project this PSF to the F160W PSF of 0.06$\arcsec$/pixel using 
{\sc reproject\_exact} {\citep{astropy}}.
Our PSFs have a format of $3\arcsec \times 3\arcsec$, which  
is much larger than the FWHM of each PSF profile (F2100W $\rm FWHM=0.67\arcsec$).
We have also tested larger PSFs (e.g., $5\arcsec \times 5\arcsec$) and the resulting 
photometry only has limited minor changes, but the {\sc tphot} run becomes significantly 
slower.

\begin{figure}
    \centering
	\includegraphics[width=\columnwidth]{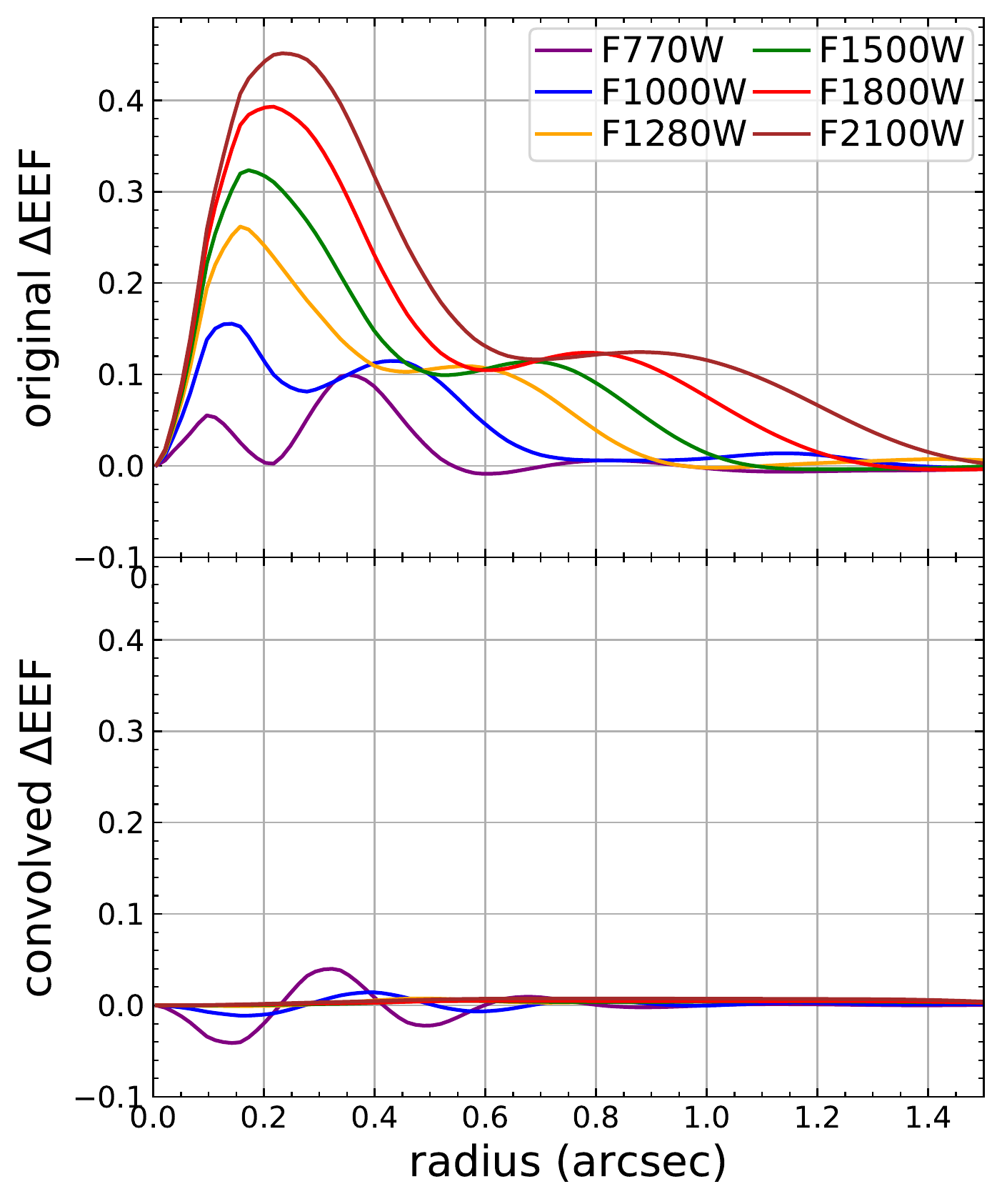}
    \caption{$\Delta$EEF (F160W$-$MIRI) as a function of redshift. 
    The top and bottom panels are for F160W before and after kernel convolution,
    respectively.
    Different colors indicate different MIRI bands, respectively.
    After convolution, $\Delta$EEF becomes significant smaller as expected. We find $\Delta$EEF is less than 4\% for all bands at all radii, and $<$1\% for all bands (except F770W).
    }
    \label{fig:eef}
\end{figure}

To derive the kernels, we utilize {\sc pypher} (version~0.6.4; \citealt{boucaud16})
which is based on Wiener filtering.
{\sc pypher} has a tunable regularization parameter ($\mu$). 
A lower $\mu$ value leads to higher fidelity at the expense of smoothness.
The fidelity can be measured as $\Delta$EEF (encircled energy fraction) 
between the convolved high-resolution PSF (F160W) and the low-resolution PSF (MIRI bands).
We normalize the total fluxes of all the input PSFs to unity before feeding them to {\sc pypher}.
We adopt a $\mu$ value of 0.03, which leads to $\Delta$EEF always below 4\% 
at any radius for all of the MIRI bands (and better than 1\% for all radii, except for F770W). 
Fig.~\ref{fig:eef} displays the $\Delta$EEF (before and after the convolution)
as a function radius for different MIRI bands. 
After the convolution, $\Delta$EEF becomes much smaller, demonstrating 
the effectiveness of the kernels constructed by {\sc pypher}.


{\sc tphot} also needs a source list and a segmentation map for the input
high-resolution image (F160W), and we run {\sc source extractor} 
{\citep{bertin96}} to generate these ingredients. 
The segmentation map describes the pixel area covered by each source, 
and the area is critical in affecting the {\sc tphot} photometry.   
A small area may miss a significant fraction of light in {\sc tphot} analysis, 
systematically leading to a lower flux.
On the other hand, a large area may include too many noise-dominated pixels, 
increasing the uncertainty of the output flux.
The segmentation map is mainly controlled by the parameters ``detect\_thresh''  
and ``detect\_minarea'' in {\sc source extractor} {\citep{bertin96}}.
We find that a setting of detect\_thresh$=0.25$ and detect\_minarea$=60$
is appropriate in producing optimal {\sc tphot} photometry 
(see \S\ref{sec:tphot_res}), and we adopt this setting.
We note that a typical configuration (i.e., higher detect\_thresh and 
lower detect\_minarea) would lead to a significant flux underestimation
due to the missing of faint wings.
Our {\sc source extractor} run detects 545 sources.
We associate these sources with the simulation input catalog (i.e., 
the CANDELS/EGS catalog of 463 sources; see \S\ref{sec:mod_flux}) using a 
$1\arcsec$ matching radius, and find 388 matches. 
Most of the 75 ($463-388$) undetected sources are faint with $H_{160}>26$, and only 
a few of them would be MIRI-detected given their expected low MIRI fluxes (the 
pure-galaxy models; \S\ref{sec:tphot_res}).
The extra 157 ($545-388$) sources detected by {\sc source extractor} are noise 
likely due to our low-detect\_thresh setting.  
This is acceptable as our main goal is to study the recovered MIRI flux densities 
for the majority of the (brighter) sources detected in the F160W image. 
We therefore exclude the sources with no associated source in the input 
CANDELS/EGS catalog.

\subsubsection{TPHOT results}
\label{sec:tphot_res}
\begin{figure*}
    \centering
	\includegraphics[width=2\columnwidth]{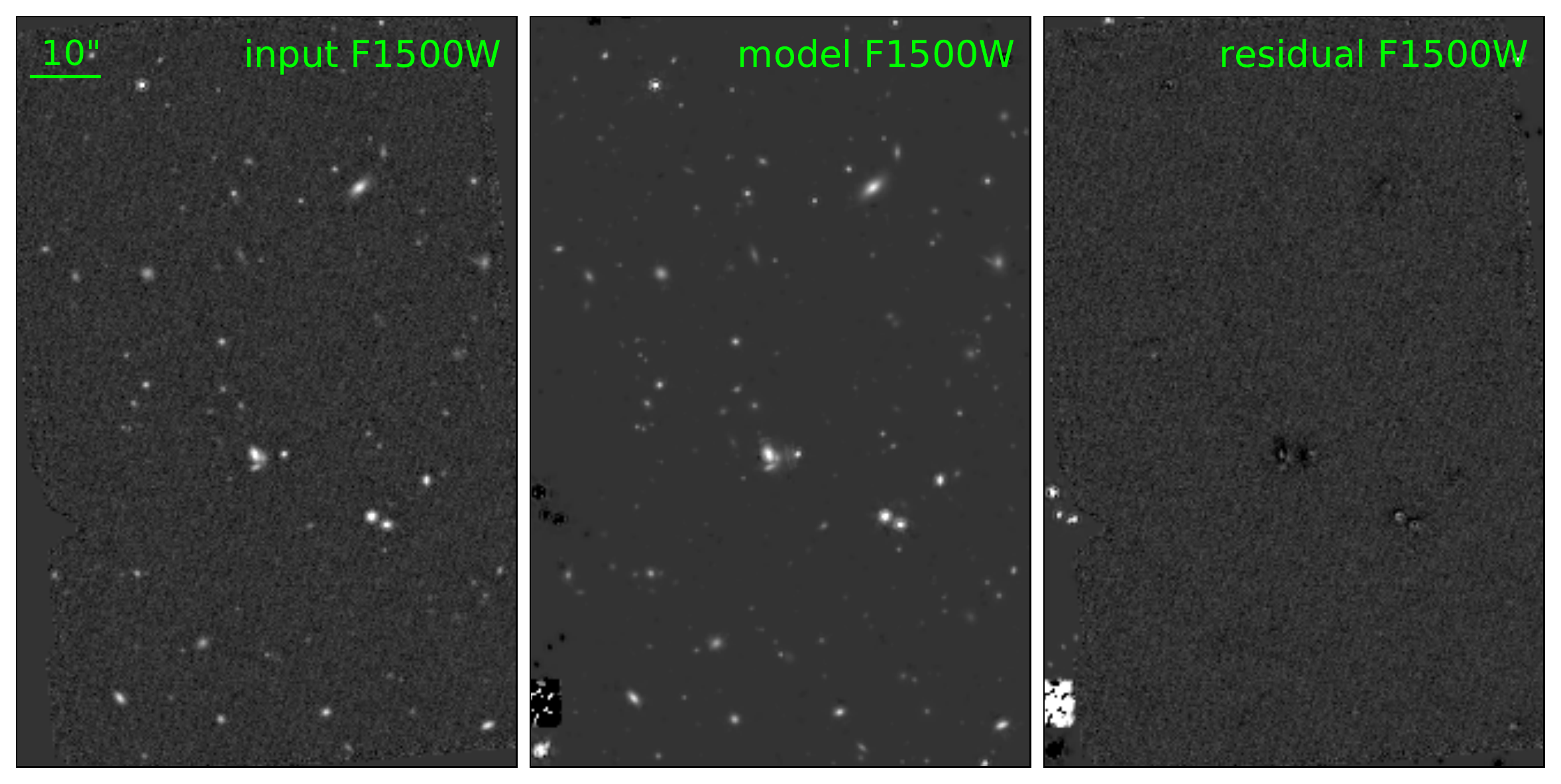}
    \caption{The {\sc tphot} input low-resolution map (left), model map (middle), 
    and residual (input~$-$~model) map (right) for the F1500W band (pure-galaxy 
    models; \S\ref{sec:pure_gal}).
    All images have the same arcsinh color scale.
    The few bright/dark spots close to the edges have poor F1500W coverage, and thus 
    are unconstrained by the MIRI data. 
    Sources in these regions are flagged by {\sc tphot} and we exclude them from the analysis.
    Except at these spots, the residual map is largely dominated by random noise, 
    indicating that the fitting of {\sc tphot} is generally acceptable.
    }
    \label{fig:tphot_maps}
\end{figure*}

We run {\sc tphot} {\citep{merlin15, merlin16}} on each MIRI 
band individually, 
while using the simulated F160W image and associated {\sc source extractor} 
results as high-resolution prior (\S\ref{sec:tphot_prep}).
Fig.~\ref{fig:tphot_maps} shows the input, model, and residual maps for F1500W
(pure-galaxy model; \S\ref{sec:pure_gal}) as an example.
Overall, the model map looks very similar to the input map, and the residual 
map is dominated by noise (except at some spots outside of MIRI detector 
coverage).
This result indicates that {\sc tphot} successfully models the MIRI galaxy 
profiles with PSF matching.
Some sources close to the image edges are poorly constrained due to the lack 
of good imaging coverage. 
These objects are flagged by {\sc tphot}, and we exclude them from further 
analysis for this reason.
%

\begin{figure*}
    \centering
	\includegraphics[width=2\columnwidth]{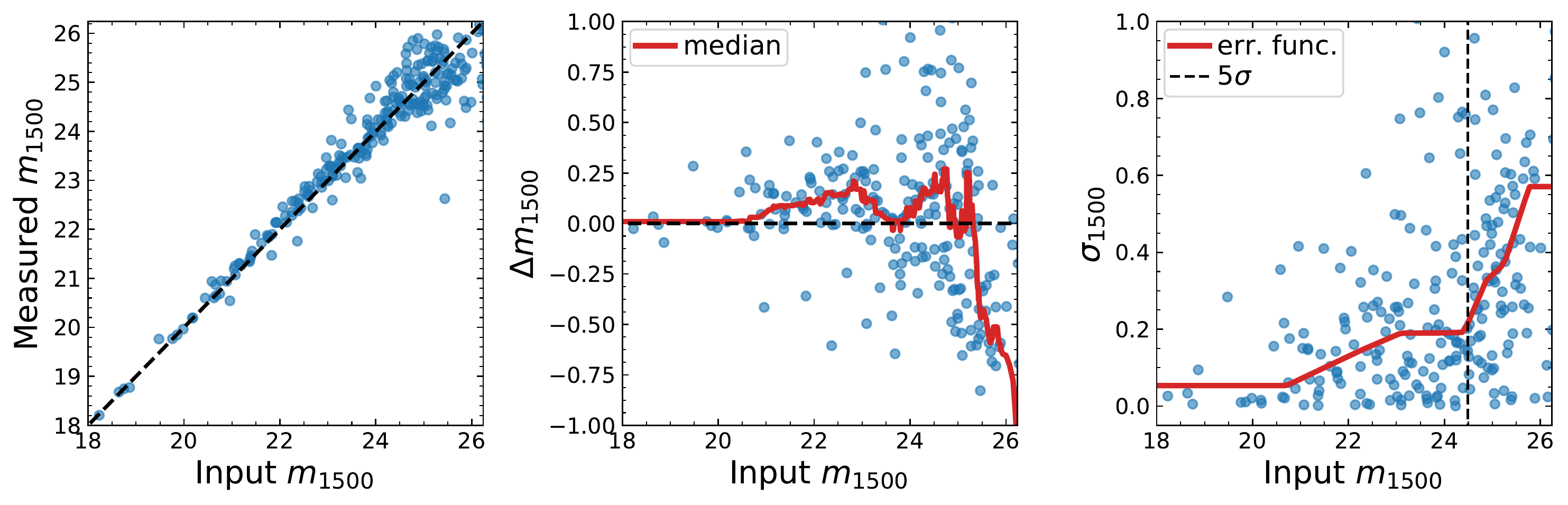}
    \caption{\textit{Left}: 
             {\sc tphot}-measured F1500W AB magnitude versus the input model 
             magnitude in \mirisim\ simulation (\S\ref{sec:mirisim}).
             The black dashed line represents the equality relation (measured = input).
            \textit{Middle}:
             Measured minus input F1500W magnitudes ($\Delta m_{1500}$) versus the 
             input magnitude.   
             The red curve represents the running median of the data points
             with 20 sources per bin.
             The black dashed line indicates an offset of zero.
            \textit{Right}: 
             The absolute value of $\sigma_{1500} = |\Delta m_{1500}|$  
             versus the input magnitude.
             The red curve represents our error function. 
             The black dashed vertical line represents our estimated $5\sigma$ 
             limiting magnitude where $f/\Delta f=5$ from our error function.
             F1500W is shown here as an example, and the full version of this plot 
             including all MIRI bands can be found in Appendix~\ref{app:tphot_mags}.
    }
    \label{fig:tphot_mags}
\end{figure*}

In Fig.~\ref{fig:tphot_mags}, we compare the {\sc tphot}-measured and input 
magnitudes (also for the F1500W of pure-galaxy model as an example). 
From Fig.~\ref{fig:tphot_mags} (middle), the measured magnitudes tend to 
be slightly fainter than the input magnitudes (except for the faint tail of
$m_{1500}\gtrsim 25$). 
The cause could be that we miss some light from the faint wing component 
of the source profiles.
To investigate this possibility, we divide the sample by (input) S\'ersic index.
We re-plot Fig.~\ref{fig:tphot_mags} (middle) for S\'ersic $n<2$ and point 
sources versus $n\geq2$ sources, respectively, in Fig.~\ref{fig:photo_vs_n} 
(see Appendix~\ref{app:tphot_mags} for the full version including all bands).
The latter have more significant wing component than the former 
(see, e.g., \citealt{haussler07}). 
From Fig.~\ref{fig:photo_vs_n}, the measured magnitudes for $n<2$ and point 
sources do not suffer from significant systematics, but those for $n\geq 2$ 
sources show a bias of up to several tenths of a magnitude.  
This result indicates that the light loss due to faint wings indeed causes 
systematic offset in photometry measurements, and care is needed to interpret 
the flux densities of these objects. 
Currently these objects constitute a minority of the sources (only 21\% of 
the $>5\sigma$ F1500W sources).  
As we do not know what the morphologies of MIRI sources will be, we defer 
more detailed analysis to a future work. 

In Fig.~\ref{fig:photo_vs_n}, we indicate crowded sources ($m_{\rm 1500, src}$) 
having at least one neighbor (within $2\arcsec$ radius) brighter than 
$m_{\rm 1500, src}+1$~mag.  
These are objects where flux residuals from the (relatively) bright companion 
can bias the flux measurements of the sources.
From Fig.~\ref{fig:photo_vs_n}, the measured photometry of these sources is 
not always deviant from the distribution, although we notice that these sources 
do have a larger scatter ($\smad=1.48\times{\rm MAD}$, where MAD is the median 
absolute deviation) in $\Delta m_{1500}$ compared to isolated sources 
(0.23 vs.\ 0.17 for the $>5\sigma$ sources in Fig.~\ref{fig:photo_vs_n} top). 
The sources with neighbors only contribute to a small fraction of the $>5\sigma$ 
population (e.g., 15\% for F1500W).
Therefore, we conclude that source crowding does not significantly affect the
quality of our measured photometry, and it is clearly superior to previous 
instruments thanks to the high angular resolution of MIRI (see, e.g., 
Fig.~\ref{fig:cutout}).
 
The photometric errors in the {\sc tphot} output are based on the
$\chi^2$ fitting of low-resolution source profiles (\S\ref{sec:photo}).
We find that these errors are often significantly smaller (up to a factor of 
$\gtrsim 10$) than the actual differences between the measured and input magnitudes.
This underestimation of uncertainties could be possibly due to the assumption
of pixel-noise is uncorrelated in {\sc tphot} {\citep{merlin15, merlin16}}.
We do not adopt the {\sc tphot} uncertainties.
Instead, we establish an empirical ``error function'' for each MIRI band based on the distribution of input-to-measured flux densities. 
First, we bin our sources on their input MIRI magnitudes with each bin containing
30 sources.
For each bin, we calculate the median input magnitude and the median difference 
between the measured and input magnitudes.  
Finally, we obtain the error function by linearly interpolating the median difference 
as a function of median magnitudes (see Fig.~\ref{fig:tphot_mags},  right). 
Although the error function for each band is derived based on the simulation set of 
pure-galaxy models (\S\ref{sec:pure_gal}), we also apply it to the other simulation
sets with non-zero $\fracA$ (\S\ref{sec:mod_agn}).
From Fig.~\ref{fig:tphot_mags} right, the uncertainty is $\approx 0.05$~mag 
at the bright end, and rises toward faint sources as expected. 
For each source, we estimate the magnitude uncertainty by evaluating the error 
function on the measured magnitude.

\begin{figure}
    \centering
	\includegraphics[width=\columnwidth]{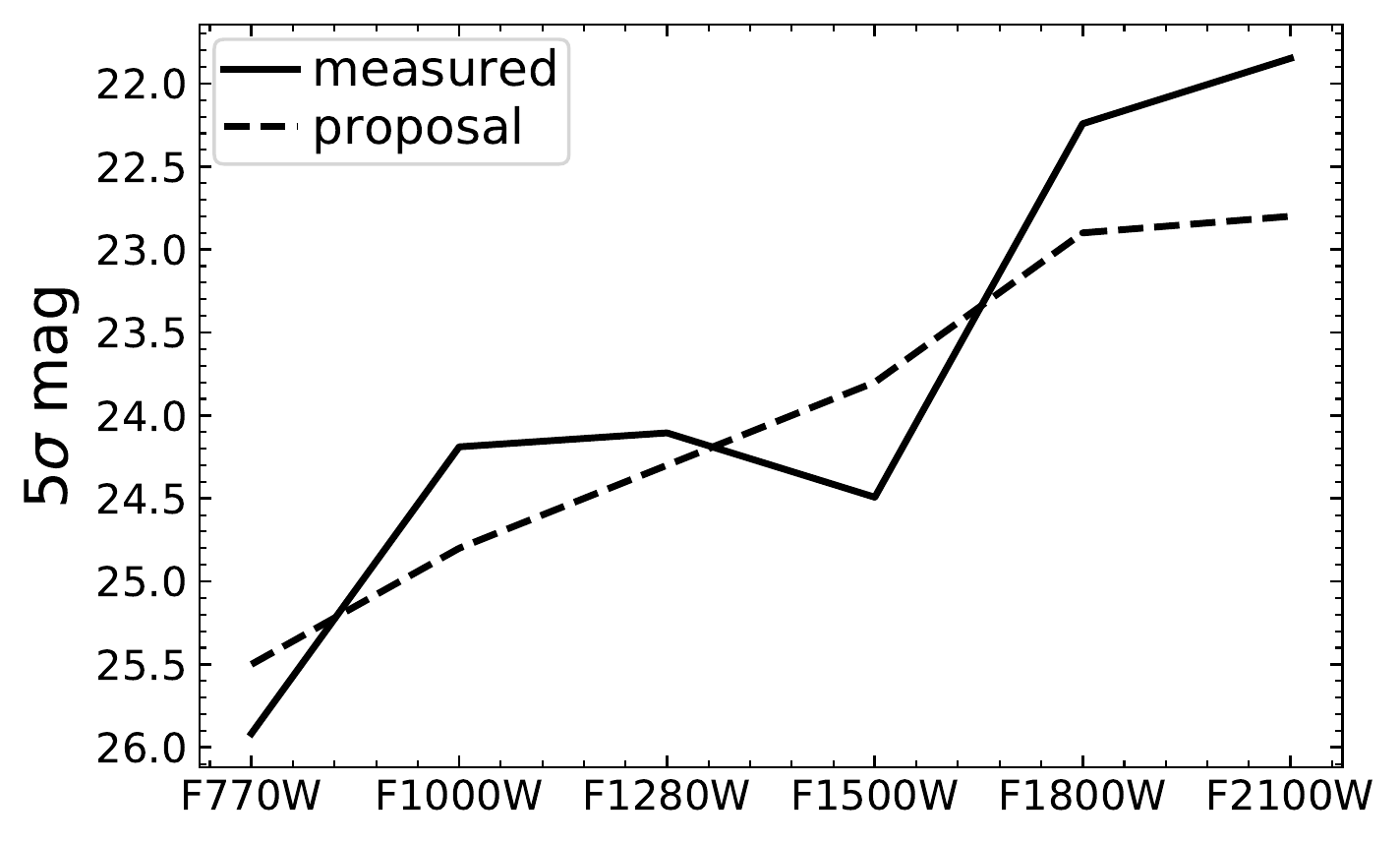}
    \caption{The $5\sigma$ limiting magnitudes for different bands 
    derived from our empirical error function.
    The dashed line represents limiting magnitudes for the CEERS configuration \citep{finkelstein17} derived from the \jwst/ETC.
    Our measured limiting magnitudes are similar to the expected limits, 
    where we interpret offsets as being a limitation of the relatively small sample. 
    }
    \label{fig:lim_mag}
\end{figure}

We estimate the 5$\sigma$ limiting magnitude for each band as the value where 
the error function equals 0.217 mag, corresponding to $\delta f/f=0.2$ 
(assuming standard error propagation, $\delta m = 1.086\times  \delta f/f$). 
Fig.~\ref{fig:lim_mag} compares these measured limiting magnitudes and those 
expected for CEERS \citep{finkelstein17} based on \jwst\ Exposure Time 
Calculator (ETC), assuming a point-source profile.
From Fig.~\ref{fig:lim_mag}, our measured $5\sigma$ depths are similar to 
those in the CEERS proposal {(differences $<1$~mag)}, where we interpret 
offsets as a result of limited sample sizes, our more realistic treatment of 
source surface brightness profiles and source crowding, {and different 
background-noise assumptions of \mirisim\ (spatially dependent; 
\S\ref{sec:mirisim}) versus ETC (uniform).
We consider the last factor is mainly responsible for the relative large 
difference between the F2100W depths of \mirisim\ and ETC (0.95~mag; 
Fig.~\ref{fig:lim_mag}), as the F2100W has the strongest background in 
our filter set. 
However, whether the background is spatially dependent or not can only be 
determined from the real MIRI imaging data after launch. 
For now, we caution that the ETC S/N could be over-optimistic for 
F2100W (and also F2550W for which background is even stronger than F2100W),
because ETC does not account for the spatial dependence of background. 
}

%

\begin{figure}
    \centering
	\includegraphics[width=\columnwidth]{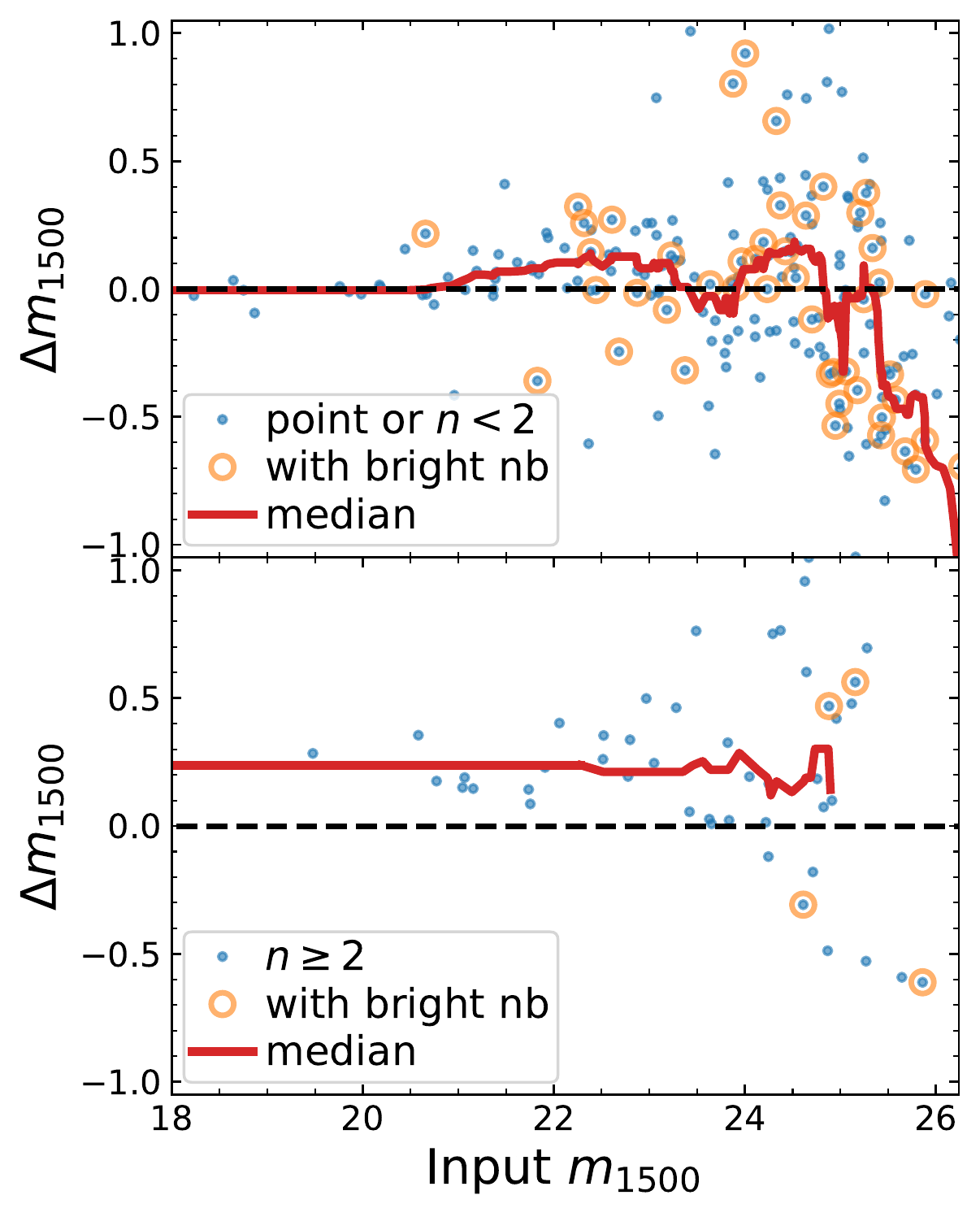}
    \caption{The difference between measured and input F1500W magnitudes 
             ($\Delta m_{1500}$) versus the input magnitude, separated by S\'ersic index. 
             The red curves represent the running median.
             The orange circles highlight the sources ($m_{\rm 1500, src}$) that 
             have at least one neighbor (``nb", within $2\arcsec$ radius) brighter 
             than $m_{\rm 1500, src}+1$~mag.
             The upper panel is for S\'ersic $n<2$ and point sources, 
             and the lower panel is for $n>2$ sources.
             For the former, the measured magnitudes do not systematically
             deviate from the input magnitudes, but this is not true for the
             latter. 
             This systematics for $n>2$ sources is likely due to their extended 
             faint wings.
             The full version of this plot including all MIRI bands can be found 
             in Appendix~\ref{app:tphot_mags}.
    }
    \label{fig:photo_vs_n}
\end{figure}

\subsection{The SED fitting}
\label{sec:sed_fit}
We again use \xcig\ to perform SED fitting of the galaxies' (simulated) 
photometry that includes measurement errors on the photometry.  
To assess the effects of MIRI data, we focus the sources having 
significant ($>5\sigma$) detections in at least two MIRI bands 
(Fig.~\ref{fig:Lir_vs_z}).
For the simulation set of pure-galaxy models (\S\ref{sec:pure_gal}), these 
MIRI sources consist of 53\% and 75\% of the $H_{160}<26$ and $H_{160}<25$ population, 
respectively. 
For the simulation sets of AGN-galaxy models (\S\ref{sec:mod_agn}), the 
detection rates are even higher due to the additional AGN contribution.
In contrast, the MIPS~$24\mu$m detections rates are only 5\% ($H_{160}<26$) and 
9\% ($H_{160}<25$) in the real data.
The high detection rates of MIRI highlight its superior flux sensitivity.

For the currently existing non-MIRI bands, we cannot use the actually observed 
photometry directly in the SED fitting.
The reason is that the observed photometry is not entirely consistent with 
that expected from the SED models used to generate model-input MIRI fluxes 
(\S\ref{sec:mod_flux}).
This is especially a serious issue for the models of high $\fracA$. 
When we add a strong {hypothetical} AGN component to the galaxy SED 
(\S\ref{sec:mod_agn}), the model-expected IRAC and MIPS fluxes could be 
significantly higher than the observed values.
{Another issue is that our adopted model-input redshifts are mostly 
photo-$z$ from the CANDELS/EGS catalog (\S\ref{sec:mod_flux}), and these
photo-$z$ are just an approximation, not the ``true'' values.
However, to assess the accuracy of our SED-fitting results, we need the true 
redshifts instead of approximated ones 
(\S\ref{sec:xcig_res_gal} and \S\ref{sec:xcig_res_agn})}.

To deal with the issues above, we use a mock catalog for each simulation set 
(\S\ref{sec:mod_flux}), 
which is built following the technique in \cite{boquien19}.
First, we convolve the model SED with the filter transmissions to obtain the 
model flux for each source.
We then perturb the model flux with a Gaussian fluctuation ($\sigma=$ observed 
flux error in the CANDELS/EGS catalog) for each non-MIRI band. 
We adopt the perturbed fluxes in the \xcig\ input, while keeping the original 
flux uncertainties.
{The mock catalog assumes the model-input SED parameters such as redshift 
and ${\fracA}$ are the ``true'' values solving the aforementioned 
issue}, while it also perturbs the photometry to account for observational 
uncertainties.
We note that our use of the mock, model photometry is only to gauge the ability 
of our methods to recover input parameters in this work.  
For real MIRI data after \jwst\ launch, the observed non-MIRI photometry rather 
than the mock one should be used.


As noted previously, one benefit of \xcig\ is that redshift and other source properties
can be modeled simultaneously, although users can also fix the redshift
at a given value such as spec-$z$.
This feature is extremely valuable for deep-field sources whose spec-$z$
are challenging to measure.

The \xcig\ run time for our simulated MIRI pointing (a few hundred sources)    
is $\lesssim 1$ hour on a typical desktop/laptop. 
The run time will still be acceptable ($\lesssim 1$ day) even for a large 
sample of a million sources, thanks to the efficient parallel algorithm 
of \xcig\ \citep{boquien19}.
%

\subsubsection{\xcig\ configurations}
\label{sec:xcig_config}
The \xcig\ parameters for our fittings are summarized in 
Table~\ref{tab:xcig_par_miri}.
The model parameters for the galaxy properties are the same as in 
Table~\ref{tab:xcig_par}. 
For the AGN component, we do not explore the full parameter space
{(9 parameters; \S\ref{sec:mod_agn})}, 
because many parameters (such as radial profile index and polar-dust 
[PD] emissivity) only have limited effects on the AGN SED. 
Also, adopting all the possible parameters would take too much 
unnecessary computational time for practical purposes. 
Therefore, we only allow multiple values for {4} key 
parameters (i.e., $\fracA$, $\tau_{9.7}$, PD $E(B-V)$, and PD 
temperature) that significantly affect the AGN SED, while fixing other 
parameters at the single default value.
Also, we allow the redshift to be a free parameter in \xcig\ over a grid 
of $z$=\hbox{0.01--6} (in steps of $\Delta \ln(1+z)=0.03$).
Therefore, the Bayesian analysis of \xcig\ also considers the
uncertainties of redshift when analyzing other source properties 
such as $\fracA$, because the analysis is performed on the entire 
parameter space.
{There are a total of 10 free parameters in our fitting
(4 for galaxy, 4 for AGN, one for redshift, and one for SED 
normalization).
We have 21 photometry data points (15 real existing bands and 
6 simulated MIRI bands).}


With the configurations above, we run \xcig\ twice.
First, we run \xcig\ with the currently existing photometry only 
(\S\ref{sec:mod_flux}). 
Second, we run \xcig\ with the photometry from the mock catalogs 
(\S\ref{sec:sed_fit}) and the simulated MIRI photometry 
(\S\ref{sec:tphot_res}).
We adopt the Bayesian (rather than best-fit) values of the source 
properties (such as redshift and $\fracA$).
Unlike the best-fit value, the Bayesian value properly 
considers all of the models weighted by their probabilities. 
The \xcig\ runs are performed for the six simulation sets of 
pure-galaxy (\S\ref{sec:xcig_res_gal}) and galaxy-AGN mixed 
models (\S\ref{sec:xcig_res_agn}).
Therefore, there are a total of $12 (=2\times 6)$ \xcig\ runs.

\begin{table*}
\centering
\caption{\xcig\ parameters for the fitting of simulated data}
\label{tab:xcig_par_miri}
\begin{tabular}{lll} \hline\hline
Module & Parameter & Values \\
\hline
    \multirow{2}{*}{\shortstack[l]{Star formation history\\
                                  $\mathrm{SFR}\propto t \exp(-t/\tau)$ }}
    & $\tau$ (Gyr) & 0.1, 0.5, 1, 5 \\
    & $t$ (Gyr) & 0.5, 1, 3, 5, 7 \\
\hline
\multirow{2}{*}{\shortstack[l]{Simple stellar population\\ \cite{bruzual03}}}
    & IMF & \cite{chabrier03} \\
    &     &  \\
\hline
\multirow{3}{*}{\shortstack[l]{Stellar extinction\\ \cite{calzetti00} \\ \cite{leitherer02} }}
    & \multirow{3}{*}{$E(B-V)$} 
        & \multirow{3}{*}{\shortstack[l]{0.1, 0.2, 0.3, 0.4, \\ 0.5, 0.7, 0.9}} \\
    \\
    & 
    \\
\hline
\multirow{2}{*}{\shortstack[l]{Galactic dust emission\\ \cite{dale14}}}
    & Radiation $\alpha$ slope 
    & 1.0, 1.5, 2.0, 2.5 \\
    \\
\hline
    \multirow{7}{*}{\shortstack[l]{AGN (UV-to-IR) \\ SKIRTOR}}
    & Torus $\tau_{9.7}$ & 3, 5, 7, 9, 11 \\
    & Torus angle & 40$^\circ$ \\
    & Viewing angle & 70$^\circ$ \\
    & $\fracA$ & 0--0.99 (step 0.1) \\
    & Polar-dust extinction law & SMC \\
    & Polar-dust $E(B-V)$ & 0--0.3 (step 0.1) \\
    & Polar-dust temperature (K) & 100, 200, 300\\
\hline
    \multirow{2}{*}{Redshifting} & $z$ & 0.01--6 (step \\ 
                                 &     & $\Delta \ln(1+z)=0.03$ \\
\hline
\end{tabular}
\begin{flushleft}
{\sc Note.} --- For parameters not listed here, we use \xcig\ default values.
\end{flushleft}
\end{table*}

\subsubsection{Fitting results for pure-galaxy models}
\label{sec:xcig_res_gal}
In this section, we present the SED-fitting results for the simulation set
of {realistic} pure-galaxy models (i.e., $\fracA=0$; \S\ref{sec:pure_gal}).
Note that ``pure-galaxy'' here refers to the input models that generate the 
MIRI photometry. 
In the fitting (\S\ref{sec:xcig_config}), we allow both zero and positive 
$\fracA$ (Table~\ref{tab:xcig_par_miri}), because we do not know if a 
MIRI-observed galaxy hosts AGN or not in realistic cases.
We remind the readers that the output source properties (such as $\fracA$) 
are continuously distributed, although the parameters in 
Table~\ref{tab:xcig_par_miri} are discrete.
This is because the Bayesian results are PDF-weighted means of the parametric 
grid values (see \S\ref{sec:xcig_config} and \S4.3 of \citealt{boquien19}).
As stated in \S\ref{sec:sed_fit}, we focus on sources that have $>5\sigma$ 
detections in $\geq 2$ MIRI bands.

\begin{figure}
    \centering
	\includegraphics[width=\columnwidth]{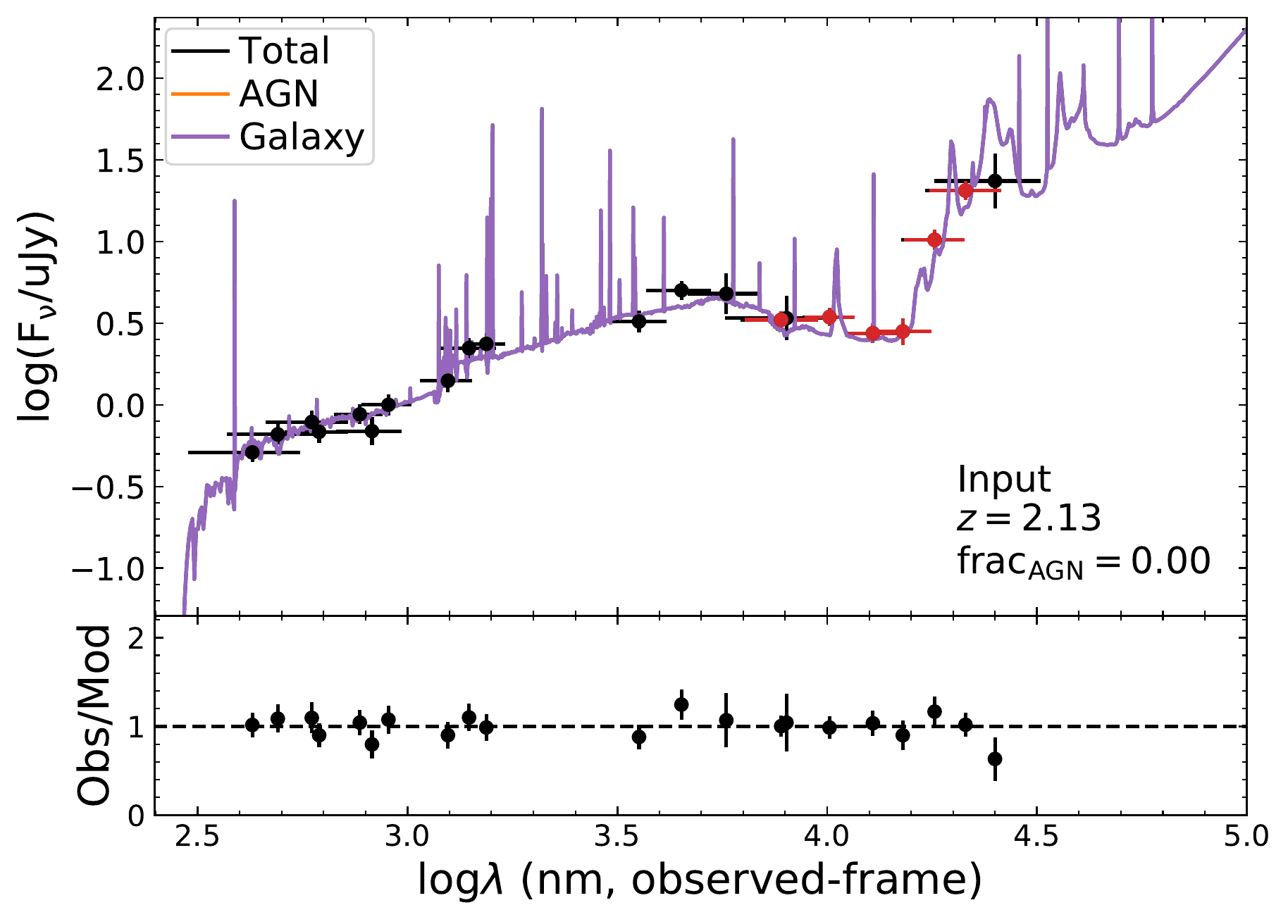}
	\includegraphics[width=\columnwidth]{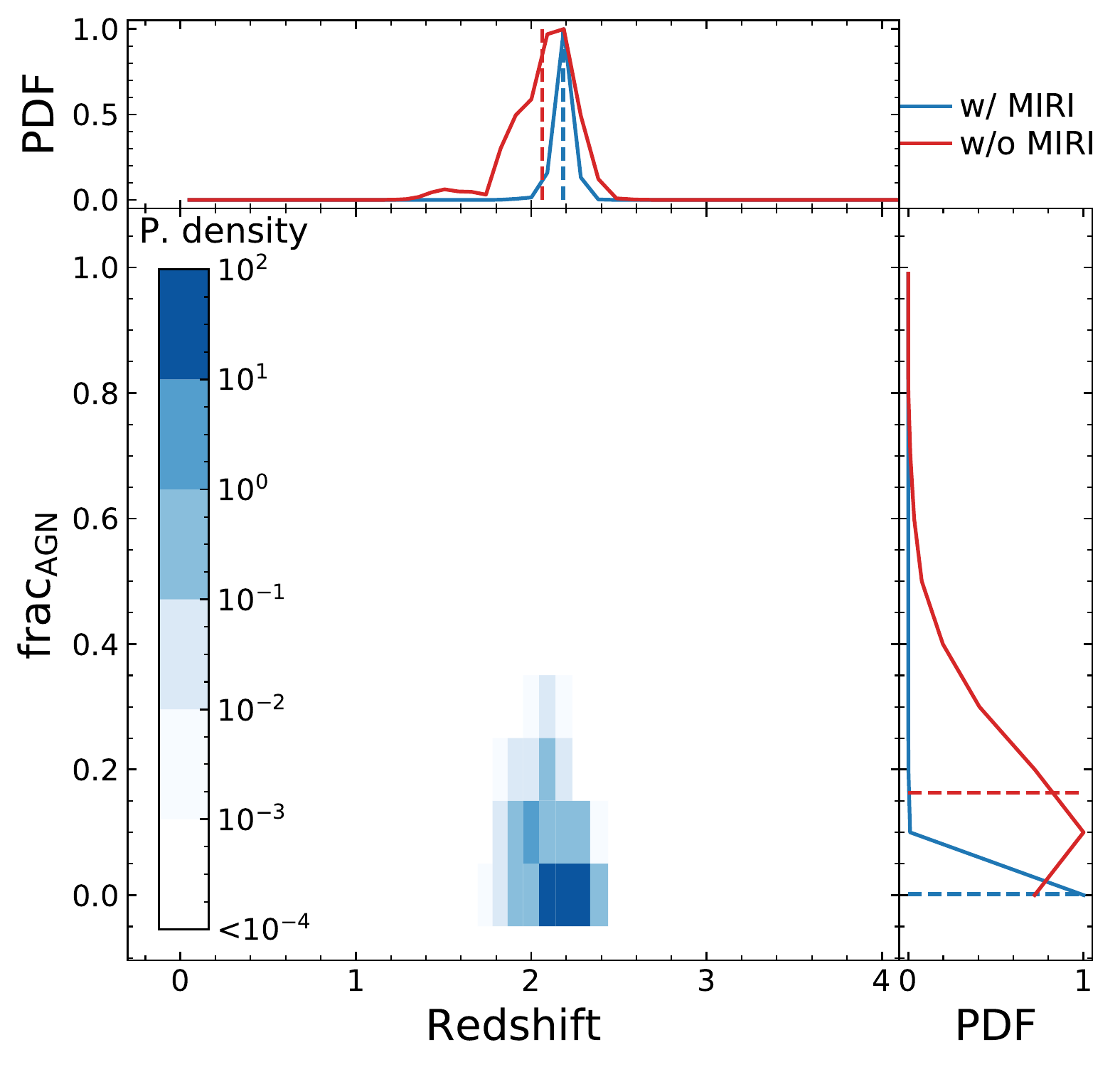}
    \caption{\textit{Top}: An example SED fitting and residual of a source in 
    the mock catalog. 
    The galaxy has (true) input $z=2.13$ and $\fracA=0$ (i.e., zero AGN component). 
    The orange and purple curves indicate AGN and galaxy components, respectively.
    Since the best-fit $\fracA$ is 0 for this source, the galaxy component overlaps 
    with the total SED.
    The MIRI data points are highlighted in red. 
    The MIRI bands cover the PAH emission features, which are redshift indicators and constrain the nature of the IR emission.
    \textit{Bottom}: The 2D PDF and 1D PDFs of redshift and $\fracA$ for the source 
    on the top panel. 
    The 2D PDF density is scaled such that the integral over the 2D plane equals 
    unity. 
    The blue and red solid curves are from the fits with and without MIRI 
    bands, respectively, as labeled.
    The dashed line indicates the Bayesian values from \xcig\ output.
    The MIRI data is helpful in constraining redshift and $\fracA$ more tightly. 
    }
    \label{fig:exm1}
\end{figure}

Fig.~\ref{fig:exm1} displays the fitting results for an example source. 
Note that the MIRI bands cover the PAH emission feature, which can be used as 
a robust redshift indicator and constrains the nature of the IR emission. 
Including the MIRI data greatly improves both redshift and $\fracA$ constraints, 
compared to the fitting results without MIRI.  
In particular, MIRI facilities a constraint on $\fracA$, where data that lack 
MIRI provide almost no information. 
This is because MIRI covers the shape of the emission associated with the 
hot-dust heated by the AGN (\S\ref{sec:intro}).

\begin{figure*}
    \centering
    \includegraphics[width=1.6\columnwidth]{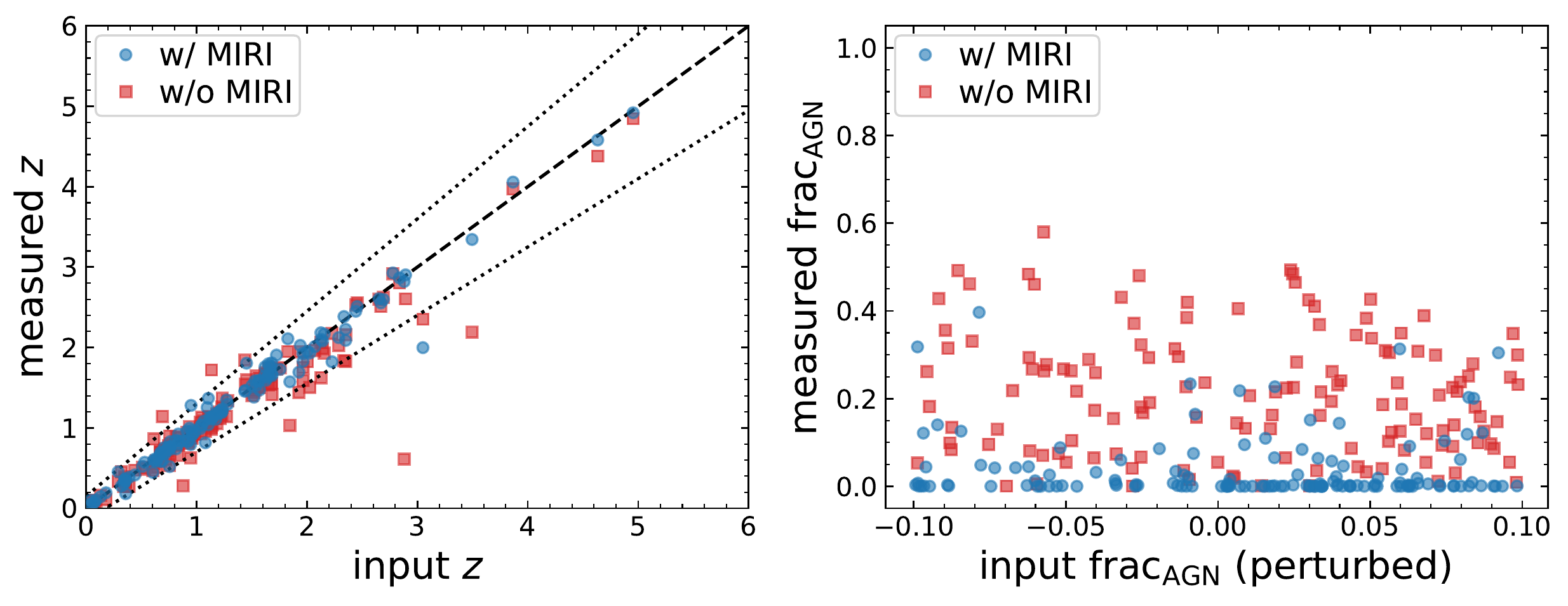} \\
    \includegraphics[width=1.6\columnwidth]{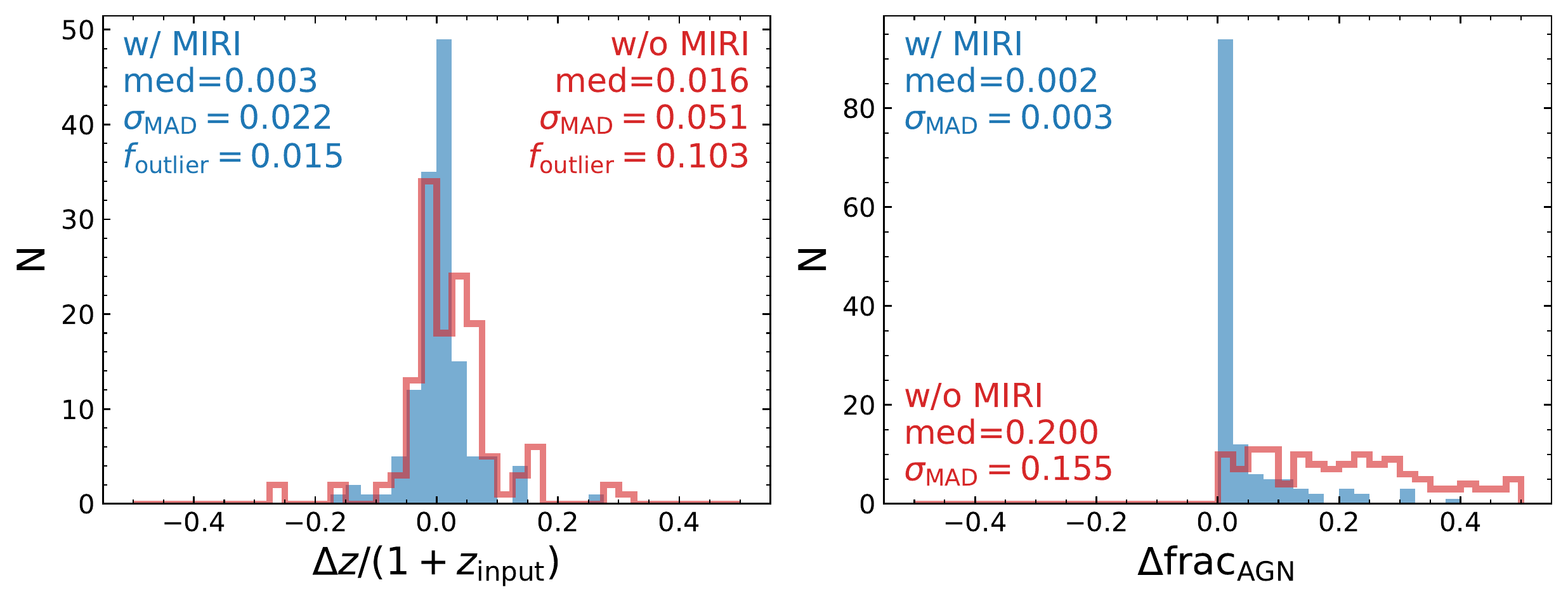}
    \caption{\textit{Top}: The measured source properties versus the input source properties 
             in the \mirisim\ simulation, where we only use galaxy component for the input 
             MIRI fluxes (\S\ref{sec:mirisim}).
             The left and right panels are for redshift and $\fracA$, respectively.
             The blue points and red squares are from the SED fits with and without MIRI 
             photometry, respectively.
             On the left panel, the black dashed line indicates equality (input $z$ = measured $z$), and the 
             black dotted lines indicate a 15\% redshift uncertainty.
             On the right panel, the $x$-axis positions are randomly perturbed for 
             display purposes only, as the input $\fracA$ is 0.
             \textit{Bottom}: The distributions of 
             $(z_{\rm fit}-z_{\rm input})/(1+z_{\rm input})$ (left) and 
             $\rm frac_{AGN, fit}-frac_{AGN, input}$ (right).
             The median and $\smad$ values of these distributions are labeled.
             On the left panel, we also label the fraction of redshift outliers
             ($|\Delta z|/(1+z) > 0.15$).
             With MIRI photometry, both of the uncertainties of redshift and 
             $\fracA$ smaller.
    }
    \label{fig:ms_vs_true}
\end{figure*}

In Fig.~\ref{fig:ms_vs_true}, we compare the \xcig\ redshift and $\fracA$ with 
the input values (\S\ref{sec:mod_flux}).
Following the convention in the literature \citep[e.g.,][]{yang14}, we adopt 
fractional redshift uncertainty, $\Delta z / (1+z_{\rm input})$, where 
$\Delta z = z_{\rm fit}-z_{\rm input}$ and $z_{\rm fit}$ and $z_{\rm input}$ are
\xcig\ output and input-model redshifts, respectively.
We calculate the median and $\smad$ of $\Delta z / (1+z_{\rm input})$ as well 
as the outlier fraction (sources with $|\Delta z| / (1+z_{\rm input}) > 0.15$).
These results are marked in Fig.~\ref{fig:ms_vs_true}.
All of the three quantities (median, $\smad$, and $f_{\rm outlier}$) are improved 
significantly after using MIRI.
Notably, $f_{\rm outlier}$ drops from 10\% to 1.5\% thanks to MIRI.

{One factor for the photo-$z$ improvement is likely 
the capture of PAH emission by MIRI (e.g., Fig.~\ref{fig:filters}).
We note that the typical PAH emission is sufficiently strong to be 
detected by our MIRI filters. 
For example, the 6.2~${\mu}$m line has a typical rest-frame equivalent 
width (EW) of ${\approx 0.5\ \mu}$m in our galaxy-dust models 
of \cite{dale14} (see, e.g., \citealt{armus07, spoon07} for similar 
values). 
This EW translates to a flux excess of 
${\Delta m_{1500} \approx 0.4}$ for a ${z=1.5}$ galaxy 
(Fig.~\ref{fig:filters}).\footnote{{$\Delta m \approx 
-2.5\log [1+\frac{W_0(1+z)}{\Delta \lambda}]$ where $W_0$ is the rest-frame
EW and $\Delta \lambda$ is the filter width \citep[e.g.,][]{papovich01}.}}
This ${\Delta m_{1500}}$ value is significantly larger than our
${>5\sigma}$ sources' photometric uncertainties 
(${\approx 0.05}$--0.22~mag; \S\ref{sec:tphot_res}).
The $7.7\ \mu$m PAH EW is typically $\approx 3$~times wider than the 
$6.2\ \mu$m one \citep[e.g.,][]{stierwalt14}, 
and thus the former will even have a stronger impact on our 
MIRI photometry than the latter.
We note that the observed EW grows with redshift as $(1+z)$ so the effects 
of PAHs on the MIRI bands are more substantive toward higher redshift.
}

We also calculate the median and $\smad$ values of $\Delta \fracA$ 
as shown in Fig.~\ref{fig:ms_vs_true}.
With MIRI, the median and $\smad$ of $\Delta \fracA$ are 0.002 and 0.003, respectively, 
both close to the 0.  
This result indicates tight constraints on the AGN component.
Without MIRI, the constraints are poor, and the median and $\smad$ are 0.200 and 0.155, 
respectively, indicating that the constraints on the AGN component are poor.
The significant role of MIRI in constraining AGN highlights its ability of sampling 
the emission from AGN-heated hot dust (e.g., Figs.~\ref{fig:filters} and \ref{fig:exm1}).
Without MIRI, there is large gap in coverage from IRAC 8~\micron\ to MIPS 24~\micron, 
which is unable to probe the hot dust heated by the AGN, and thus the constraints on 
the AGN component are much weaker (e.g., Fig.~\ref{fig:exm1}).  
Specifically, surveys that include multiple MIRI bands will be very effective at 
identifying galaxies \textit{without} AGN emission. 

\begin{figure}
    \centering
    \includegraphics[width=\columnwidth]{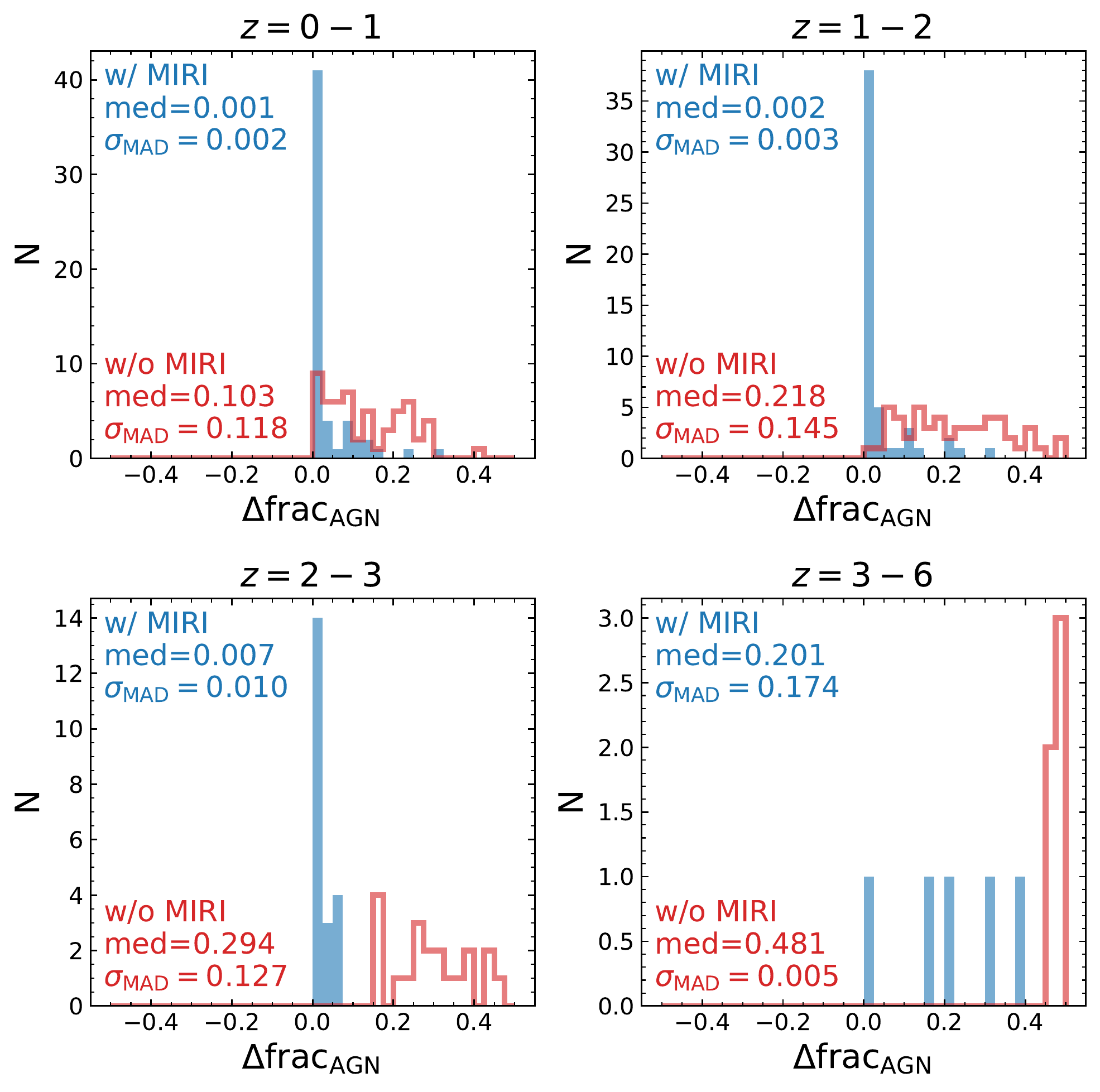} \\
    \caption{Same format as Fig.~\ref{fig:ms_vs_true} bottom right, 
    but dividing into different redshift bins as labeled.
    Values in blue (``w/MIRI'') correspond to results that include the MIRI data, 
    and values in red (``w/o MIRI'') exclude the MIRI data.
    With MIRI data, the constraint on $\fracA$ is robust at $z\lesssim 3$. 
    But the constraint becomes weaker at $z\gtrsim 3$, because the bulk of AGN 
    hot-dust emission shifts out of MIRI coverage.
    }
    \label{fig:det_dif_z}
\end{figure}

Because the AGN hot-dust emission typically peaks at rest-frame $\sim 10\mu$m 
(e.g., Fig.~\ref{fig:sed_templates}), it is possible that the constraints 
on $\fracA$ become weaker at high redshifts, when the bulk of AGN emission 
shifts out of the MIRI coverage.
To investigate this redshift dependence, we show the distributions of 
$\Delta \fracA$ for different redshift bins in Fig.~\ref{fig:det_dif_z}.
The constraints on $\fracA$ are excellent up to $z\approx 3$, with median 
and $\smad$ both below $\approx 2\%$.
At higher redshifts ($z \gtrsim 3$), the constraints become substantially weaker as expected as the mid-IR features associated with hot-dust from the AGN shift to higher wavelengths than can be probed by even MIRI. 
The photo-$z$ quality also appears to drop at $z\gtrsim 3$, although the high-$z$
sample size is not sufficiently large for a solid conclusion.
The relatively poor photo-$z$ quality at high-$z$, if true, could result from 
the fact that the PAH features also shift to wavelength beyond those covered by MIRI.

\subsubsection{Fitting results for galaxy-AGN mixed models}
\label{sec:xcig_res_agn}
\begin{figure}
    \centering
	\includegraphics[width=\columnwidth]{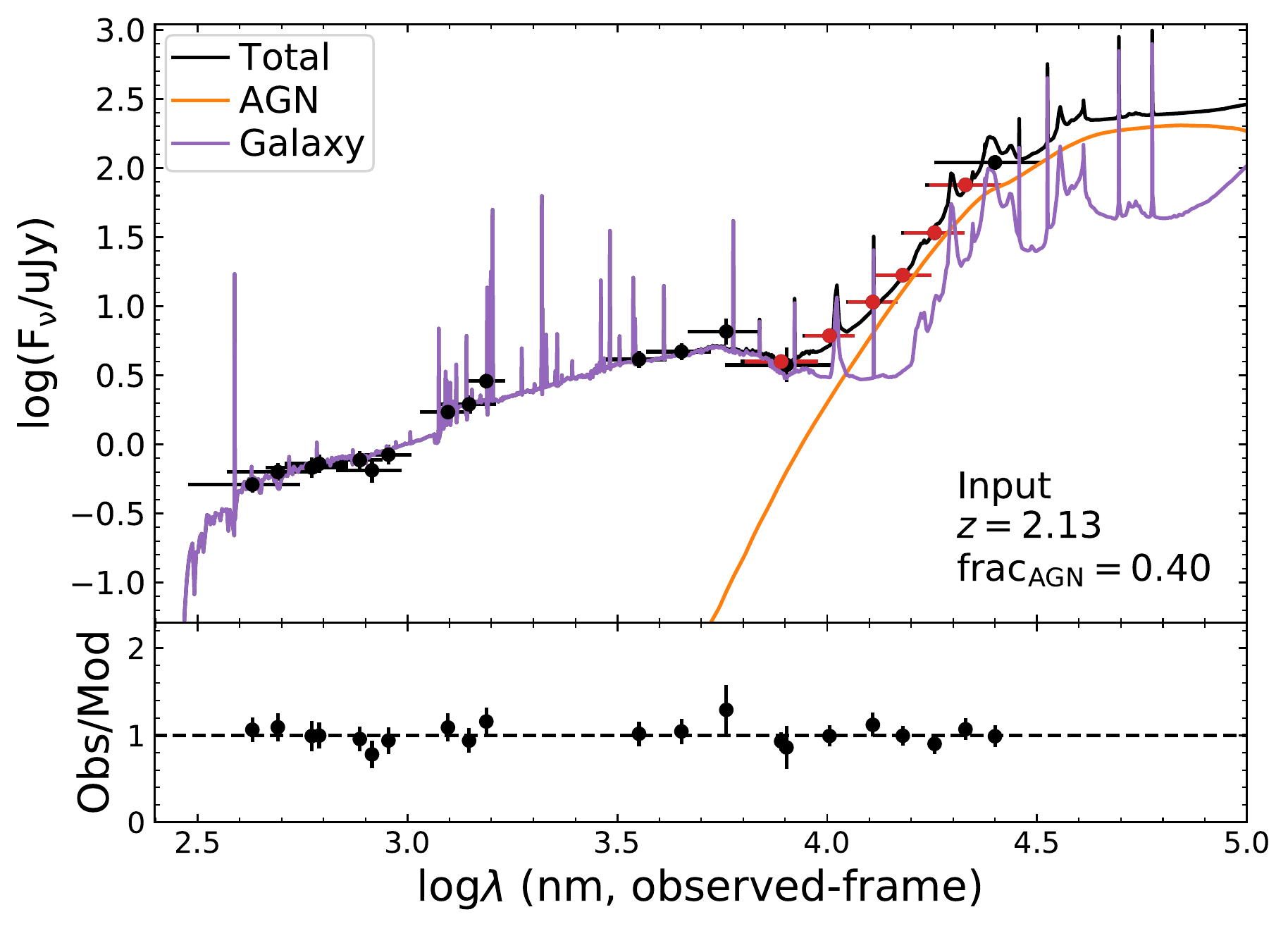}
	\includegraphics[width=\columnwidth]{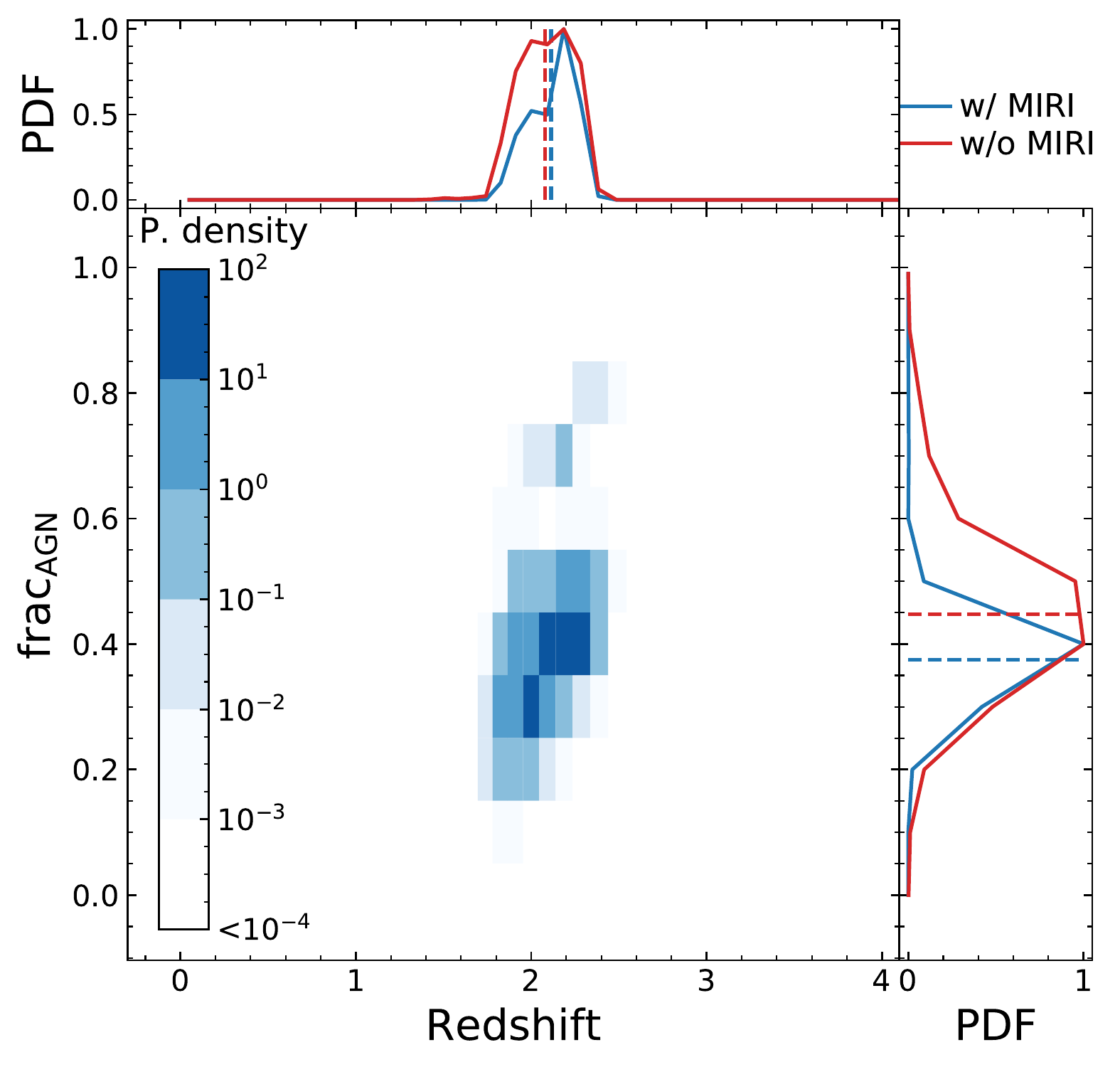}
    \caption{Same format and source as Fig.~\ref{fig:exm1}, but in the 
    simulation with model-input $\fracA=0.4$.
    Although the $\fracA$ Bayesian values (dashed lines) of from fitting with 
    and without MIRI are similar, the PDF of the fitting with MIRI is much narrower  
    than that without MIRI.
    }
    \label{fig:exm2}
\end{figure}

In this section, we present the SED-fitting results for the simulation sets
of {hypothetical} galaxy-AGN mixed models with different positive values 
of input $\fracA$ (\S\ref{sec:mod_agn}).
Fig.~\ref{fig:exm2} displays an example SED fitting and PDFs for a source with 
model $\fracA=0.4$.
After using MIRI data, the projected 1D PDFs of both redshift and $\fracA$ 
become narrower.

\begin{figure*}
    \centering
    \includegraphics[width=2\columnwidth]{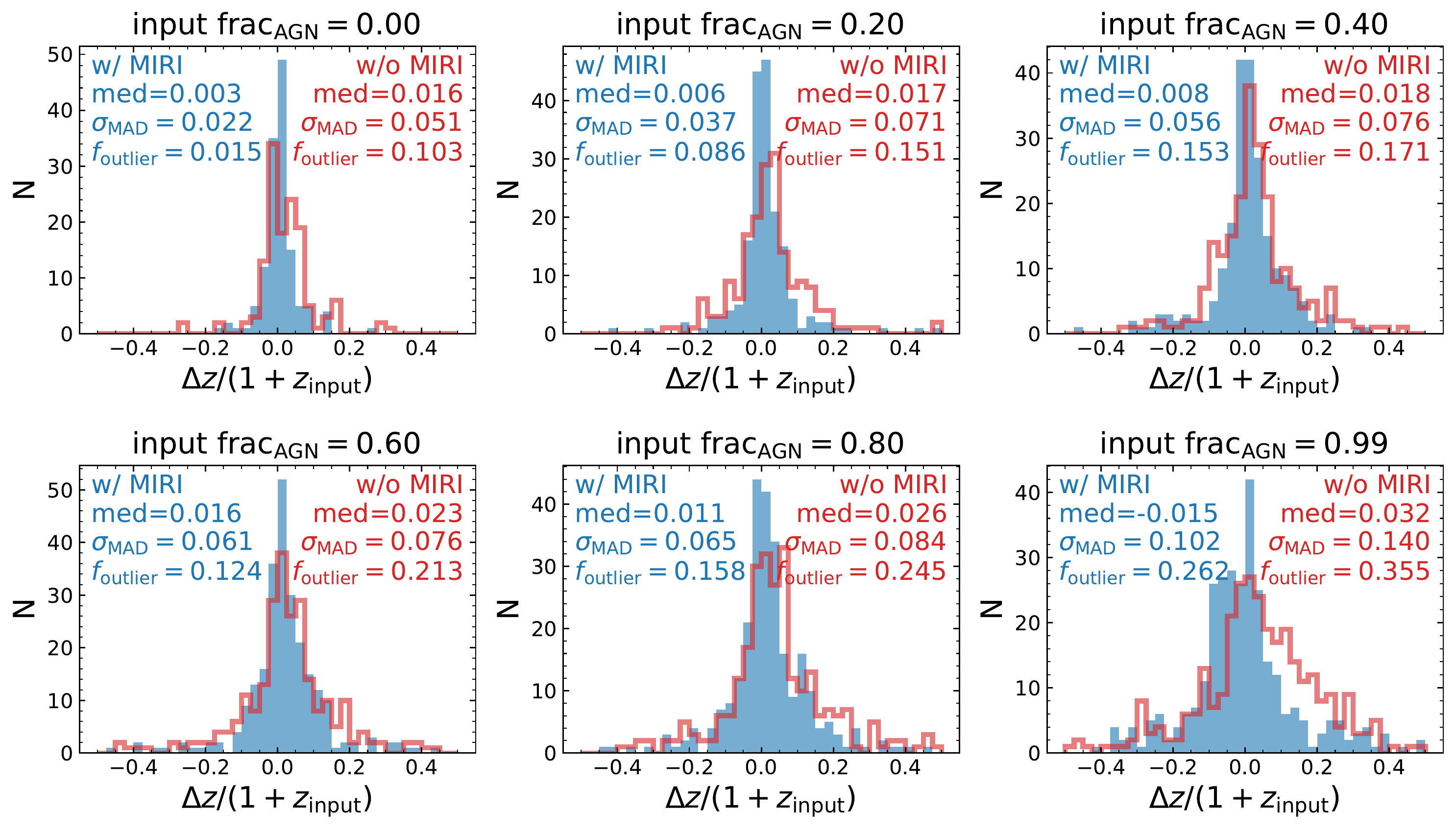} \\
    \caption{Same format as Fig.~\ref{fig:ms_vs_true} bottom left but 
    for the simulations with different model-input $\fracA$ as labeled.
    Note that model $\fracA=0.00$ corresponds to pure-galaxy models (i.e., 
    Fig.~\ref{fig:ms_vs_true} bottom).
    Values in blue (``w/MIRI'') correspond to results that include the MIRI data, 
    and values in red (``w/o MIRI'') exclude the MIRI data.
    The redshift constraints are significantly tighter after adding MIRI 
    data to the SED fitting.
    }
    \label{fig:det_hist_z}
\end{figure*}

In Fig.~\ref{fig:det_hist_z}, we compare the distribution of redshift 
uncertainties from the fittings with and without MIRI data for different 
model-input $\fracA$.
The redshift accuracy is improved after using MIRI data for each $\fracA$ 
case, as both the scatter and outlier fraction improve with the inclusion of MIRI data.
The photo-$z$ scatter, $\sigma_\mathrm{MAD}$, increases as the input $\fracA$ increases.
This is because, as AGN strength rises, the PAH features (used as redshift 
indicator) becomes weaker in the SED (see Fig.~\ref{fig:sed_templates}).
Another reason is that we select more optically faint objects for higher 
$\fracA$, as our selection is based on mid-IR (detected in $\geq 2$ MIRI 
bands; \S\ref{sec:sed_fit}). 
The optical flux uncertainties are larger for these optically faint sources. Regardless, there is appreciable gain when including the MIRI bands. 

\begin{figure}
    \centering
    \includegraphics[width=\columnwidth]{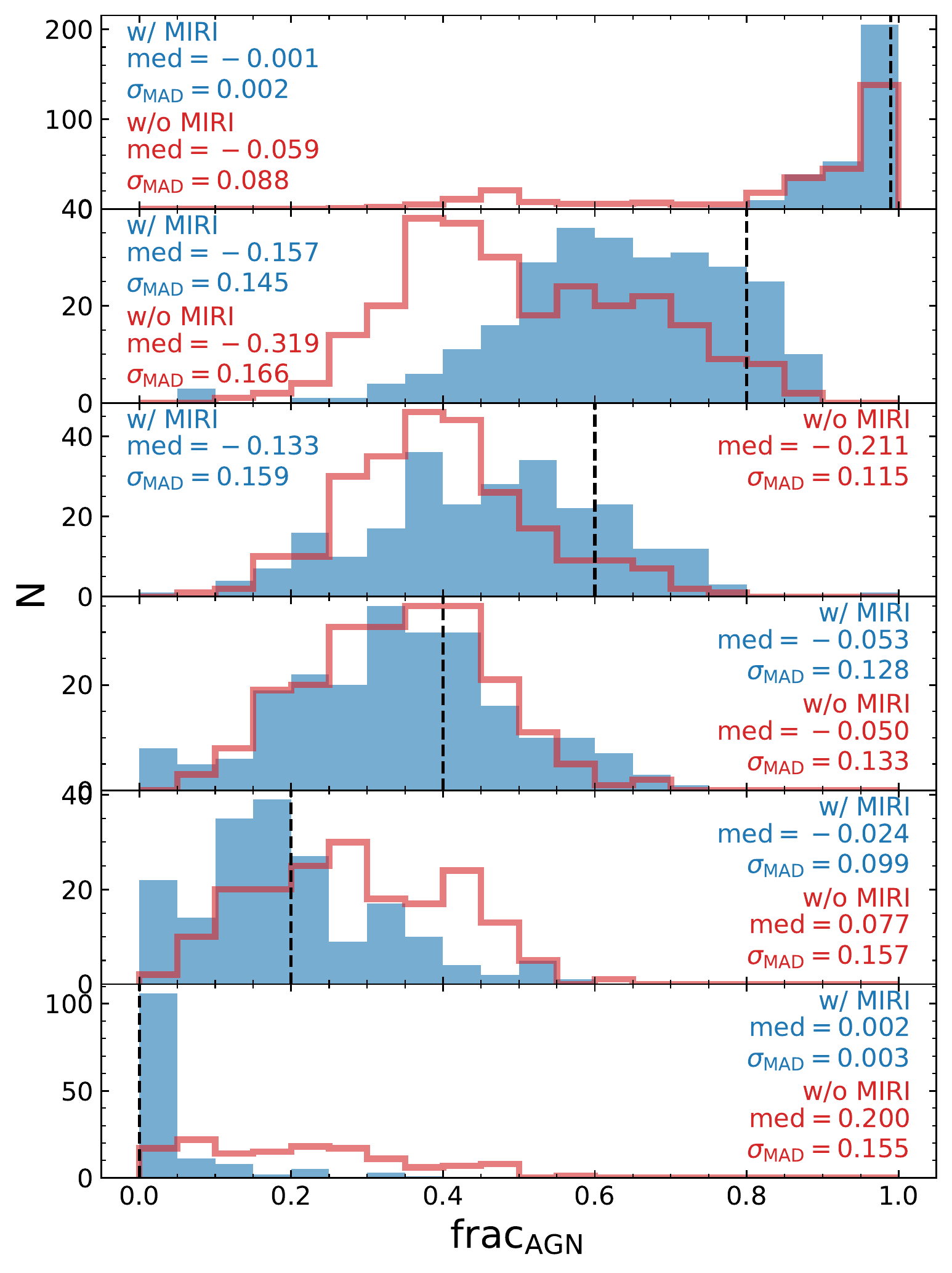} \\
    \caption{The distributions of $\fracA$ fitting results.
    From bottom to top, the model-input $\fracA$ values are 0, 0.2, 
    0.4, 0.6, 0.8, and 0.99, respectively, as marked by the vertical 
    black dashed lines.
    Values in blue (``w/MIRI'') correspond to results that include the MIRI data, 
    and values in red (``w/o MIRI'') exclude the MIRI data.
    The $\fracA$ constraints becomes generally tighter after adding MIRI 
    data to the SED fitting.
    }
    \label{fig:det_hist_frac}
\end{figure}

Fig.~\ref{fig:det_hist_frac} shows the fitting results of $\fracA$. 
With MIRI, the dispersion of $\Delta \fracA$ is remarkably small ($\lesssim 0.003$) 
for the cases of pure-galaxy ($\fracA=0$) and AGN-dominant ($\fracA=0.99$), 
and it is much larger ($\gtrsim 0.1$) for the intermediate cases.
This is expected, as in the two extreme cases ($\fracA=0$ or 0.99), the SED features of galaxy
or AGN are dominant.
In contrast, when $\fracA$ is intermediate, the total SED has mixed features of 
galaxy and AGN, and the SED decomposition is challenging.    
These sources would likely be considered ``composites'' in previous studies \citep[e.g.,][]{kirkpatrick17}.

\begin{figure*}
    \centering
    \includegraphics[width=2\columnwidth]{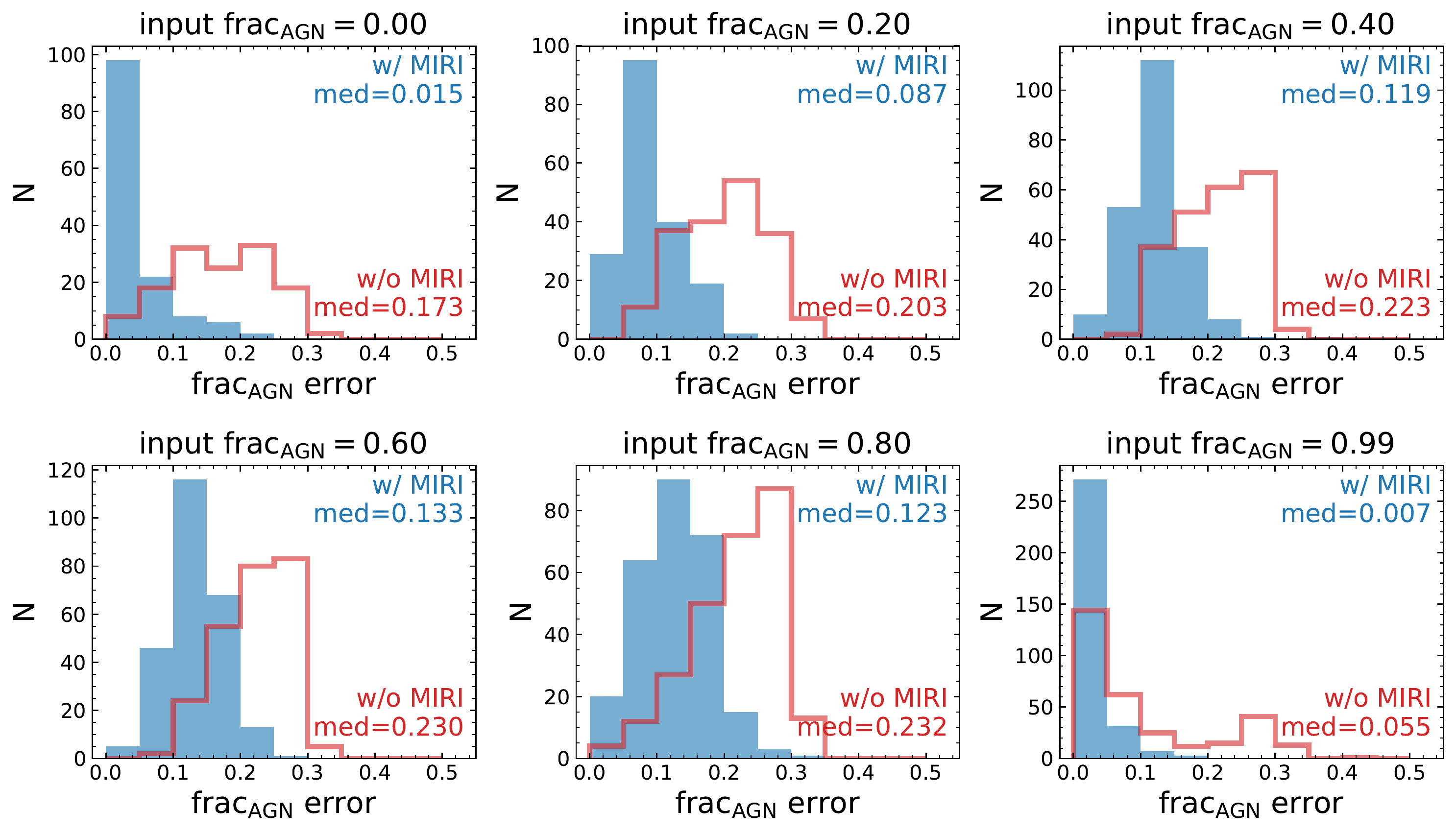} \\
    \caption{The distribution of $\fracA$ errors from \xcig. 
    Different panels are for different model $\fracA$ as labeled.  
    Values in blue (``w/MIRI'') correspond to results that include the MIRI data, and values in red (``w/o MIRI'') exclude the MIRI data.
    The blue and red colors indicate the SED fittings with and without 
    MIRI data, respectively.
    The median values of the distributions are labeled on each panel.
    For each model $\fracA$, the uncertainties become smaller after adding
    MIRI data to the SED fitting.
    }
    \label{fig:bayes_err_frac}
\end{figure*}

At first glance, for model $\fracA=0.4$ and $\fracA=0.6$, it may appear that 
the accuracy of $\fracA$ does not change significantly after using MIRI data.
However, this is misleading. 
This is because the Bayesian output value in \xcig\ is calculated as the 
PDF-weighted mean (see \S4.3 of \citealt{boquien19}), and this produces a median that tends 
to be located near the center of the parametric range (0.5 for $\fracA$) in the case that the constraints are weak.  Considering the extreme case when the model $\fracA=0.5$ and the PDF is totally 
flat, the Bayesian $\fracA$ will be exactly the same as the model value, 
although the constraints on $\fracA$ is none.
The Bayesian $\fracA$ values (dashed lines) are similar for the fitting with and 
without MIRI, but the PDFs for individual objects is much narrower after using MIRI data.
Therefore, sometimes it is not sufficient to use the mean values from the PDF only, and 
the errors (PDF-weighted standard deviation; \S4.3 of \citealt{boquien19})
calculated by \xcig\ may serve as a necessary diagnostic.
Fig.~\ref{fig:bayes_err_frac} displays the error distributions for all model 
$\fracA$ configurations. 
Indeed, for all model $\fracA$ (including $\fracA=0.4$ and $\fracA=0.6$), 
the median of $\fracA$ uncertainties always becomes smaller after using the MIRI 
data, showing that the constraint on $\fracA$ becomes tighter with MIRI.
Quantitatively, MIRI data improves the $\fracA$ accuracy by a factor of $\approx 2$ 
for the case of AGN-galaxy composite input (see Fig.~\ref{fig:bayes_err_frac}).

\subsubsection{Constraints on AGN accretion power with MIRI}
\label{sec:Lbol}
In \S\ref{sec:xcig_res_agn}, we assess the SED-fitting quality of $\fracA$ 
for the cases where an {hypothetical} AGN is present in the input models. 
$\fracA$ describes the relative luminosities of AGN vs.\ galaxy in terms
of total IR luminosity.
It is understandable that the fitted $\fracA$ still has significant scatter
even using MIRI data, because we do not have far-IR photometry to tightly 
constrain the galaxy cold-dust emission, which also affects $\fracA$.

However, it is often useful to obtain absolute AGN luminosities in AGN 
studies, as black-hole (BH) accretion rates can be estimated 
from the absolute luminosities \citep[e.g.,][]{ni19, ni20, yang19}.
\xcig\ is able to estimate the intrinsic AGN accretion disk luminosity, 
$\ld$ (i.e., ``agn.accretion\_power'' in the output; \citealt{yang20}).
The SKIRTOR AGN model in \xcig\ adopts anisotropic disk emission, and 
$\ld$ is calculated averaging over all viewing angles. 
Note that the $\ld$ is numerically equivalent to the angle-averaged obscured 
disk $+$ dust luminosity, as SKIRTOR is a physical model obeying energy 
conservation. 

\begin{figure*}
    \centering
    \includegraphics[width=2\columnwidth]{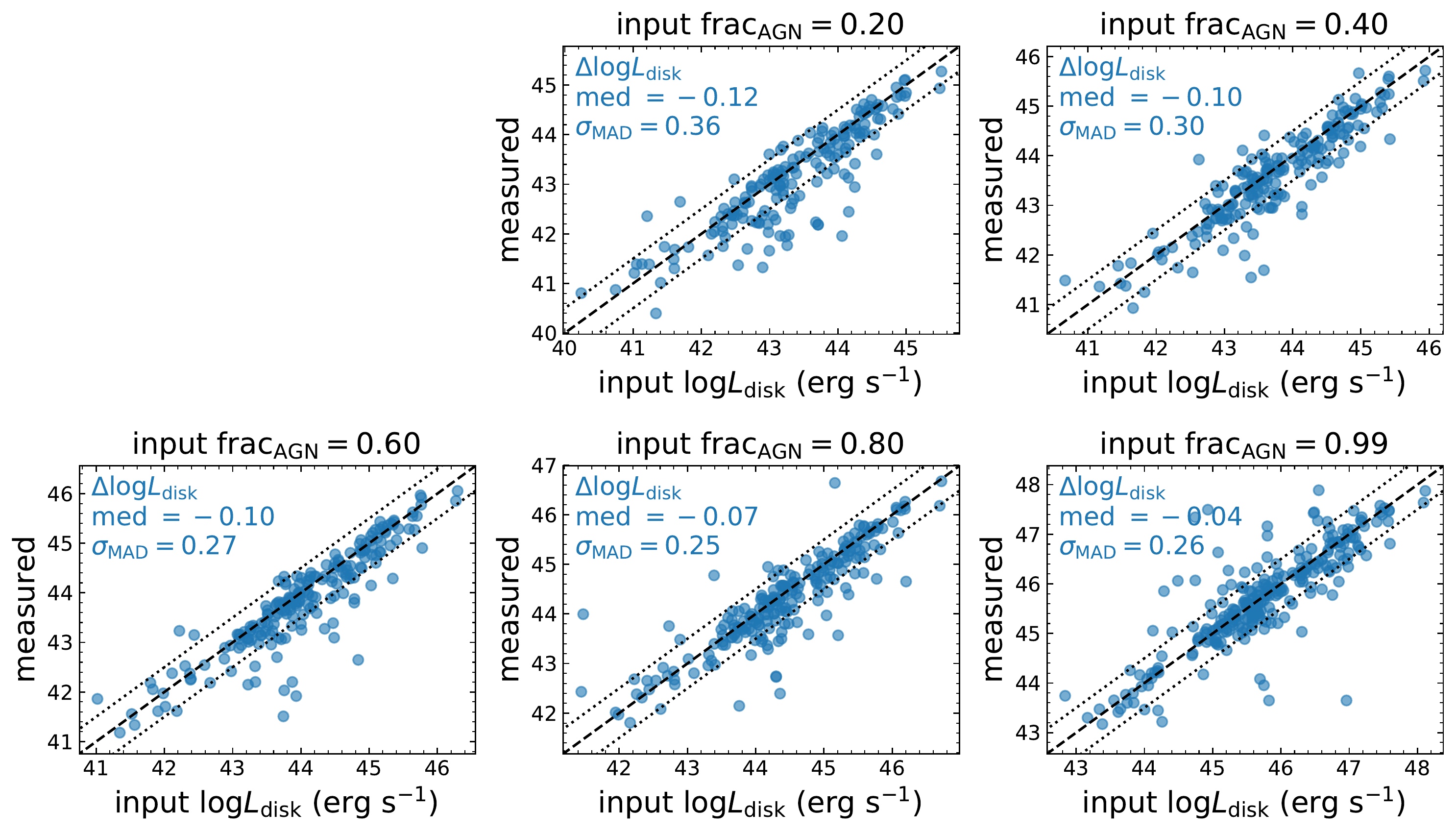} \\
    \caption{Measured vs.\ input AGN intrinsic disk luminosities. 
    The measured $\ld$ is based on the SED fitting with the 
    simulated MIRI data, and different panels are for simulation
    sets of different input $\fracA$ as labeled.
    On each panel, the black dashed line represents the 1:1 relation, 
    and the dotted lines represent $\pm 0.5$~dex from the 1:1 relation.
    The median and $\smad$ values of $\Delta \ld$ (measured $-$ input)
    are marked on each panel.
    }
    \label{fig:Lbol_vs_Lbol}
\end{figure*}

We compare the measured $\ld$ and model-input $\ld$ in 
Fig.~\ref{fig:Lbol_vs_Lbol} for the simulation sets of different input 
$\fracA$ (\S\ref{sec:mod_agn}).
The median and $\smad$ for all $\fracA$ cases are $\approx -0.1$~dex 
and $\approx 0.3$~dex, respectively (see Fig.~\ref{fig:Lbol_vs_Lbol}).
Therefore, we can reliably recover $\ld$ from SED fitting of the 
photometric data of MIRI (and other bands).
In the future, MIRI photometric surveys can be widely used in the 
studies of BH accretion and evolution across cosmic history
(\S\ref{sec:comp_xray}).
From Fig.~\ref{fig:Lbol_vs_Lbol}, the quality of $\ld$ becomes slightly
better toward higher $\fracA$, indicating that the constraint on AGN power
is better for sources whose AGN SED component is more dominant in infrared.

Therefore, in addition to the constraints on $\fracA$, the \xcig\ results 
also provide reliable constraints on the BH accretion power which is a 
useful quantity for AGN studies.

\section{Discussion}
\label{sec:discuss}

\subsection{Comparison to \citet{kirkpatrick17}} 
\label{sec:compare_k17}
\cite{kirkpatrick17} were one of the first to evaluate the use of MIRI to 
classify and characterize AGN and star-formation in distant galaxies.
Their classification scheme is based on MIRI color-color diagrams, and 
demonstrated that MIRI can identify objects whose IR emission is powered 
by star-formation, AGN, and composite cases.  
The method here uses \xcig\ and SED-fitting and improves on this previous 
work as it uses all the available MIRI bands, and measures simultaneously 
the photometric redshift and galaxy properties simultaneously, providing 
PDF for each parameter.  
%

We provide here a qualitative comparison between the two methods. 
\citealt{kirkpatrick17} derived a parameter $\fracM$ (fractional AGN contribution to the 
rest-frame \hbox{5--15 $\mu$m} SED, where the superscript $K$ denotes these are derived by \citealt{kirkpatrick17}).  They then divided their 
source SEDs into three classes, i.e., galaxy ($\fracM<0.3$), 
composite ($0.3\leq \fracM < 0.7$), and AGN ($\fracM\geq 0.7$),
and define an empirical color-color scheme to classify simulated sources 
into these classes. 
They find that their color-color scheme is able to reach a high 
accuracy level of $\approx 79\%$ (galaxy), 76\% (composite), and 
87\% (AGN).

The thresholds $\fracM=0.3$ and $\fracM=0.7$ above roughly correspond to our 
$\fracA \approx 0.1$ and $\fracA \approx 0.4$, respectively (the exact conversions vary from model to model).
Therefore, to compare with \cite{kirkpatrick17}, we choose 
$\fracA<0.1$, $0.1\leq \fracA<0.4$, and $\fracA>0.4$ as the criteria 
for the galaxy, composite, and AGN classes in our analysis.  
Under these criteria, for our simulation set of pure-galaxy input models 
(\S\ref{sec:pure_gal}), 88\% sources are correctly classified as a (star-formation-dominated) ``galaxy''.
(i.e., the $\fracA$ in \xcig\ output below 0.1; Fig.~\ref{fig:ms_vs_true}).
For the input models of $\fracA=0.2$ (i.e., the composite models,
\S\ref{sec:mod_agn}), the classification accuracy is 72\% 
(Fig.~\ref{fig:det_hist_z}). 
For the input models of $\fracA=0.6$, 0.8, and 0.99 (i.e., the AGN models), 
the success rates are 77\%, 90\%, and 99\%, respectively. 
Therefore, there is a high degree of overlap between the method of 
\cite{kirkpatrick17} and our SED-fitting method here.  %

Again, we note that the classifications above are for the comparison with 
\cite{kirkpatrick17} only. 
The method of \cite{kirkpatrick17} is straightforward: one just needs to 
apply a suitable color-color scheme depending on the source's redshift
which can be spec-$z$ (if available) or photo-$z$.
The qualitative classification results can be obtained instantaneously.
Our SED fitting of \xcig\ is a quantitative method that yields PDF-based 
property estimation as well as best-fit SEDs.  

Another advantage of our \xcig\ method is that the redshift does need to 
be known \textit{a priori}, and \xcig\ will fit simultaneously for the 
redshift and other parameters (including $\fracA$). 
This feature is extremely valuable for deep-field sources whose spec-$z$
are challenging to measure.

The \xcig\ run time for our simulated MIRI pointing (a few hundred sources)  
is $\lesssim 1$ hour on a typical desktop/laptop. 
The run time will still be acceptable ($\lesssim 1$ day) even for a large 
sample of a million sources, thanks to the efficient parallel algorithm 
of \xcig\ \citep{boquien19}.

\subsection{Comparison with X-ray AGN selection}
\label{sec:comp_xray}
X-ray observations are effective in AGN identification 
\citep[e.g.,][]{brandt15, xue17}.
Strong \xray\ emission is almost a universal property of the BH
accretion process, and galactic processes (e.g., \xray\ binaries and hot gas)
can only reach low \xray\ luminosities typically below 
$\lx \sim 10^{42}$~erg~s$^{-1}$.
Thanks to these strengths, \xray\ surveys have detected numerous AGNs in the 
distant universe, significantly deepening our understanding of BH
evolution across cosmic history \citep[e.g.,][]{civano16, luo17, yang18, yang18b}.
Like \xray\ observations, MIRI can also reliably constrain the AGN accretion 
power (see \S\ref{sec:Lbol}). 
To evaluate the effectiveness of MIRI AGN selection, we compare the 
sensitivities of MIRI vs. \xray\ observations below.

\begin{figure}
    \centering
	\includegraphics[width=\columnwidth]{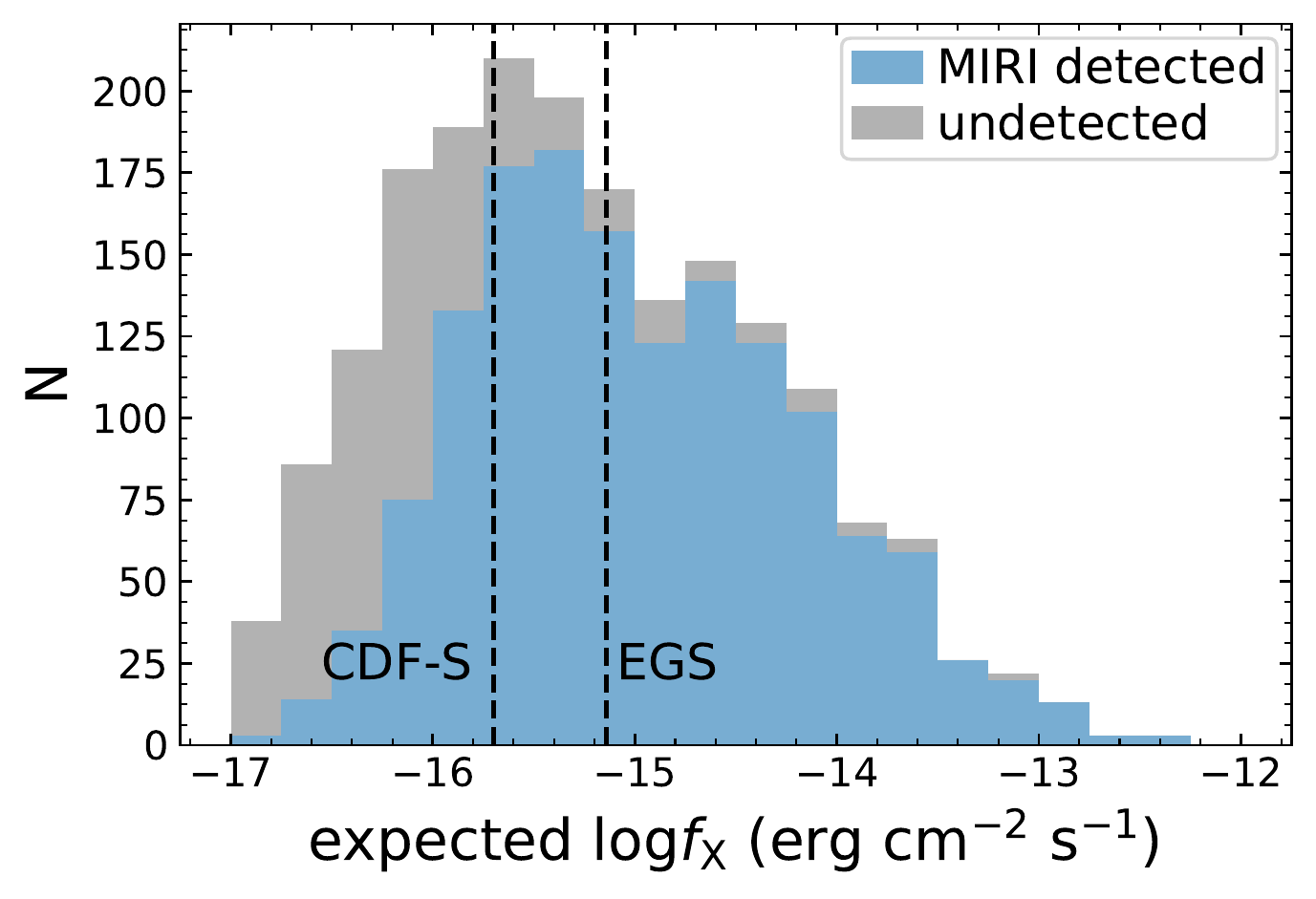}
    \caption{The distributions of the expected \hbox{0.5--7~keV} \xray\ fluxes for 
    our simulated AGNs.
    The blue and grey colors represent MIRI detected and undetected sources, 
    respectively.
    The $f_{\rm X}$ values are estimated based on the empirical $\lsix$-$\lx$ 
    relation \citep{stern15} assuming \xray\ obscuration decreases fluxes by 
    a factor of 2.
    The vertical dashed black lines indicate the sensitivities for the \hbox{CDF-S} 
    \citep{luo17} and EGS \citep{nandra15} \xray\ surveys as labeled.  
    Given the relative accurate measures of the AGN disk luminosity 
    ($L_\mathrm{disk}$) over a wide range of $\fracA$, this shows the analysis 
    of MIRI data for galaxies will provide accurate measurements of the 
    AGN accretion power.  
    }
    \label{fig:fx_hist}
\end{figure}

To estimate the equivalent \xray\ flux limit in the MIRI selections, 
first, we calculate model-input AGN 6$\mu$m luminosities ($\lsix$) for 
all {hypothetical sources} in all of the $\fracA>0$ simulations 
(\S\ref{sec:mod_agn}).
We then convert $\lsix$ to $\lx$ based on the empirical $\lsix$-$\lx$ relation 
in \cite{stern15}.
We derive the \xray\ fluxes ($f_{\rm X}$, observed-frame \hbox{0.5--7 keV}) 
from $\lx$ assuming an \xray\ spectral photon index of $\Gamma=1.8$ 
\citep[e.g.,][]{yang16,liu17}.
Finally, we decrease these $f_{\rm X}$ by a factor of 2, which represents the 
typical \xray\ obscuration effect \citep[e.g.,][]{luo17}.
The $f_{\rm X}$ distributions of MIRI detected and undetected sources are displayed 
in Fig.~\ref{fig:fx_hist}.
From Fig.~\ref{fig:fx_hist}, the MIRI detection becomes significantly incomplete 
($<50\%$) as the flux drops below $f_{\rm X} \approx 10^{-16}$~erg~cm$^{-2}$~s$^{-1}$.
This $f_{\rm X}$ is slightly lower than the 50\%-completeness limit of \hbox{CDF-S}
(\citealt{luo17}; see Fig.~\ref{fig:fx_hist}).
Therefore, the deep MIRI exposures (like the multi-hour CEERS/MIRI2) have the potential to identify fainter AGN than even the deepest, highest sensitivity \xray\ data achieved so far.

MIRI can even go beyond \xray\ for AGN studies, as \xray\ AGN selection has a 
significant weakness. 
When the obscuration level is high ($N_{\rm H} \gtrsim 10^{24}$~cm$^{-2}$), 
even hard \hbox{X-rays} can be easily scattered/absorbed, escaping the census 
of \xray\ surveys \citep[e.g.,][]{brandt15, hickox18}. 
Such heavily obscured AGNs are often called ``Compton-thick'' AGNs because of 
the strong Compton-scattering effect in this high-$N_{\rm H}$ regime. 
The analyses of cosmic \xray\ background (CXB) suggest that Compton-thick sources 
may contribute to a large fraction of the AGN population (up to $\approx 50\%$; 
e.g., \citealt{gilli07, akylas12, ueda14}). 
However, the population of Compton-thick AGNs is poorly understood due to the lack 
of effective selection methods \citep[e.g.,][]{buchner15, li19, li20}.
The forthcoming MIRI surveys will be a game-changer. 
The mid-IR emission comes from AGN-heated hot dust, and Compton-thick 
AGNs likely have abundant obscuring dust \citep[e.g.,][]{georgantopoulos11}. 
Therefore, MIRI should be able to detect the missing population of Compton-thick 
AGNs, and \xcig\ can serve as a reliable tool to identify their AGN nature 
\citep[e.g.,][]{li20, pouliasis20}.

Assuming the $\lsix$ versus intrinsic $\lx$ relation \citep{stern15} also holds for 
Compton-thick AGNs, MIRI should be able to detect even many low-luminosity 
Compton-thick sources according to our sensitivity estimation above. 
For example, MIRI can sample Compton-thick AGNs with intrinsic $\lx$ down to
$\sim 10^{42.5}$~erg~s$^{-1}$ at $z\sim 2$ (the peak of cosmic AGN activity).
This $\lx$ sensitivity is $\sim 100$ times below the break $\lx$ of the known AGN
luminosity function, which is derived based on non-Compton-thick AGNs 
\citep[e.g.,][]{aird10, ueda14}.
Therefore, future MIRI surveys will allow us to infer a completely new AGN 
luminosity function for Compton-thick AGNs, shedding light on this mysterious 
population. 

In summary, with a moderate amount of exposure time (such as the 3.6~hours for
CEERS/MIRI2), MIRI can already reach a sensitivity level similar to the deepest 
\xray\ survey.
MIRI should also be able to identify the Compton-thick population, which is largely 
missed in \xray\ surveys. 
In the future, MIRI surveys will provide a complete census of the entire population 
of accreting BHs, enabling unbiased studies of BH evolution across the 
cosmic history (\S\ref{sec:Lbol}).

\subsection{{The effects of different MIRI bands}}
\label{sec:eff_band}
{Our analyses in \S\ref{sec:sed_fit} utilizes the full set of simulated 
MIRI bands from F770W to F2100W. 
Using all these six bands takes full advantage of the data, and thus yields the
tightest possible constraints on source properties. 
However, there are likely to be MIRI observations in the future that do not have full band coverage as we expect for CEERS. 
Therefore, understanding the effects on redshifts and galaxy diagnostics using different combinations of MIRI bands is helpful for 
future survey design and analysis. 
Below, we investigate the effects of the bluer (F770W, F1000W, and F1280W) 
and redder (F1500W, F1800W, and F2100W) MIRI bands, respectively. 
This test is an example of using different band 
combinations.\footnote{Readers may need to know the 
effects of other band sets for purposes of, e.g.,
survey planning.
We are willing to perform similar assessment
for any specific band combination upon e-mail 
request.} }

{We re-perform the SED fittings in \S\ref{sec:xcig_res_gal} (pure-galaxy 
inputs) where we only add the three bluer MIRI bands (F770W, F1000W, F1280W) or the three redder MIRI bands (F1500W, F1800W, F2100W) to the other CANDELS multiband catalogs. 
These combinations result in photometric accuracies of $\smad=0.029/0.042$ and outlier fractions of $f_{\rm outlier}=3.7\%/5.9\%$ when using the blue/red MIRI bands, respectively.
These values are intermediate between those obtained by using none of the MIRI bands versus all MIRI bands 
(Fig.~\ref{fig:ms_vs_true}), as expected.
The blue MIRI bands improve the photo-$z$ accuracy more significantly than 
the red bands. 
We attribute the reason to that the blue bands generally have higher 
sensitivity than the red bands. 
For example, the median S/Ns in F1000W and F1800W are 6.6 and 3.9, 
respectively, for our analyzed sources.
Notably, the blue bands significantly reduce $f_{\rm outlier}$ by a factor 
of $\approx 3$, compared to the case when using only non-MIRI bands.
This result is consistent with the findings of \cite{bisigello16}, 
who concluded blue MIRI bands (F560W and F770W in their case) can significantly 
reduce the photo-$z$ outliers.  
Therefore, future photo-$z$ surveys could consider including a few MIRI blue 
bands in their data if possible.}

{Lastly, we discuss the effects of blue/red MIRI bands on $\fracA$.
We re-perform the SED fittings for different $\fracA$ inputs 
(\S\ref{sec:xcig_res_gal} and \S\ref{sec:xcig_res_agn}) but using the same combinations of only 
the blue/red MIRI bands, respectively, combined with the other CANDELS multiwavelength bands.  
As expected, the resulting accuracy of $\fracA$ is between those when 
using none and all MIRI bands (Fig.~\ref{fig:det_hist_frac}), 
similar to our finding for the photometric redshifts above.
However, in contrast to the photo-$z$ results, the $\fracA$ 
accuracy is similar when using either the combination of blue or red MIRI bands.
For example, for input $\fracA=0.6$, the fitted $\fracA$ qualities 
are $\mathrm{med}=-0.155/-0.185$ and $\smad=0.170/0.165$ when using 
the blue/red bands, respectively.  
We conclude that this is because  AGN emission is typically stronger in the red bands than in the blue  bands (e.g., Figs.~\ref{fig:filters} and \ref{fig:exm2}), which enables similar constraining power on the AGN emission and this overcomes the fact that the bluer MIRI bands typically have higher flux sensitivity. }

\section{Summary and Future Prospects}\label{sec:summary}
In this work, we simulate the \jwst/MIRI photometric data, and investigate 
its ability to constrain source properties with SED fitting.  
Specifically we focus on the ability of surveys with multi-band MIRI imaging 
to constrain the AGN properties of distant galaxies.  
Our data processing and results are summarized below.

\begin{enumerate}

\item Based on the currently existing broad-band photometry from CANDELS/EGS, 
      we perform SED fitting for the 463 sources within the CEERS/MIRI2 pointing 
      (\S\ref{sec:mod_flux}).
      We employ \xcig\ to realize the fitting, using pure-galaxy models. 
      In addition, we also add {a hypothetical AGN} component to the 
      best-fit galaxy SEDs, assuming different fractions of the AGN to the total 
      IR emission ($\fracA$).
      We obtain model MIRI fluxes for the cases when there is no AGN and when 
      AGN is present, by convolving the model SEDs with the MIRI filters.

\item We simulate the MIRI imaging data with {\sc mirisim}, using the predicted 
      flux densities above, and adopting S\'ersic profiles for the galaxy 
      morphologies (\S\ref{sec:mirisim}).
      We take the raw (Stage 0) data products from the simulation, and reduce them 
      using the {\sc jwst calibration pipeline} applying custom corrections to 
      the background.   
      We then obtain the dither-merged images for each MIRI band (\S\ref{sec:pipeline}).  
      We apply scaling and aperture corrections to the data based on a set of simulated 
      bright point sources (Appendix~\ref{app:cor}).
      Theses corrections can be refined empirically for real observations of bright
      calibration stars after launch.
      We perform PSF-matched photometry with {\sc tphot} using the existing CANDELS/EGS 
      \hst\ F160W source catalog.
      We show this achieves $5\sigma$ depths in the MIRI data similar to those expected 
      in the CEERS using the \jwst/ETC(\S\ref{sec:photo}). 
      The detection rate of sources in the \hst\ catalog by MIRI is high:
      75\% of the $H_{160}<25$ sources are significantly ($>5\sigma$) detected in at least
      two MIRI bands (for the default case where SEDs assume pure-galaxy models).

\item We perform \xcig\ SED fitting, with and without the addition of the MIRI data 
      (\S\ref{sec:sed_fit}).  
      We focus on the ability of the SED fitting to constrain the photometric 
      redshift and $\fracA$ (these are fit simultaneously, and we marginalize of 
      the probability density functions to derive constraints on these parameters).
      The accuracies of both redshift and $\fracA$ are improved significantly after 
      using MIRI data, thanks to the capture of PAH features and AGN hot-dust 
      emission by MIRI.
      Notably, for the simulation set of pure-galaxy models (which is likely the 
      case for most MIRI sources), $\fracA$ can be constrained to the level of 
      $\approx 0.2\%$ with MIRI data, which is $\approx 100$ times better than 
      the fitting without MIRI data. 
      At the same time, the photo-$z$ scatter and outlier fraction are improved 
      by a factor of $\approx 2$ and $\approx 7$, respectively, thanks to MIRI's 
      capture of PAH emission.

\item Our SED-fitting method can reliably recover the source types
      (galaxy/composite/AGN; \S\ref{sec:compare_k17}), similar to results from 
      previous studies that based on MIRI color-color diagnostics to characterize 
      sources \citep[e.g.,][]{kirkpatrick17}. 
      
\item We assess the AGN-detection sensitivity of MIRI vs.\ \xray\
      (\S\ref{sec:comp_xray}) and find that, for a deep MIRI exposures (such as 
      the CEERS-depth, 3.6-hour data simulated in this work), MIRI can already 
      reach a sensitivity level even slightly higher than the deepest \xray\ survey, 
      \hbox{CDF-S}.
      MIRI should also detect Compton-thick AGNs for which \xray\ selection is 
      highly incomplete.
      Therefore, we conclude that MIRI will provide a complete census of the 
      entire population of accreting BHs, enabling unbiased studies of BH evolution 
      across cosmic history.

  \item {We discuss the effects of using different MIRI bands in SED 
      fitting (\S\ref{sec:eff_band}). 
      In particular, we focus on comparing the bluer (F770W, F1000W, and F1280W) 
      and redder (F1500W, F1800W, and F2100W) bands. 
      We find that the blue bands are more helpful for photo-$z$ improvement than
      the red bands, because the former are more sensitive than the latter.
      However, the blue and red bands have similar effects in terms of 
      constraining $\fracA$, likely due to the fact that AGN emission is 
      typically stronger in the red bands than in the blue ones.
      }
\end{enumerate}

Although this work focuses on the CEERS/MIRI2 observational strategy, it has general 
implications for other MIRI extragalactic surveys. 
For the surveys with the same filters (F770W to F2100W) but different depths, the
qualitative conclusions in this paper are likely to hold in general as any MIRI data with similar depth will achieve similar results.  
The only change is that different exposures will yield different detection limits 
(e.g., Fig.~\ref{fig:Lir_vs_z}). 
For example, deeper exposures will be able to detect more faint mid-IR sources at higher fidelity,
thereby constraining their source properties (such as photo-$z$ and $\fracA$).
For surveys with similar exposures as CEERS/MIRI2 but different band coverage, 
{we have performed an example assessment in \S\ref{sec:eff_band}. 
We are willing to repeat the assessment for any other specific band set upon
e-mail request, to satisfy the realistic needs of, e.g., survey planning.
}


\section*{Acknowledgements}
We thank the helpful discussions with Jacqueline Antwi-Danso, 
Emiliano Merlin, Karl Gordon, and the STScI \jwst\ team.  {We also thank our collaborators on CEERS for their work and contributions to the project.}
The authors acknowledge the Texas A\&M University Brazos HPC cluster 
and Texas A\&M High Performance Research Computing Resources (HPRC,
http://hprc.tamu.edu) that contributed to the research reported
here.  
We also acknowledge our collaborators within CEERS for their input 
in the project.
This work acknowledges support from the NASA/ESA/CSA James Webb 
Space Telescope through the Space Telescope Science Institute, 
which is operated by the Association of Universities for Research 
in Astronomy, Incorporated, under NASA contract NAS5-03127. 
Support for program number JWST-ERS-01345 was provided through a 
grant from the STScI under NASA contract NAS5-03127.  
PGP-G acknowledges support from Spanish Government research 
grant PGC2810-093499-BI00.

\appendix
In this appendix, we provide addition information about the simulations, 
modeling, and tests on the MIRI data. 

\section{Simulated MIRI Data Validation and Aperture Corrections}
\label{app:cor}
Both \mirisim\ and \textsc{pipeline} are still in development.  This work uses current versions of the available software, but we acknowledge these could be updated prior to the acquisition of \jwst\ data post-launch. 
%
%
One known issue is that the the \mirisim\ and \textsc{pipeline} adopt different MIRI calibration files, which might 
also affect our simulated MIRI imaging data.
Therefore, to investigate potential issues related to \mirisim\ and \textsc{pipeline}, 
we applied additional tests to validate the imaging data before the photometry-extraction 
process (\S\ref{sec:photo}). 
Below we perform data validation and correction processes based on a set of 
simulated bright point sources. 

\begin{figure}
    \centering
	\includegraphics[width=\columnwidth]{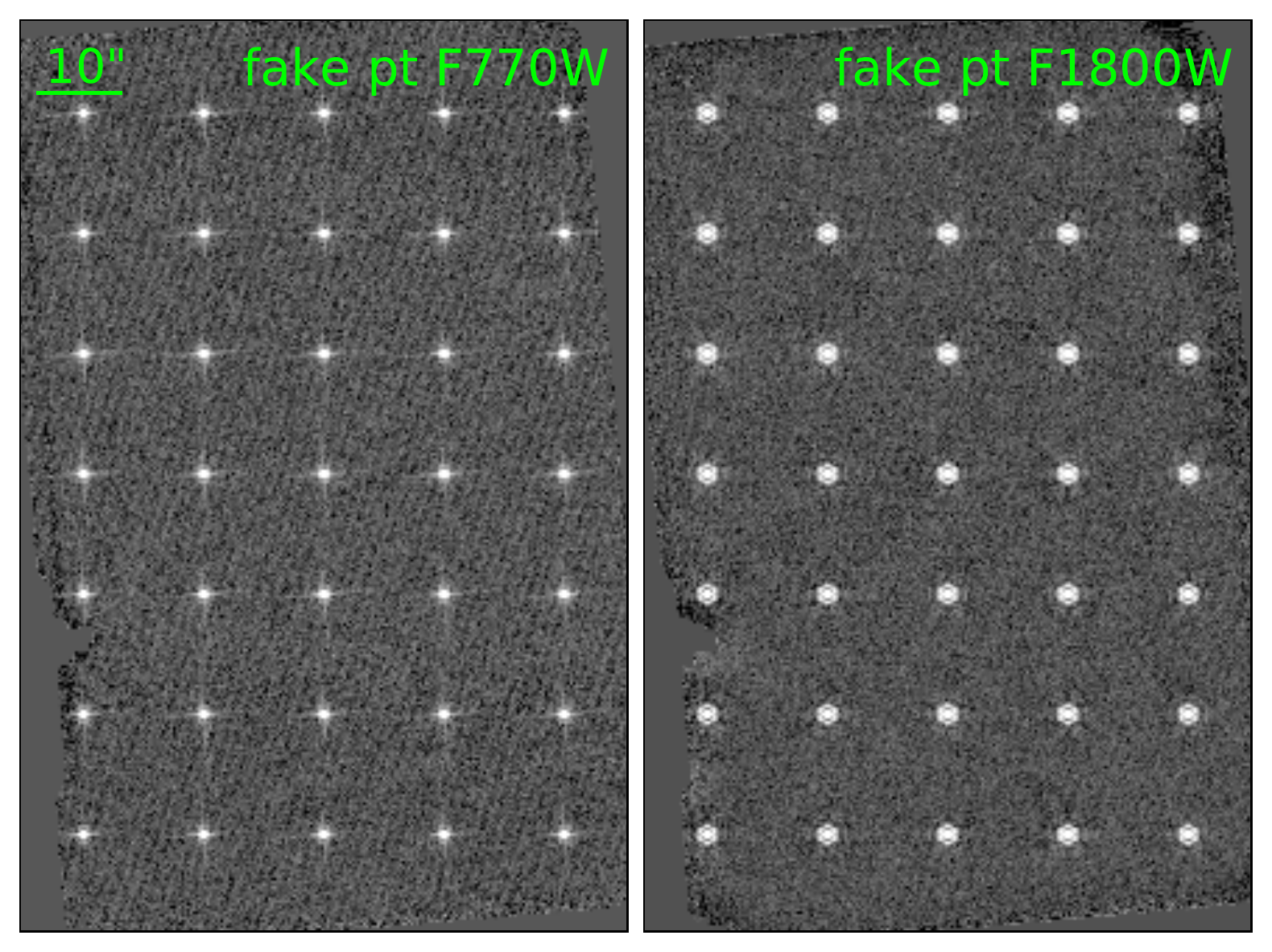}
    \caption{The simulated point sources (after \textsc{pipeline} reduction) for 
    the purposes of data validation and corrections. 
    The left and right panels are for F770W and F1800W, respectively.
    The sources at different positions appear to have similar brightness, 
    suggesting that the flat-field correction in \textsc{pipeline} is reliable.
    }
    \label{fig:fake_pts}
\end{figure}

Using \mirisim\ (\S\ref{sec:mirisim}), we simulate a grid of bright (but 
non-saturated) point sources with a signal-to-noise (S/N) of $\approx 1000$ for 
each MIRI band.
The separation between the neighboring sources is $15\arcsec$, which is 
sufficiently large to avoid light pollution from neighbors.
Fig.~\ref{fig:fake_pts} displays example simulated images for the MIRI F770W and F1800W bands
(where these images have been fully reduced following the procedures in \S\ref{sec:pipeline}), where we have repeated this process for all MIRI bands to be obtained for the CEERS/MIRI2 field.
These sources have a high S/N of 1000, and thus are less affected by the details of noise in the data-reduction process 
%
%
compared to the realistic faint sources.

\begin{figure}
    \centering
	\includegraphics[width=\columnwidth]{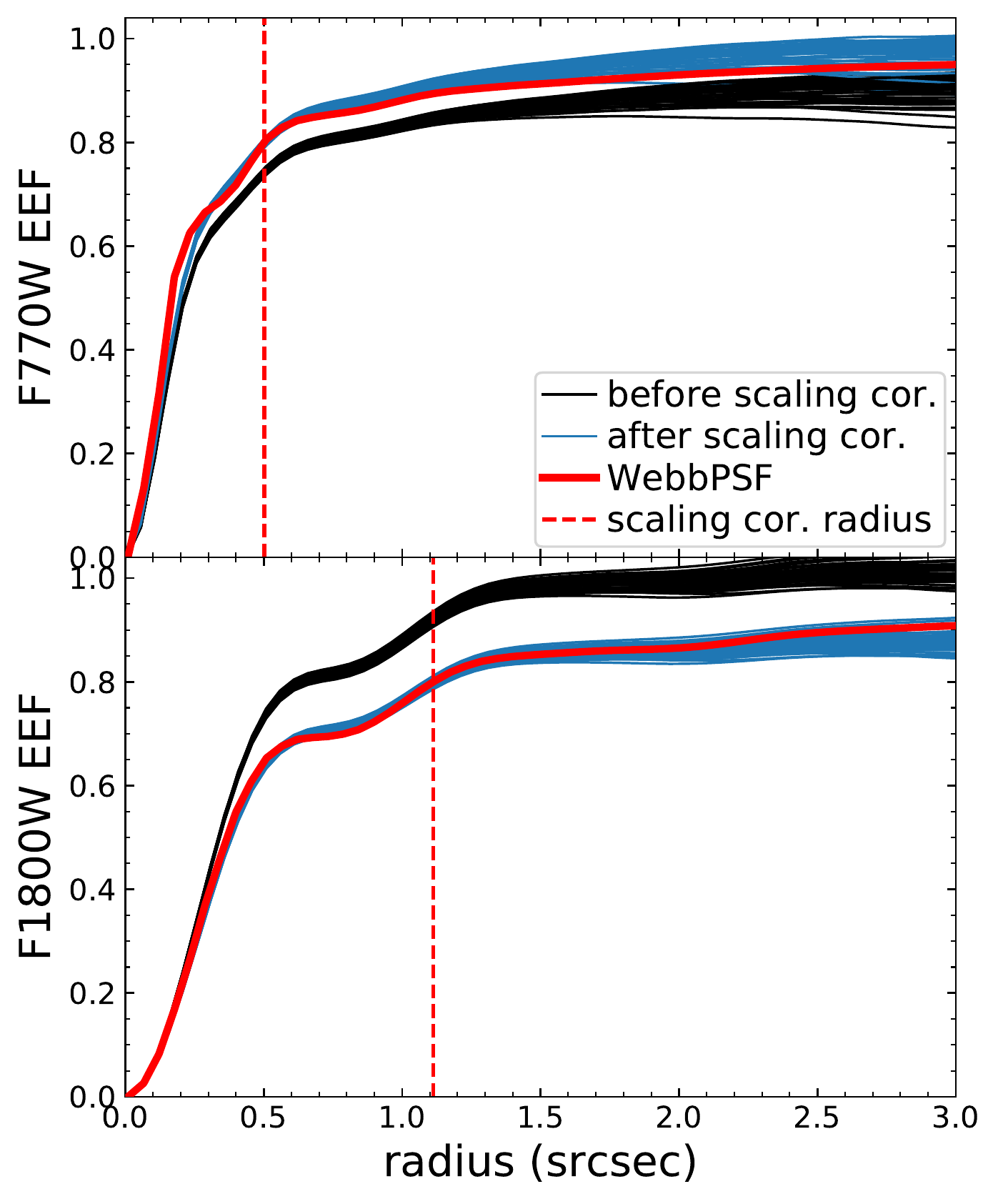}
    \caption{The EEFs for the simulated bright point sources of Fig.~\ref{fig:fake_pts}
    in F770W (top) and F1800W (bottom).
    The black and blue curves represent the EEFs before and after 
    the scaling correction (Appendix~\ref{app:cor}).
    The red solid curve is from the PSF generated by {\sc webbpsf}.
    The red dashed vertical line indicates the radius corresponding 
    to PSF EEF$=0.8$, where the scaling-correction factor is estimated. 
    }
    \label{fig:calib}
\end{figure}

We measure the EEF as a function of radius for 
each simulated point source at different positions.
The EEF at a given radius is calculated as the encircled flux within the 
radius divided by the model-input flux. 
The results for F770W and F1800W are displayed in Fig.~\ref{fig:calib} as 
examples.
The EEFs for the point sources at different positions show slight variations, which is expected given the spatial variations in the background (\S\ref{sec:pipeline}).  but the scatter is small ($\approx 1\%$ at $1\arcsec$). 
This small scatter indicates that the flat-field correction in \textsc{pipeline}
functions well.
However, we find systematic offsets between the measured EEFs for these point sources compared to the theoretical PSF EEF from 
{\sc webbpsf} (version 0.90).\footnote{https://www.stsci.edu/jwst/science-planning/proposal-planning-toolbox/psf-simulation-tool}
The specific reason is possibly due to the inconsistencies between the 
MIRI calibration files adopted by \mirisim\ and \textsc{pipeline} (K.~Gordon 2020, private communication).  It is also possible that one of the simulation steps are truncating the wings of the PSF.  As these possibilities are associated with the production of the simulations, they are unlikely to be present in the real \jwst\ data.  Here, for our purposes we measure the systematic offset as a ``scaling correction'', which we then apply to the flux densities in MIRI that we measure.   This scaling ``correction factor'' is calculated as,  
\begin{equation}
\label{eq:scaling}
    f_{\rm scaling} = \frac{0.8}{\rm med(EEF_{PT, 0.8})},
\end{equation}
where $\rm med(EEF_{PT, 0.8})$ is the median of point-source EEF at 
$r_{0.8}$ (the radius where PSF EEF$=0.8$; see Fig.~\ref{fig:calib}).
The value 0.8 is chosen, because most of the light is encircled within 
$r_{0.8}$ and the pixels are still source-dominated instead of 
background-dominated at $r_{0.8}$.
The values of $f_{\rm scaling}$ are all within $\simeq$10\% of unity and range from 0.87 (F1800W) to 1.08 (F770W).
We apply the scaling correction by multiplying all pixel values 
by a factor of $f_{\rm scaling}$ for each MIRI band. 
After this procedure, the point-source EEFs become similar to the PSF 
EEF as displayed in Fig.~\ref{fig:calib}.

Besides the scaling correction above, we also need to consider aperture
corrections. 
This is because the PSF faint wings extend to large radii, and can contribute to the background (Fig.~\ref{fig:calib}).
We calculate the aperture-correction by comparing the total (input/true) flux density to the averaged measured flux density.  The aperture-correction factor is,
\begin{equation}
\label{eq:aper}
    f_{\rm aper} = \frac{1}{\rm EEF_{PSF, 1.5\arcsec}},
\end{equation}
where $\rm EEF_{PSF, 1.5\arcsec}$ is the PSF EEF at 1.5$\arcsec$. 
We choose the radius of $1.5\arcsec$ as a fiducial value as beyond this radius the EEF has little 
growth (see Fig.~\ref{fig:calib}).
We apply the aperture correction by multiplying the measured source
fluxes by a factor of $f_{\rm aper}$. 
As expected, $f_{\rm aper}$ is higher for longer wavelengths which
correspond to more extended PSF. 
The values of $f_{\rm aper}$ increase with PSF FWHM (which scales with wavelength) and range from 1.10 (F770W) to 1.19 (F2100W).

We note that the correction procedures above are not limited to the 
simulated MIRI data and can be applied to real MIRI data.
After the launch of \jwst, some bright ``standard stars'' with 
known mid-IR fluxes can play the same role as the simulated point 
sources in this work.
In fact, some key calibration programs focusing on standard stars
have already been scheduled for the upcoming 
Cycle~1.\footnote{https://jwst-docs.stsci.edu/data-processing-and-calibration-files/absolute-flux-calibration} 

\begin{figure}
    \centering
    \includegraphics[width=\columnwidth]{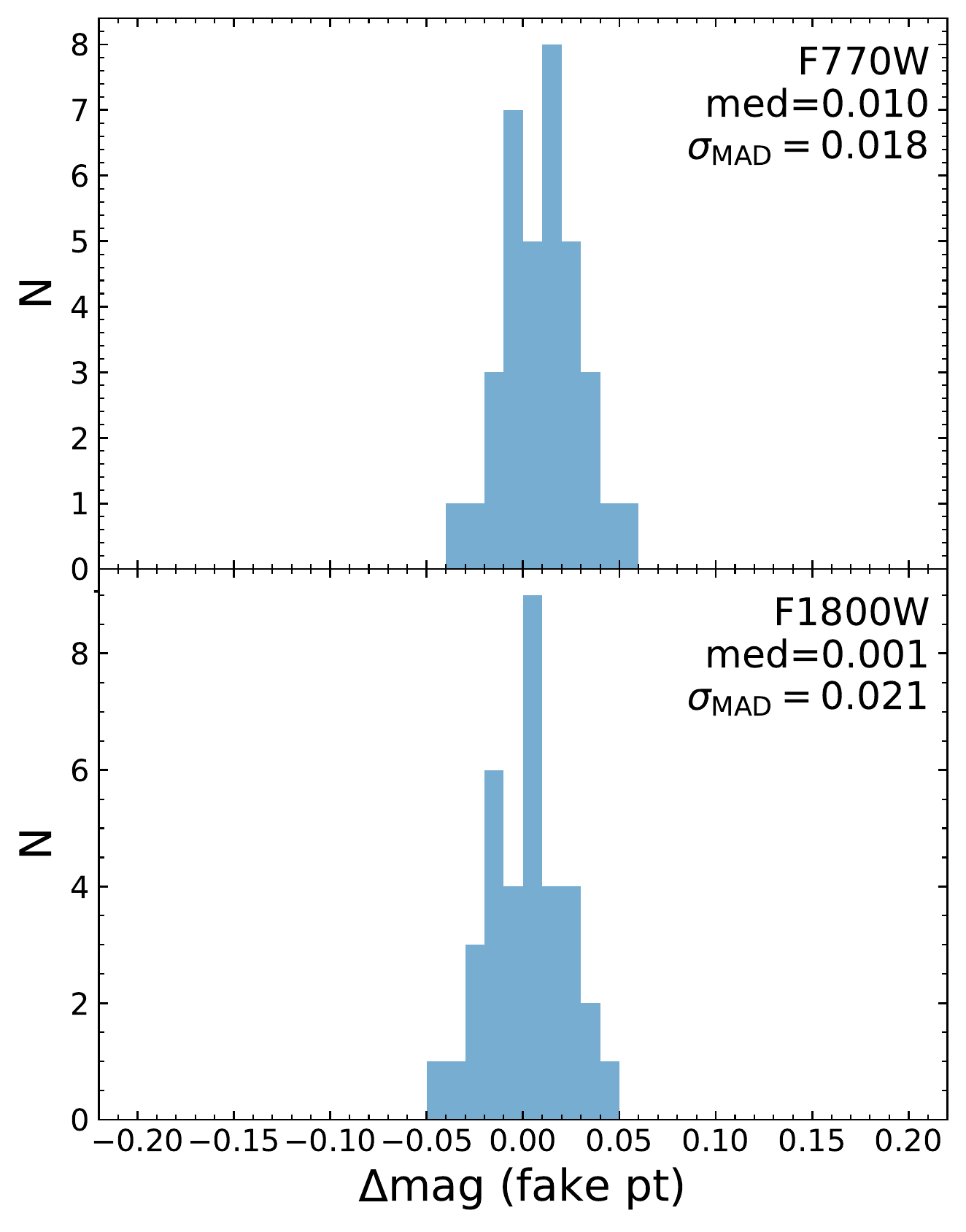}
    \caption{The stacked-histogram distribution of $\Delta $mag 
    ({\sc tphot}-measured $-$ model-input) of F770W (top) and F1800W (bottom)
    for the simulated bright point sources of Fig.~\ref{fig:fake_pts} (where ``fake pt" stands for ``fake point sources").
    The median and $\smad$ values are labeled.
    The small systematics and scatters indicate that our photometric 
    corrections and {\sc tphot} measurements are reliable.
    }
    \label{fig:det_mag_fk_pt}
\end{figure}

After the corrections above, we extract photometry for the simulated
point sources using {\sc tphot} (see \S\ref{sec:photo}).
We assess the photometry quality by comparing the measured magnitudes 
and the model-input magnitudes.  
Fig.~\ref{fig:det_mag_fk_pt} shows the distribution of the magnitude 
offsets for F770W and F1800W as examples.
The systematics and scatters are both $\lesssim 0.02$~mag for all MIRI 
bands, indicate that our photometric corrections above and {\sc tphot} 
measurements are reliable.

\section{The simulation of F160W imaging data}
\label{app:f160w}
\begin{figure}
    \centering
	\includegraphics[width=\columnwidth]{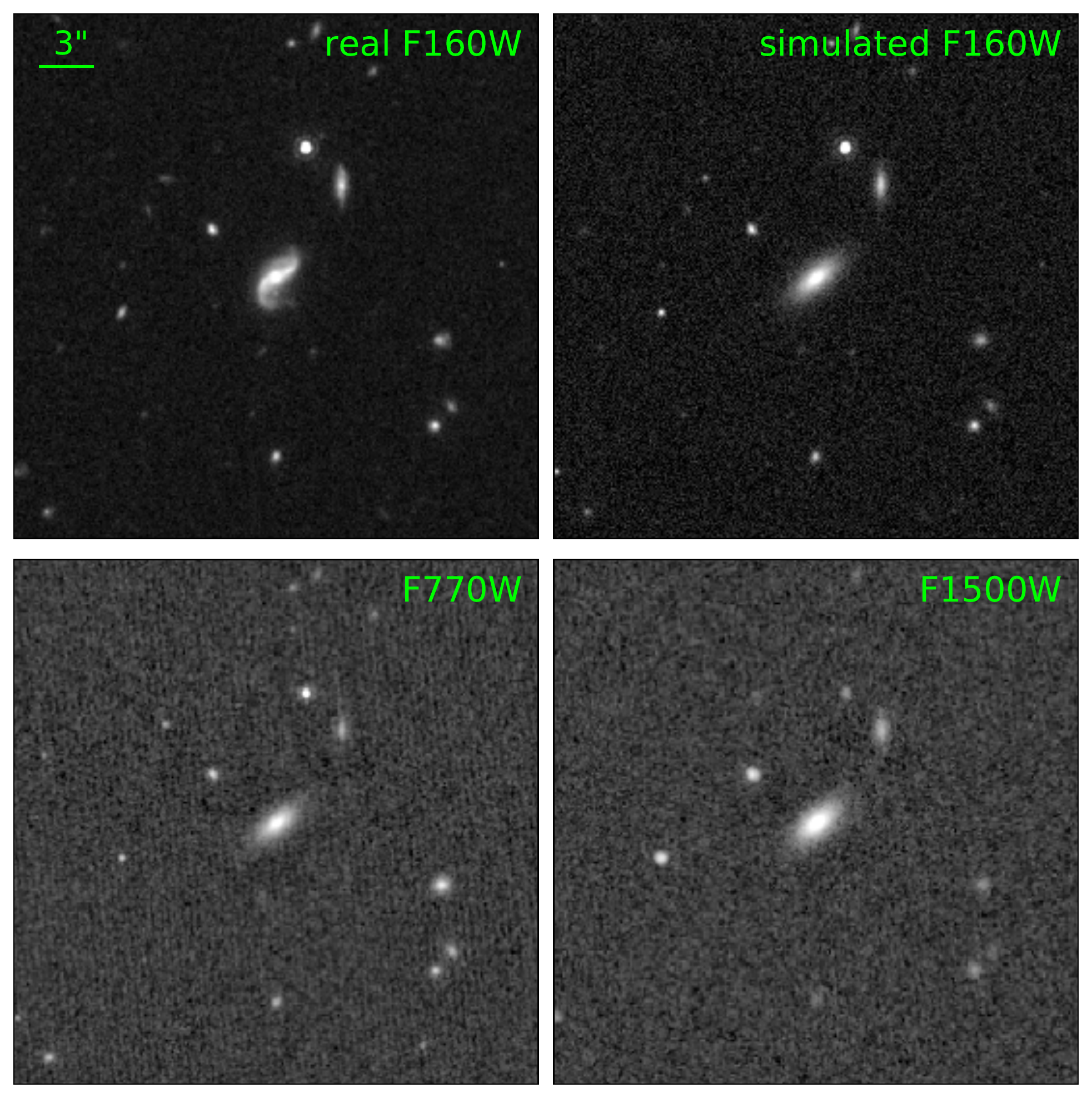}
    \caption{\textit{Top}: The real (left) and simulated (right) \hst\ F160W 
    $30\arcsec \times 30\arcsec$ cutouts with the same arcsinh color scale.  
    The galaxies morphologies in the real and simulated images are broadly similar 
    but not exactly the same.
    For example, the simulation does not reproduce the spiral arms of the galaxy 
    at the center, since the simulation is limited to S\'ersic profiles. 
    \textit{Bottom}:
    The simulated (pure-galaxy models) MIRI F770W (left) and F1500W (right) images 
    of the same sky region as the top panels. 
    The MIRI morphologies are more consistent with that of the simulated F160W 
    than the real F160W by design.
    This morphological consistency is the reason why we need to use the simulated
    F160W in photometry extraction.
    }
    \label{fig:f160w}
\end{figure}

To extract the photometry on the simulated MIRI images, we also need to 
simulated F160W images that have source morphologies consistent with MIRI 
morphologies (see \S\ref{sec:tphot_prep}). 
Note that we use the simulated \hst\ image only to measure MIRI flux 
densities from TPHOT (we use the real F160W source fluxes and errors 
from the CANDELS/EGS catalog when fitting the SEDs in \S\ref{sec:pure_gal}). 
First, we create an empty image covering the same region as the simulated MIRI 
fields, with pixel size set to $0.06\arcsec$ (the same as CANDELS survey).
We then position all sources \citep{stefanon17} within the FOV on the image. 
The fluxes are set to the observed F160W fluxes in the \cite{stefanon17} 
catalog, and the morphologies use the same S\'ersic model parameters (from the \citealt{van_der_wel12} catalog) that are used for the \mirisim\ input above. 
In addition, for a point-like sources, we allow slight shifts to center the pixel at the pixel nearest to the source 
position;
for a S\'ersic source, we generate its profile utilizing the {\sc sersic2d} 
function of {\sc astropy} \citep{astropy}, which allows fractional pixel 
positions.
Next, we convolve the image with the F160W point spread function 
(PSF) derived by the 3DHST team \citep{momcheva16}.
We perform this convolution process with the {\sc convolve\_fft} function 
of {\sc astropy}.
Finally, we add a random Gaussian noise to each pixel with an amplitude from 
the RMS map produced by the CANDELS team \citep{grogin11, koekemoer11}.
Fig.~\ref{fig:f160w} compares the cutouts of the original/simulated F160W images
and two MIRI-band images.  
For the galaxy near the center, the simulated and original F160W profiles 
are obviously different, since a S\'ersic profile, as a proxy, cannot reproduce 
the complex morphological feature of spiral arms.
As expected, the MIRI images have the similar profiles as the simulated F160W 
image, because they share the same input S\'ersic profile.

\section{{Comparison between source extractor and tphot photometry}}
\label{app:tphot_vs_se}
{{\sc source extractor} \citep{bertin96} is widely used for photometry
extraction \citep[e.g.,][]{caputi13, yang14, kauffmann20}.
Therefore, it is instructive to test the performance of {\sc source extractor}
on our simulated MIRI images.
Using F1500W as an example, we perform {\sc source extractor} photometry and 
compare the resulting quality with that from {\sc tphot} below.
The results are similar for other MIRI bands.
} 

{The ``AUTO'' mode photometry in {\sc source extractor} provides an 
estimation of the bulk of the total flux based on an adaptive-aperture 
algorithm \citep{kron80}.
The AUTO method automatically adjusts the size and shape of an elliptical 
aperture and it works best for bright resolved sources with high S/Ns. 
However, for a faint low-S/N source, the aperture size (and thus the resulting
flux) could be systematically underestimated \citep{bertin96}.
This issue can be alleviated by adopting a circular photometry aperture of 
a fixed size for faint sources.
Therefore, we adopt a strategy of choosing the brighter one of the AUTO and 
fixed-aperture fluxes (after corrections) for each source.}

{We run {\sc source extractor} on the same F1500W image as in our 
{\sc tphot} run (\S\ref{sec:tphot_res}; pure-galaxy inputs).
Our {\sc source extractor} run adopts the parameters listed in 
Table.~\ref{tab:se}. 
The aperture size (PHOTO\_APERTURES $=31$) corresponds to 80\% 
point-source EEF in F1500W, and thus we correct the resulting aperture 
fluxes by a factor of 1/0.8 to estimate the total fluxes.
Based on our choice of ``PHOTO\_AUTOPARAMS'', the AUTO flux is 
expected to recover $\approx 94\%$ of the total flux on average \citep{bertin96}. 
Therefore, we correct the resulting AUTO fluxes by a factor of 1/0.94
to obtain the total.
The run detects 379 F1500W sources.
We associate them with the input source catalog using a $1''$ matching 
radius, and this procedure results in 177 matches.
We assess the {\sc source extractor} photometry quality using these 177 
sources below.}

\begin{figure*}
    \centering
	\includegraphics[width=2\columnwidth]{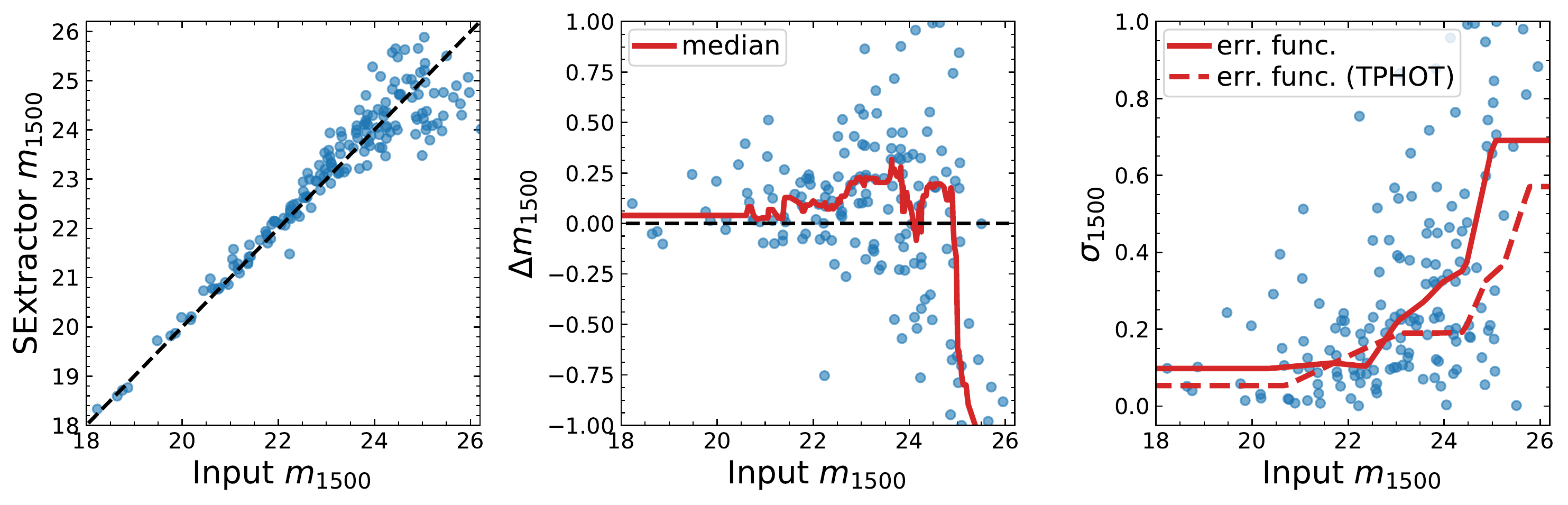}
    \caption{{Similar format as Fig.~\ref{fig:tphot_mags} but for 
    {\sc source extractor} photometry. 
    In the right panel, we also plot the {\sc tphot} error function
    for comparison.
    The {\sc source extractor} uncertainties are larger than {\sc tphot}
    uncertainties in general.}
    }
    \label{fig:se_mags}
\end{figure*}

{Fig.~\ref{fig:se_mags} compares the resulting {\sc source extractor} 
magnitudes with the input magnitudes. 
The right panel of Fig.~\ref{fig:se_mags} displays the error function 
of both {\sc source extractor} and {\sc tphot} photometry.
The {\sc tphot} photometric uncertainties are generally smaller than the 
{\sc source extractor} uncertainties at a given magnitude, indicating 
that {\sc tphot} photometry has superior quality than {\sc source extractor} 
photometry. 
Another advantage of {\sc tphot} is that it can estimate the fluxes of all sources
provided in the high-resolution prior catalog (see \S\ref{sec:tphot_prep}), and this provides valuable information for the mid-IR emission in these sources.
However, {\sc source extractor} could miss many sources that are faint in the 
photometric band.
For example, our {\sc source extractor} run misses 53\% of the sources in the
input catalog.
These sources are all present in our {\sc tphot} catalog.
}

\begin{figure}
    \centering
	\includegraphics[width=\columnwidth]{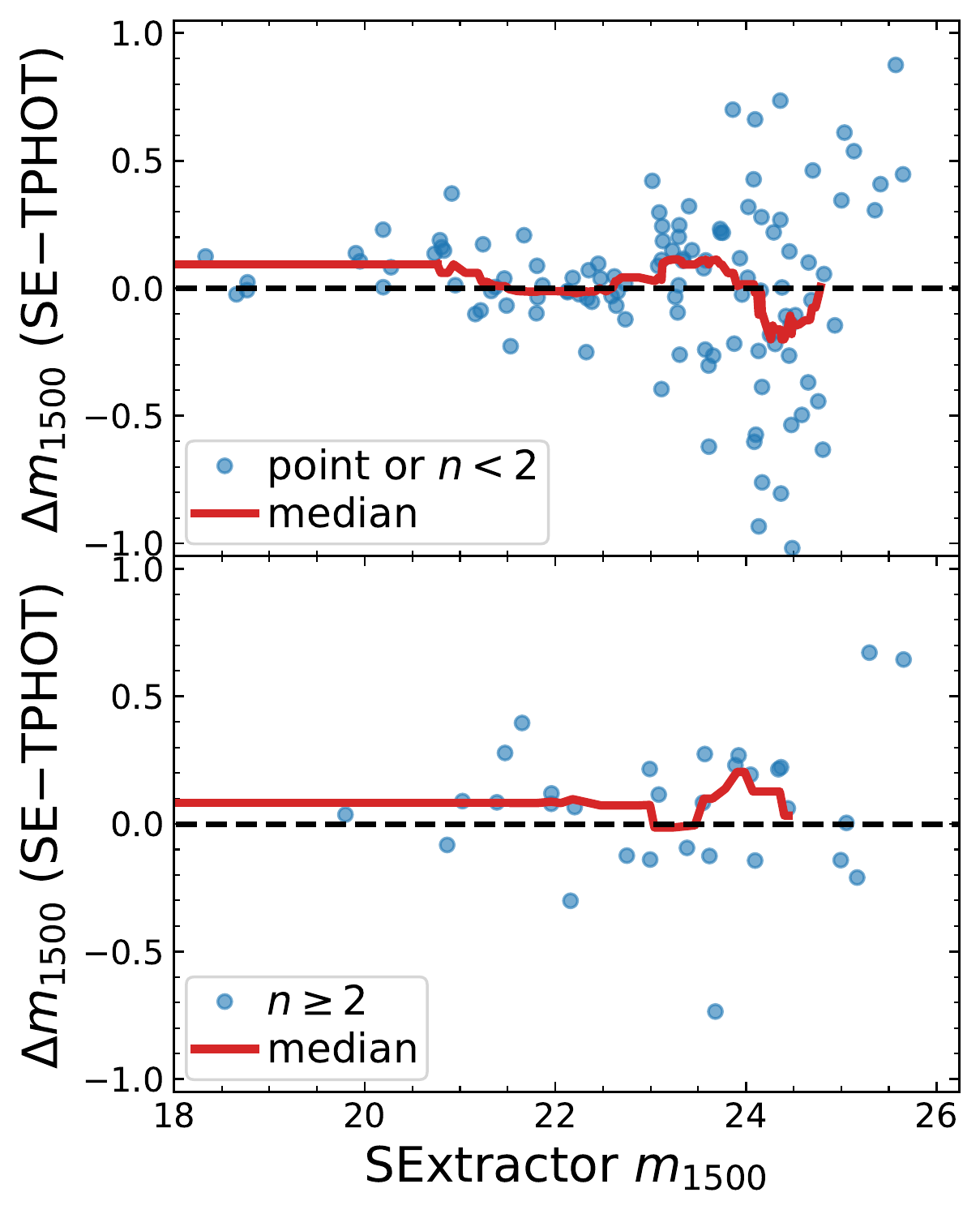}
    \caption{{The difference between {\sc source extractor} and 
             TPHOT F1500W magnitudes ($\Delta m_{1500}$) versus the 
             {\sc source extractor} magnitude. 
             The red curves represent the running median.
             The upper panel is for S\'ersic $n<2$ and point sources, 
             and the lower panel is for $n>2$ sources.
             For $n\geq 2$ sources, the median $\Delta m_{1500}$ is 
             $\approx 0$, indicating that, like {\sc tphot}, 
             {\sc source extractor} also misses the light from the 
             faint extended wings (see \S\ref{sec:tphot_res}).}
    }
    \label{fig:se_vs_tphot}
\end{figure}

{Fig.~\ref{fig:se_vs_tphot} directly compares {\sc source extractor} 
vs.\ {\sc tphot} magnitudes for point/$n<2$ and $n\geq2$ sources (similar as 
Fig.~\ref{fig:photo_vs_n}). 
The systematic differences (as indicated by the median) between 
{\sc source extractor} and {\sc tphot} magnitudes are generally 
small ($\lesssim 0.05$~mag).
This systematic consistency indicates that, like {\sc tphot}, 
{\sc source extractor} also misses the light from the 
faint extended wings for $n\geq2$ sources (see \S\ref{sec:tphot_res}).
}

\begin{table}
\centering
\caption{{The adopted {\sc source extractor} parameter values for the 
        F1500W photometry}}
\label{tab:se}
\begin{tabular}{ll} \hline\hline
Parameter & Value \\
\hline
DETECT\_MINAREA & 5 \\
FILTER & Y \\
FILTER\_NAME & gauss\_5.0\_9x9.conv \\
PHOTO\_APERTURES & 31 \\
PHOTO\_AUTOPARAMS & 2.5, 3.5 \\
BACK\_SIZE & 64 \\
BACK\_FILTERSIZE & 6 \\
BACKPHOTO\_TYPE & LOCAL \\
BACKPHOTO\_THICK & 40 \\
\hline
\end{tabular}
\begin{flushleft}
    {{\sc Note.} --- For parameters not listed here, we use default values.}
\end{flushleft}
\end{table}

\section{Complete Comparisons between Input and Measured Flux Densities 
for all MIRI bands}
\label{app:tphot_mags}
\begin{figure*}
    \centering
	\includegraphics[width=0.65\columnwidth]{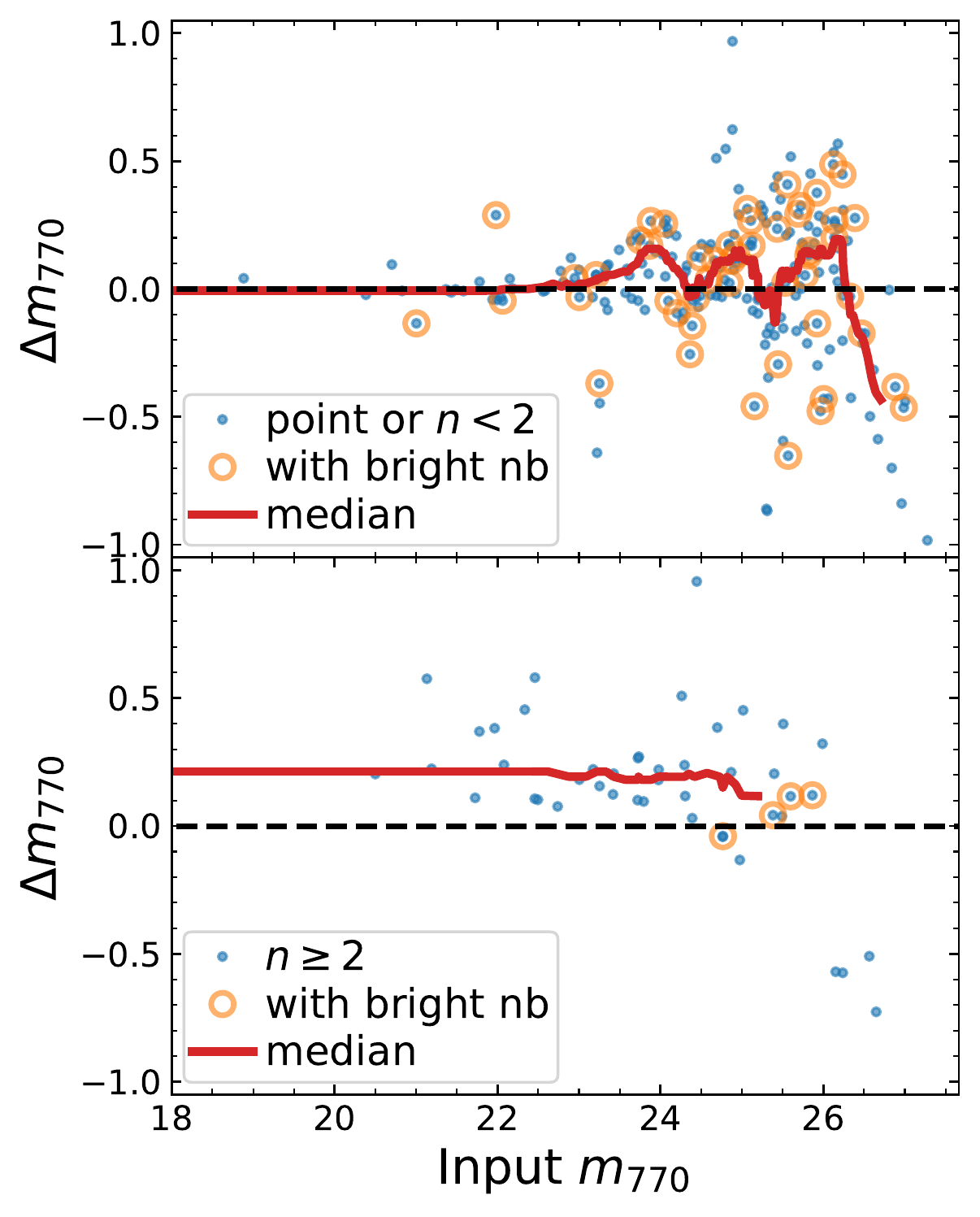}
	\includegraphics[width=0.65\columnwidth]{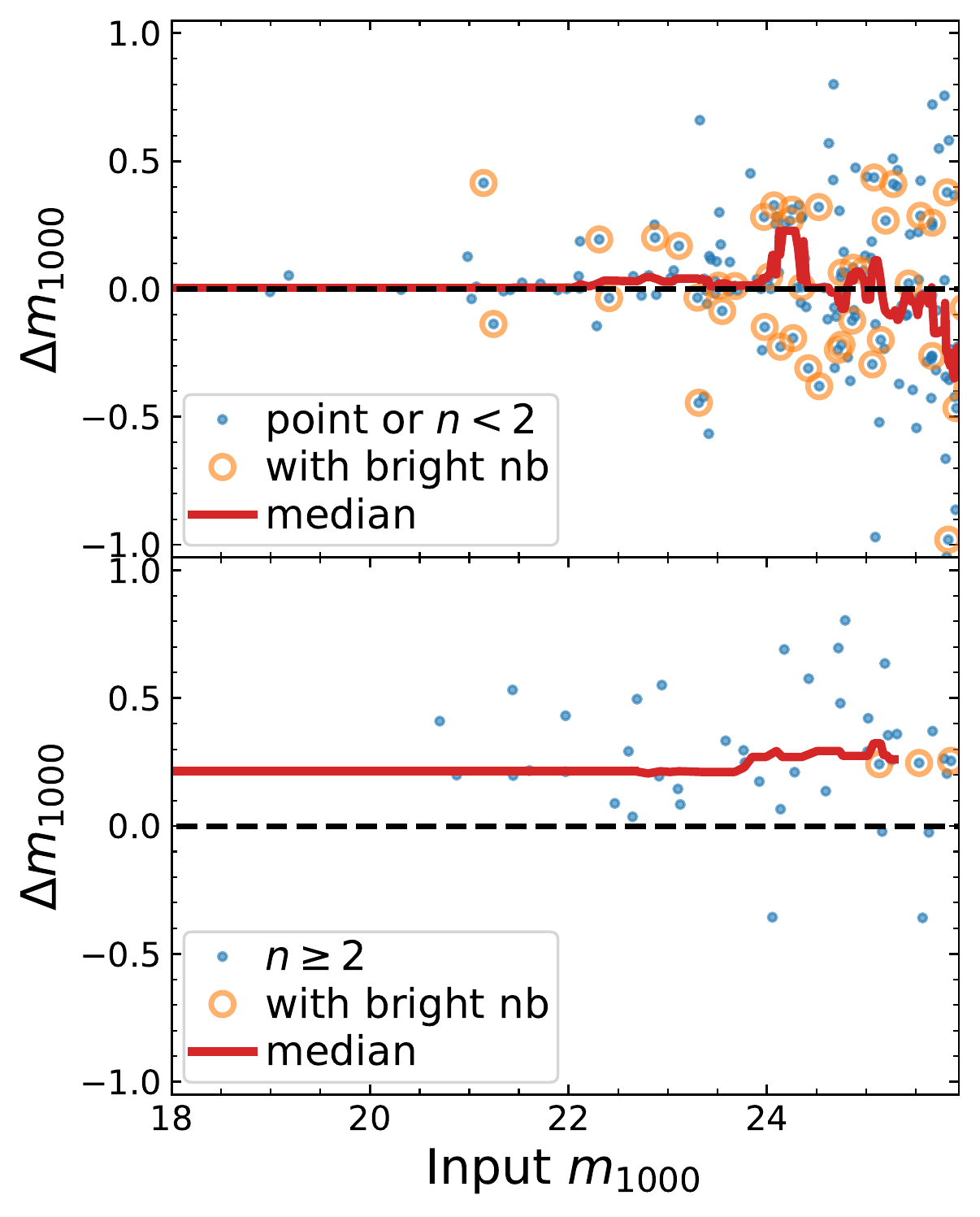}
	\includegraphics[width=0.65\columnwidth]{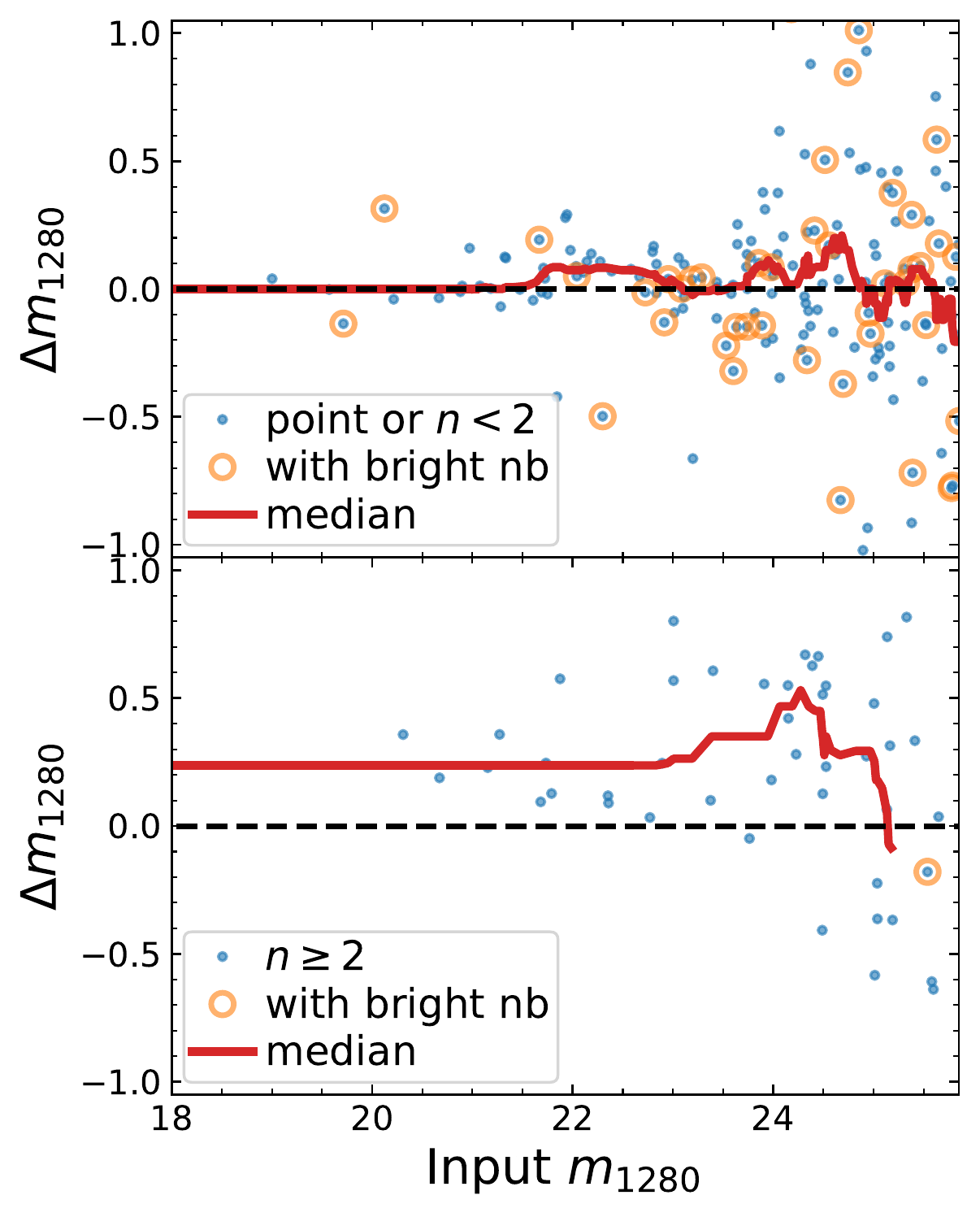}
	\includegraphics[width=0.65\columnwidth]{photo_vs_n_F1500W.pdf}
	\includegraphics[width=0.65\columnwidth]{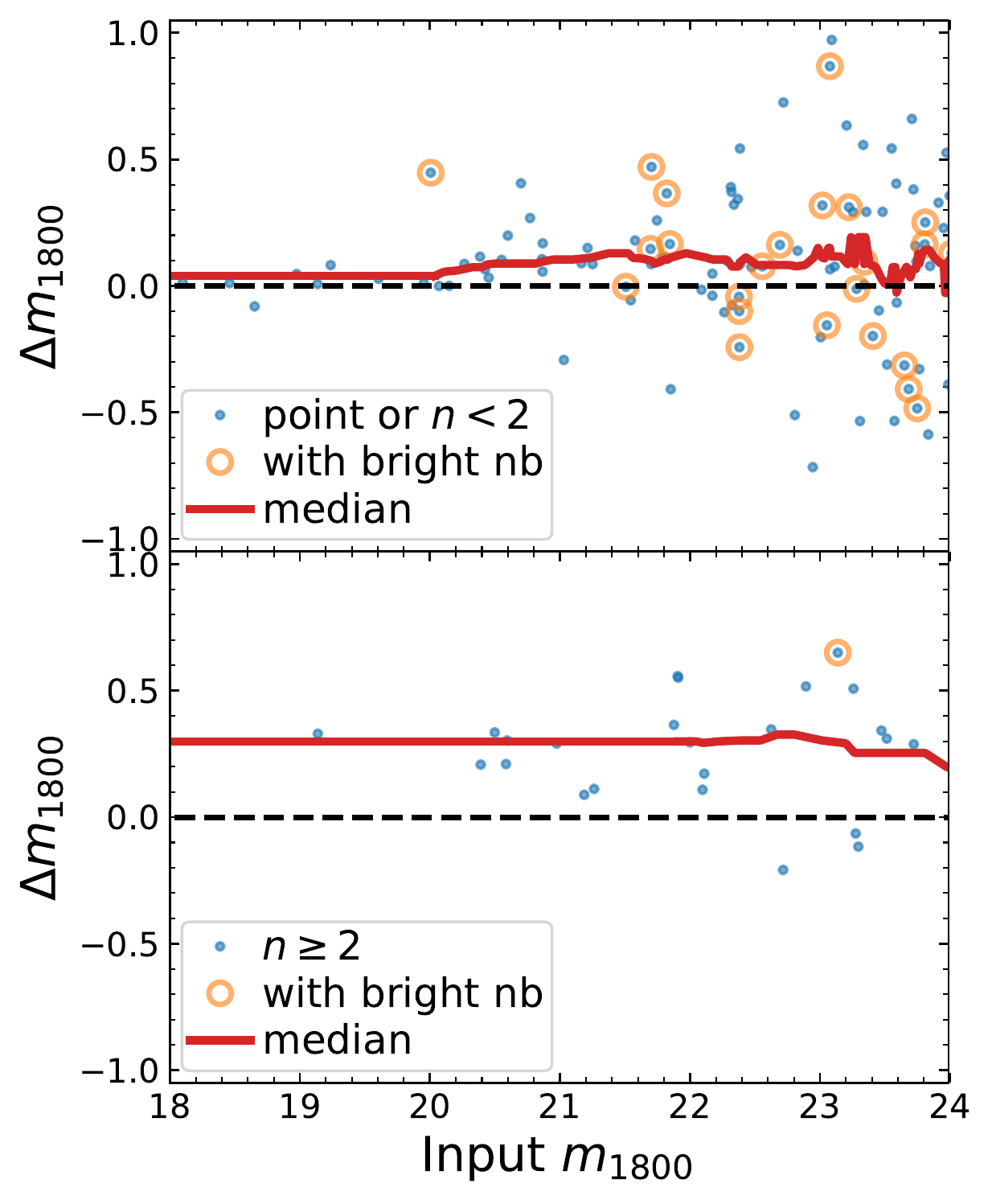}
	\includegraphics[width=0.65\columnwidth]{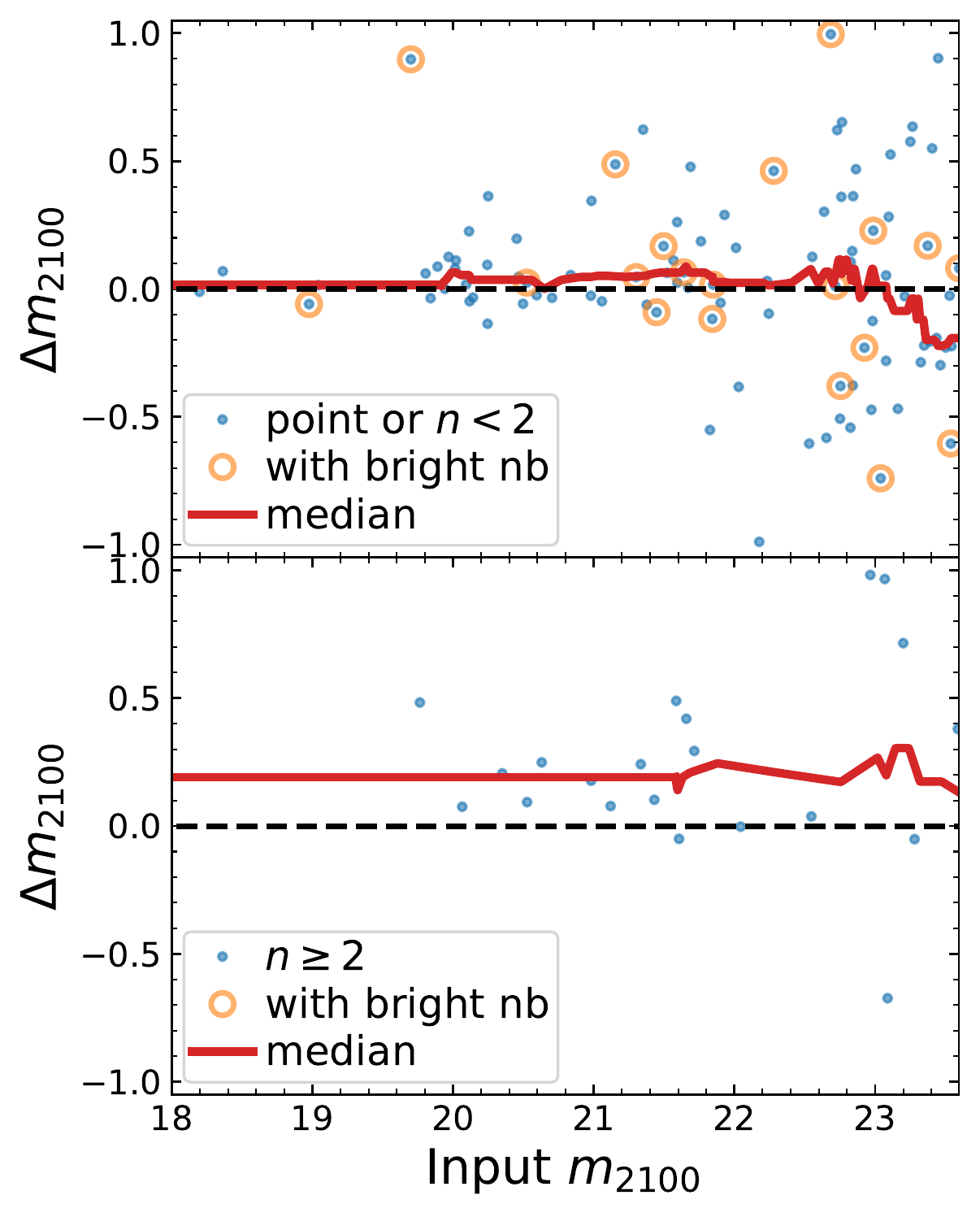}
    \caption{Same format as Fig.~\ref{fig:photo_vs_n} but for all MIRI bands.
    }
    \label{fig:photo_vs_n_full}
\end{figure*}

In this appendix, Fig.~\ref{fig:tphot_mags_full} shows the comparison between 
input and measured flux densities for all the MIRI bands (and is similar to 
Fig.~\ref{fig:tphot_mags} above).  
Figure~\ref{fig:photo_vs_n_full} compares the results dividing the sample by 
morphology (S\'ersic index) for all bands (and is similar to 
Fig.~\ref{fig:photo_vs_n} above).

\begin{figure*}
    \centering
	\includegraphics[width=2\columnwidth]{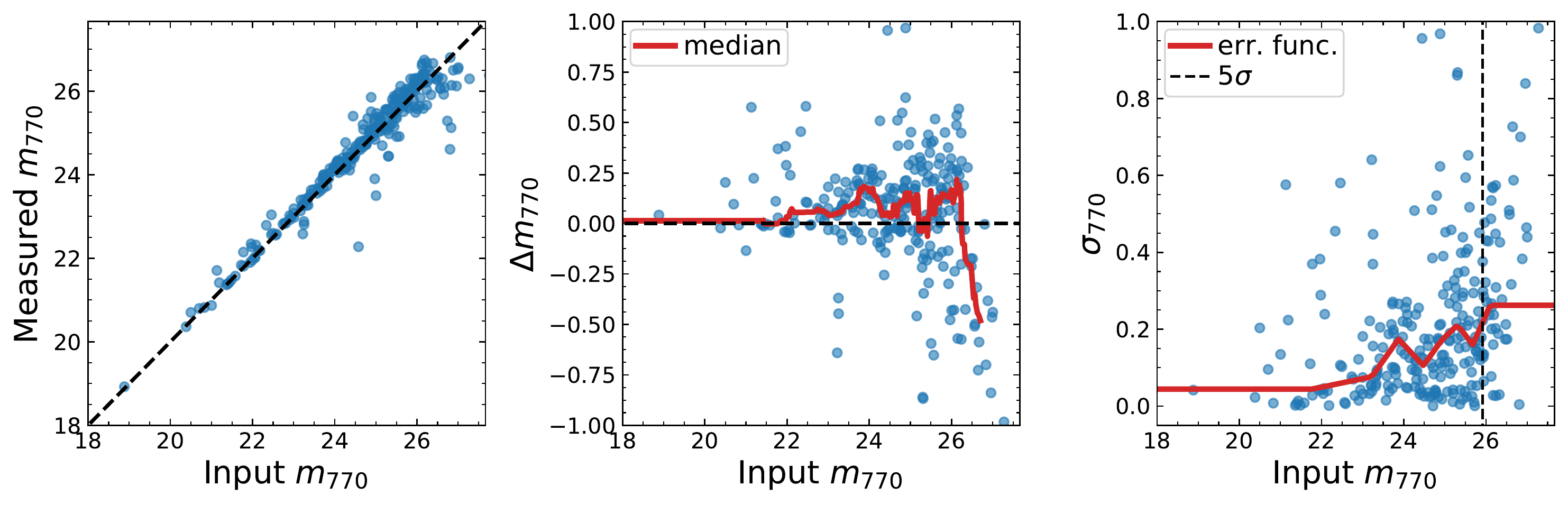}
	\includegraphics[width=2\columnwidth]{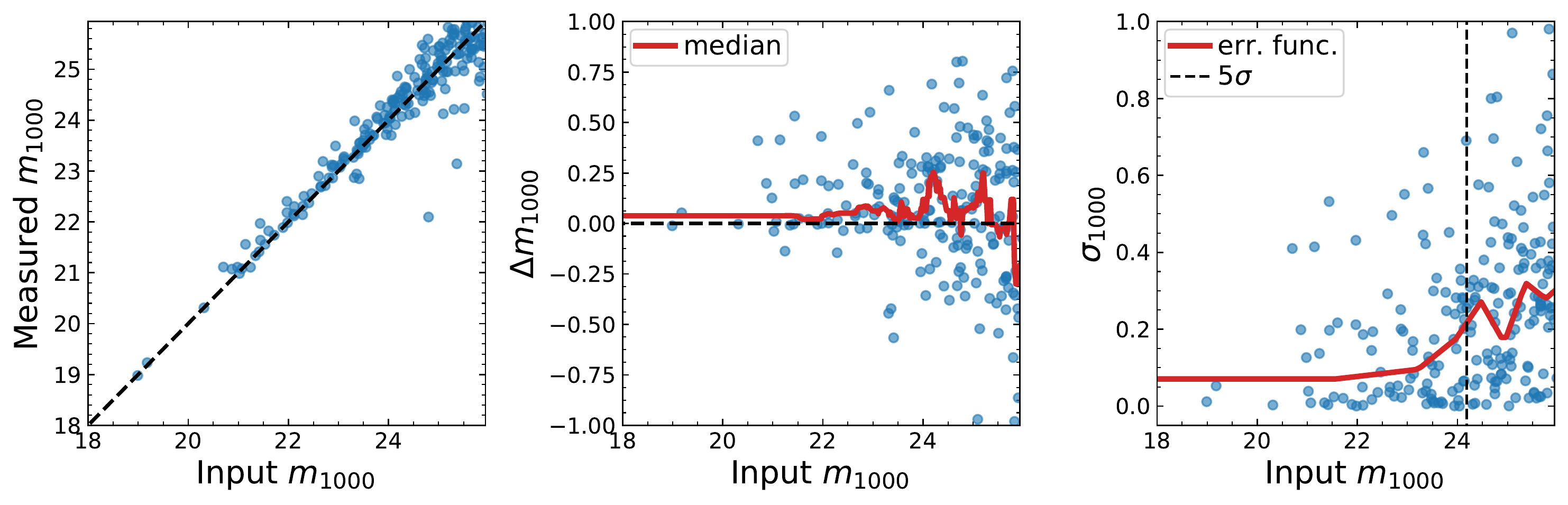}
	\includegraphics[width=2\columnwidth]{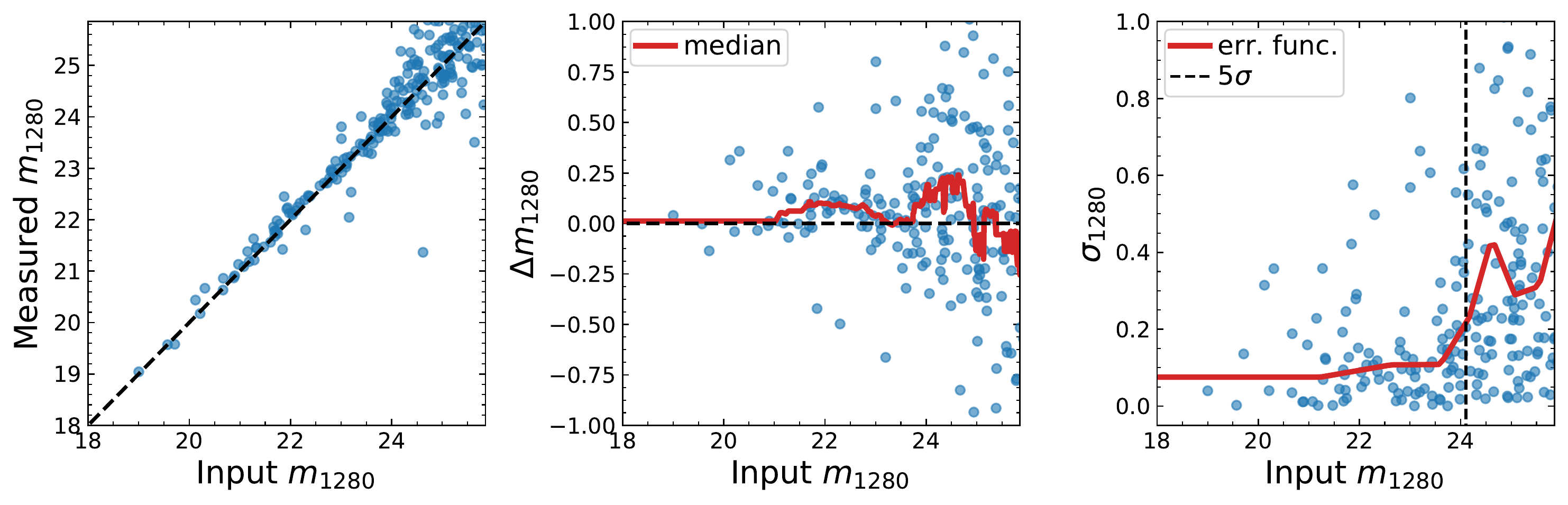}
	\includegraphics[width=2\columnwidth]{photo_det_vs_mag_F1500W.pdf}
    \caption{Same format as Fig.~\ref{fig:tphot_mags} but for all MIRI bands.
    }
    \label{fig:tphot_mags_full}
\end{figure*}

\begin{figure*}
    \setcounter{figure}{22}
    \centering
	\includegraphics[width=2\columnwidth]{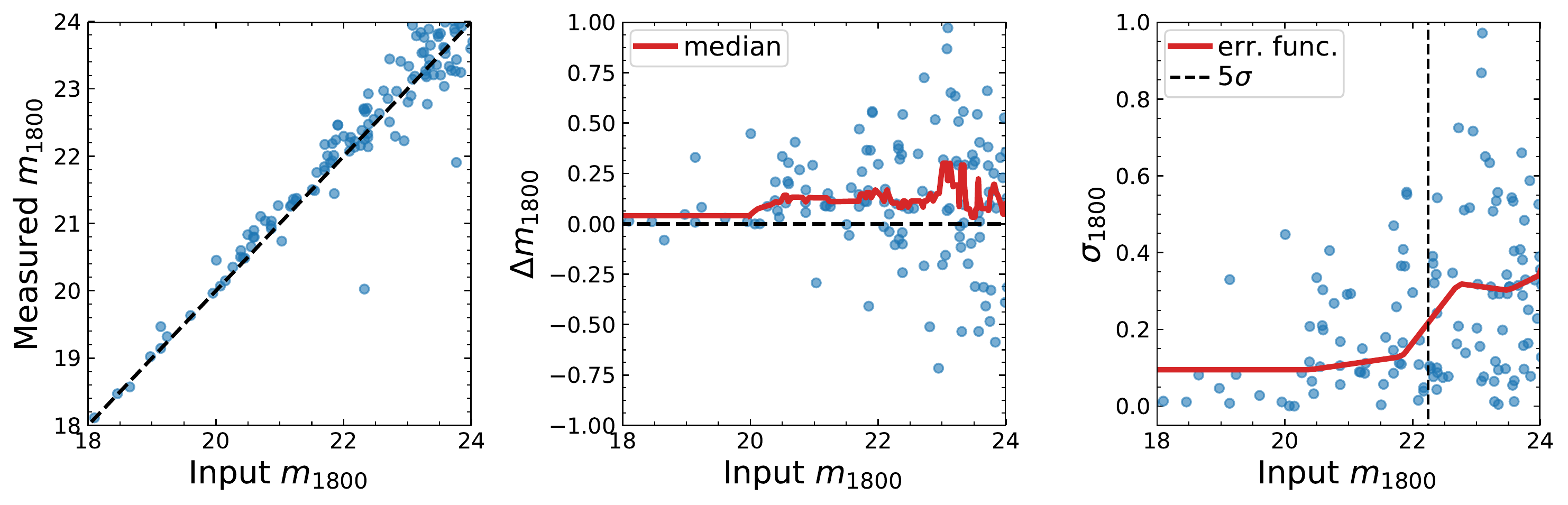}
	\includegraphics[width=2\columnwidth]{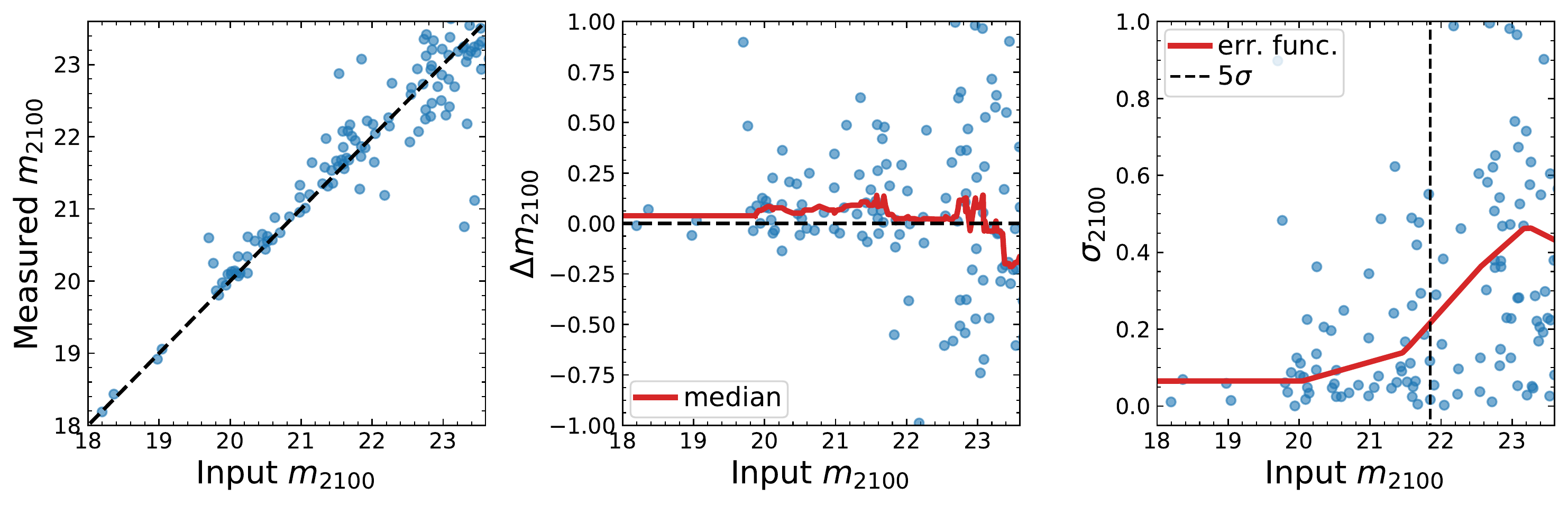}
    \caption{Continued.
    }
\end{figure*}

\bibliography{all}{}

\begin{thebibliography}{}
\expandafter\ifx\csname natexlab\endcsname\relax\def\natexlab#1{#1}\fi
\providecommand{\url}[1]{\href{#1}{#1}}
\providecommand{\dodoi}[1]{doi:~\href{http://doi.org/#1}{\nolinkurl{#1}}}
\providecommand{\doeprint}[1]{\href{http://ascl.net/#1}{\nolinkurl{http://ascl.net/#1}}}
\providecommand{\doarXiv}[1]{\href{https://arxiv.org/abs/#1}{\nolinkurl{https://arxiv.org/abs/#1}}}

\bibitem[{{Aird} {et~al.}(2010){Aird}, {Nandra}, {Laird}, {Georgakakis},
  {Ashby}, {Barmby}, {Coil}, {Huang}, {Koekemoer}, {Steidel}, \&
  {Willmer}}]{aird10}
{Aird}, J., {Nandra}, K., {Laird}, E.~S., {et~al.} 2010, \mnras, 401, 2531,
  \dodoi{10.1111/j.1365-2966.2009.15829.x}

\bibitem[{{Akylas} {et~al.}(2012){Akylas}, {Georgakakis}, {Georgantopoulos},
  {Brightman}, \& {Nandra}}]{akylas12}
{Akylas}, A., {Georgakakis}, A., {Georgantopoulos}, I., {Brightman}, M., \&
  {Nandra}, K. 2012, \aap, 546, A98, \dodoi{10.1051/0004-6361/201219387}

\bibitem[{{Alberts} {et~al.}(2020){Alberts}, {Rujopakarn}, {Rieke},
  {Jagannathan}, \& {Nyland}}]{alberts20}
{Alberts}, S., {Rujopakarn}, W., {Rieke}, G.~H., {Jagannathan}, P., \&
  {Nyland}, K. 2020, \apj, 901, 168, \dodoi{10.3847/1538-4357/abb1a0}

\bibitem[{{Alexander} {et~al.}(2008){Alexander}, {Chary}, {Pope}, {Bauer},
  {Brandt}, {Daddi}, {Dickinson}, {Elbaz}, \& {Reddy}}]{alexander08}
{Alexander}, D.~M., {Chary}, R.-R., {Pope}, A., {et~al.} 2008, \apj, 687, 835,
  \dodoi{10.1086/591928}

\bibitem[{{Alonso-Herrero} {et~al.}(2006){Alonso-Herrero},
  {P{\'e}rez-Gonz{\'a}lez}, {Alexand er}, {Rieke}, {Rigopoulou}, {Le Floc'h},
  {Barmby}, {Papovich}, {Rigby}, {Bauer}, {Brandt}, {Egami}, {Willner}, {Dole},
  \& {Huang}}]{alonso_herrero06}
{Alonso-Herrero}, A., {P{\'e}rez-Gonz{\'a}lez}, P.~G., {Alexand er}, D.~M.,
  {et~al.} 2006, \apj, 640, 167, \dodoi{10.1086/499800}

\bibitem[{{Antonucci}(1993)}]{antonucci93}
{Antonucci}, R. 1993, \araa, 31, 473,
  \dodoi{10.1146/annurev.aa.31.090193.002353}

\bibitem[{{Armus} {et~al.}(2007){Armus}, {Charmandaris}, {Bernard-Salas},
  {Spoon}, {Marshall}, {Higdon}, {Desai}, {Teplitz}, {Hao}, {Devost}, {Brand
  l}, {Wu}, {Sloan}, {Soifer}, {Houck}, \& {Herter}}]{armus07}
{Armus}, L., {Charmandaris}, V., {Bernard-Salas}, J., {et~al.} 2007, \apj, 656,
  148, \dodoi{10.1086/510107}

\bibitem[{{Ashby} {et~al.}(2018){Ashby}, {Caputi}, {Cowley}, {Deshmukh},
  {Dunlop}, {Milvang-Jensen}, {Fynbo}, {Muzzin}, {McCracken}, {Le F{\`e}vre},
  {Huang}, \& {Zhang}}]{ashby18}
{Ashby}, M.~L.~N., {Caputi}, K.~I., {Cowley}, W., {et~al.} 2018, \apjs, 237,
  39, \dodoi{10.3847/1538-4365/aad4fb}

\bibitem[{{Asmus}(2019)}]{asmus19}
{Asmus}, D. 2019, arXiv e-prints, arXiv:1908.03552.
\newblock \doarXiv{1908.03552}

\bibitem[{{Assef} {et~al.}(2013){Assef}, {Stern}, {Kochanek}, {Blain},
  {Brodwin}, {Brown}, {Donoso}, {Eisenhardt}, {Jannuzi}, {Jarrett}, {Stanford},
  {Tsai}, {Wu}, \& {Yan}}]{assef13}
{Assef}, R.~J., {Stern}, D., {Kochanek}, C.~S., {et~al.} 2013, \apj, 772, 26,
  \dodoi{10.1088/0004-637X/772/1/26}

\bibitem[{{Astropy Collaboration} {et~al.}(2018){Astropy Collaboration},
  {Price-Whelan}, {Sip{\H{o}}cz}, {G{\"u}nther}, {Lim}, {Crawford}, {Conseil},
  {Shupe}, {Craig}, \& {Dencheva}}]{astropy}
{Astropy Collaboration}, {Price-Whelan}, A.~M., {Sip{\H{o}}cz}, B.~M., {et~al.}
  2018, \aj, 156, 123, \dodoi{10.3847/1538-3881/aabc4f}

\bibitem[{{Azadi} {et~al.}(2018){Azadi}, {Coil}, {Aird}, {Shivaei}, {Reddy},
  {Shapley}, {Kriek}, {Freeman}, {Leung}, {Mobasher}, {Price}, {Sanders},
  {Siana}, \& {Zick}}]{azadi18}
{Azadi}, M., {Coil}, A., {Aird}, J., {et~al.} 2018, \apj, 866, 63,
  \dodoi{10.3847/1538-4357/aad3c8}

\bibitem[{{Barbary}(2016)}]{barbary16}
{Barbary}, K. 2016, The Journal of Open Source Software, 1, 58,
  \dodoi{10.21105/joss.00058}

\bibitem[{{Bertin} \& {Arnouts}(1996)}]{bertin96}
{Bertin}, E., \& {Arnouts}, S. 1996, \aaps, 117, 393

\bibitem[{{Bisigello} {et~al.}(2016){Bisigello}, {Caputi}, {Colina}, {Le
  F{\`e}vre}, {N{\o}rgaard-Nielsen}, {P{\'e}rez-Gonz{\'a}lez}, {Pye}, {van der
  Werf}, {Ilbert}, {Grogin}, \& {Koekemoer}}]{bisigello16}
{Bisigello}, L., {Caputi}, K.~I., {Colina}, L., {et~al.} 2016, \apjs, 227, 19,
  \dodoi{10.3847/0067-0049/227/2/19}

\bibitem[{{Bisigello} {et~al.}(2017){Bisigello}, {Caputi}, {Colina}, {Le
  F{\`e}vre}, {N{\o}rgaard-Nielsen}, {P{\'e}rez-Gonz{\'a}lez}, {van der Werf},
  {Ilbert}, {Grogin}, \& {Koekemoer}}]{bisigello17}
---. 2017, \apjs, 231, 3, \dodoi{10.3847/1538-4365/aa7a14}

\bibitem[{{Bonato} {et~al.}(2017){Bonato}, {Sajina}, {De Zotti}, {McKinney},
  {Baronchelli}, {Negrello}, {Marchesini}, {Roebuck}, {Shipley}, {Kurinsky},
  {Pope}, {Noriega-Crespo}, {Yan}, \& {Kirkpatrick}}]{bonato17}
{Bonato}, M., {Sajina}, A., {De Zotti}, G., {et~al.} 2017, \apj, 836, 171,
  \dodoi{10.3847/1538-4357/aa5c85}

\bibitem[{{Boquien} {et~al.}(2019){Boquien}, {Burgarella}, {Roehlly}, {Buat},
  {Ciesla}, {Corre}, {Inoue}, \& {Salas}}]{boquien19}
{Boquien}, M., {Burgarella}, D., {Roehlly}, Y., {et~al.} 2019, \aap, 622, A103,
  \dodoi{10.1051/0004-6361/201834156}

\bibitem[{{Boucaud} {et~al.}(2016){Boucaud}, {Bocchio}, {Abergel}, {Orieux},
  {Dole}, \& {Hadj-Youcef}}]{boucaud16}
{Boucaud}, A., {Bocchio}, M., {Abergel}, A., {et~al.} 2016, \aap, 596, A63,
  \dodoi{10.1051/0004-6361/201629080}

\bibitem[{{Brandt} \& {Alexander}(2015)}]{brandt15}
{Brandt}, W.~N., \& {Alexander}, D.~M. 2015, \aapr, 23, 1,
  \dodoi{10.1007/s00159-014-0081-z}

\bibitem[{{Bruzual} \& {Charlot}(2003)}]{bruzual03}
{Bruzual}, G., \& {Charlot}, S. 2003, \mnras, 344, 1000,
  \dodoi{10.1046/j.1365-8711.2003.06897.x}

\bibitem[{{Buchner} {et~al.}(2015){Buchner}, {Georgakakis}, {Nandra},
  {Brightman}, {Menzel}, {Liu}, {Hsu}, {Salvato}, {Rangel}, {Aird}, {Merloni},
  \& {Ross}}]{buchner15}
{Buchner}, J., {Georgakakis}, A., {Nandra}, K., {et~al.} 2015, \apj, 802, 89,
  \dodoi{10.1088/0004-637X/802/2/89}

\bibitem[{{Calzetti} {et~al.}(2000){Calzetti}, {Armus}, {Bohlin}, {Kinney},
  {Koornneef}, \& {Storchi-Bergmann}}]{calzetti00}
{Calzetti}, D., {Armus}, L., {Bohlin}, R.~C., {et~al.} 2000, \apj, 533, 682,
  \dodoi{10.1086/308692}

\bibitem[{{Caputi}(2013)}]{caputi13}
{Caputi}, K.~I. 2013, \apj, 768, 103, \dodoi{10.1088/0004-637X/768/2/103}

\bibitem[{{Chabrier}(2003)}]{chabrier03}
{Chabrier}, G. 2003, \apjl, 586, L133, \dodoi{10.1086/374879}

\bibitem[{{Chang} {et~al.}(2017){Chang}, {Le Floc'h}, {Juneau}, {da Cunha},
  {Salvato}, {Civano}, {Marchesi}, {Ilbert}, {Toba}, {Lim}, {Tang}, {Wang},
  {Ferraro}, {Urry}, {Griffiths}, \& {Kartaltepe}}]{chang17}
{Chang}, Y.-Y., {Le Floc'h}, E., {Juneau}, S., {et~al.} 2017, \apjs, 233, 19,
  \dodoi{10.3847/1538-4365/aa97da}

\bibitem[{{Chary} {et~al.}(2007){Chary}, {Teplitz}, {Dickinson}, {Koo}, {Le
  Floc'h}, {Marcillac}, {Papovich}, \& {Stern}}]{chary07}
{Chary}, R.-R., {Teplitz}, H.~I., {Dickinson}, M.~E., {et~al.} 2007, \apj, 665,
  257, \dodoi{10.1086/519243}

\bibitem[{{Civano} {et~al.}(2016){Civano}, {Marchesi}, {Comastri}, {Urry},
  {Elvis}, {Cappelluti}, {Puccetti}, {Brusa}, {Zamorani}, {Hasinger},
  {Aldcroft}, {Alexander}, {Allevato}, {Brunner}, {Capak}, {Finoguenov},
  {Fiore}, {Fruscione}, {Gilli}, {Glotfelty}, {Griffiths}, {Hao}, {Harrison},
  {Jahnke}, {Kartaltepe}, {Karim}, {LaMassa}, {Lanzuisi}, {Miyaji}, {Ranalli},
  {Salvato}, {Sargent}, {Scoville}, {Schawinski}, {Schinnerer}, {Silverman},
  {Smolcic}, {Stern}, {Toft}, {Trakhenbrot}, {Treister}, \&
  {Vignali}}]{civano16}
{Civano}, F., {Marchesi}, S., {Comastri}, A., {et~al.} 2016, \apj, 819, 62,
  \dodoi{10.3847/0004-637X/819/1/62}

\bibitem[{{Clements} {et~al.}(2011){Clements}, {Bendo}, {Pearson}, {Khan},
  {Matsuura}, \& {Shirahata}}]{clements11}
{Clements}, D.~L., {Bendo}, G., {Pearson}, C., {et~al.} 2011, \mnras, 411, 373,
  \dodoi{10.1111/j.1365-2966.2010.17689.x}

\bibitem[{{Daddi} {et~al.}(2007{\natexlab{a}}){Daddi}, {Dickinson}, {Morrison},
  {Chary}, {Cimatti}, {Elbaz}, {Frayer}, {Renzini}, {Pope}, {Alexander},
  {Bauer}, {Giavalisco}, {Huynh}, {Kurk}, \& {Mignoli}}]{daddi07b}
{Daddi}, E., {Dickinson}, M., {Morrison}, G., {et~al.} 2007{\natexlab{a}},
  \apj, 670, 156, \dodoi{10.1086/521818}

\bibitem[{{Daddi} {et~al.}(2007{\natexlab{b}}){Daddi}, {Alexander},
  {Dickinson}, {Gilli}, {Renzini}, {Elbaz}, {Cimatti}, {Chary}, {Frayer},
  {Bauer}, {Brandt}, {Giavalisco}, {Grogin}, {Huynh}, {Kurk}, {Mignoli},
  {Morrison}, {Pope}, \& {Ravindranath}}]{daddi07}
{Daddi}, E., {Alexander}, D.~M., {Dickinson}, M., {et~al.} 2007{\natexlab{b}},
  \apj, 670, 173, \dodoi{10.1086/521820}

\bibitem[{{Dale} {et~al.}(2014){Dale}, {Helou}, {Magdis}, {Armus},
  {D{\'\i}az-Santos}, \& {Shi}}]{dale14}
{Dale}, D.~A., {Helou}, G., {Magdis}, G.~E., {et~al.} 2014, \apj, 784, 83,
  \dodoi{10.1088/0004-637X/784/1/83}

\bibitem[{{Del Moro} {et~al.}(2016){Del Moro}, {Alexander}, {Bauer}, {Daddi},
  {Kocevski}, {McIntosh}, {Stanley}, {Brandt}, {Elbaz}, {Harrison}, {Luo},
  {Mullaney}, \& {Xue}}]{del_moro16}
{Del Moro}, A., {Alexander}, D.~M., {Bauer}, F.~E., {et~al.} 2016, \mnras, 456,
  2105, \dodoi{10.1093/mnras/stv2748}

\bibitem[{{Dickinson} {et~al.}(2006){Dickinson}, {Alexander}, {Bell}, {Brand
  t}, {Calzetti}, {Casertano}, {Chapman}, {Chary}, {Daddi}, {Davis}, {Dole},
  {Dunlop}, {Eisenhardt}, {Elbaz}, {Faber}, {Fazio}, {Ferguson}, {Frayer},
  {Giavalisco}, {Halpern}, {Huang}, {Huynh}, {Ivison}, {Koekemoer}, {Le
  Floc'h}, {Morrison}, {Moustakas}, {Papovich}, {Pope}, {Renzini}, {Rieke},
  {Rix}, {Scott}, {Smail}, {Yan}, {van Dokkum}, \& {van der
  Werf}}]{dickinson06}
{Dickinson}, M., {Alexander}, D., {Bell}, E., {et~al.} 2006, {A Deep-Wide
  Far-Infrared Survey of Cosmological Star Formation and AGN Activity}, Spitzer
  Proposal

\bibitem[{{Dole} {et~al.}(2004){Dole}, {Rieke}, {Lagache}, {Puget},
  {Alonso-Herrero}, {Bai}, {Blaylock}, {Egami}, {Engelbracht}, {Gordon},
  {Hines}, {Kelly}, {Le Floc'h}, {Misselt}, {Morrison}, {Muzerolle},
  {Papovich}, {P{\'e}rez-Gonz{\'a}lez}, {Rieke}, {Rigby}, {Neugebauer},
  {Stansberry}, {Su}, {Young}, {Beichman}, \& {Richards}}]{dole04}
{Dole}, H., {Rieke}, G.~H., {Lagache}, G., {et~al.} 2004, \apjs, 154, 93,
  \dodoi{10.1086/422690}

\bibitem[{{Donley} {et~al.}(2012){Donley}, {Koekemoer}, {Brusa}, {Capak},
  {Cardamone}, {Civano}, {Ilbert}, {Impey}, {Kartaltepe}, {Miyaji}, {Salvato},
  {Sanders}, {Trump}, \& {Zamorani}}]{donley12}
{Donley}, J.~L., {Koekemoer}, A.~M., {Brusa}, M., {et~al.} 2012, \apj, 748,
  142, \dodoi{10.1088/0004-637X/748/2/142}

\bibitem[{{Draine} \& {Li}(2007)}]{draine07}
{Draine}, B.~T., \& {Li}, A. 2007, \apj, 657, 810, \dodoi{10.1086/511055}

\bibitem[{{Finkelstein} {et~al.}(2017){Finkelstein}, {Dickinson}, {Ferguson},
  {Grazian}, {Grogin}, {Kartaltepe}, {Kewley}, {Kocevski}, {Koekemoer}, {Lotz},
  {Papovich}, {Pentericci}, {Perez-Gonzalez}, {Pirzkal}, {Ravindranath},
  {Somerville}, {Trump}, \& {Wilkins}}]{finkelstein17}
{Finkelstein}, S.~L., {Dickinson}, M., {Ferguson}, H.~C., {et~al.} 2017, {The
  Cosmic Evolution Early Release Science (CEERS) Survey}, JWST Proposal ID
  1345. Cycle 0 Early Release Scienc

\bibitem[{{Georgantopoulos} {et~al.}(2011){Georgantopoulos}, {Rovilos},
  {Akylas}, {Comastri}, {Ranalli}, {Vignali}, {Balestra}, {Gilli}, \&
  {Cappelluti}}]{georgantopoulos11}
{Georgantopoulos}, I., {Rovilos}, E., {Akylas}, A., {et~al.} 2011, \aap, 534,
  A23, \dodoi{10.1051/0004-6361/201117400}

\bibitem[{{Gilli} {et~al.}(2007){Gilli}, {Comastri}, \& {Hasinger}}]{gilli07}
{Gilli}, R., {Comastri}, A., \& {Hasinger}, G. 2007, \aap, 463, 79,
  \dodoi{10.1051/0004-6361:20066334}

\bibitem[{{Glasse} {et~al.}(2015){Glasse}, {Rieke}, {Bauwens},
  {Garc{\'\i}a-Mar{\'\i}n}, {Ressler}, {Rost}, {Tikkanen}, {Vandenbussche}, \&
  {Wright}}]{glasse15}
{Glasse}, A., {Rieke}, G.~H., {Bauwens}, E., {et~al.} 2015, \pasp, 127, 686,
  \dodoi{10.1086/682259}

\bibitem[{{Grogin} {et~al.}(2011){Grogin}, {Kocevski}, {Faber}, {Ferguson},
  {Koekemoer}, {Riess}, {Acquaviva}, {Alexander}, {Almaini}, {Ashby}, {Barden},
  {Bell}, {Bournaud}, {Brown}, {Caputi}, {Casertano}, {Cassata}, {Castellano},
  {Challis}, {Chary}, {Cheung}, {Cirasuolo}, {Conselice}, {Roshan Cooray},
  {Croton}, {Daddi}, {Dahlen}, {Dav{\'e}}, {de Mello}, {Dekel}, {Dickinson},
  {Dolch}, {Donley}, {Dunlop}, {Dutton}, {Elbaz}, {Fazio}, {Filippenko},
  {Finkelstein}, {Fontana}, {Gardner}, {Garnavich}, {Gawiser}, {Giavalisco},
  {Grazian}, {Guo}, {Hathi}, {H{\"a}ussler}, {Hopkins}, {Huang}, {Huang},
  {Jha}, {Kartaltepe}, {Kirshner}, {Koo}, {Lai}, {Lee}, {Li}, {Lotz}, {Lucas},
  {Madau}, {McCarthy}, {McGrath}, {McIntosh}, {McLure}, {Mobasher},
  {Moustakas}, {Mozena}, {Nandra}, {Newman}, {Niemi}, {Noeske}, {Papovich},
  {Pentericci}, {Pope}, {Primack}, {Rajan}, {Ravindranath}, {Reddy}, {Renzini},
  {Rix}, {Robaina}, {Rodney}, {Rosario}, {Rosati}, {Salimbeni}, {Scarlata},
  {Siana}, {Simard}, {Smidt}, {Somerville}, {Spinrad}, {Straughn}, {Strolger},
  {Telford}, {Teplitz}, {Trump}, {van der Wel}, {Villforth}, {Wechsler},
  {Weiner}, {Wiklind}, {Wild}, {Wilson}, {Wuyts}, {Yan}, \& {Yun}}]{grogin11}
{Grogin}, N.~A., {Kocevski}, D.~D., {Faber}, S.~M., {et~al.} 2011, \apjs, 197,
  35, \dodoi{10.1088/0067-0049/197/2/35}

\bibitem[{{H{\"a}ussler} {et~al.}(2007){H{\"a}ussler}, {McIntosh}, {Barden},
  {Bell}, {Rix}, {Borch}, {Beckwith}, {Caldwell}, {Heymans}, {Jahnke}, {Jogee},
  {Koposov}, {Meisenheimer}, {S{\'a}nchez}, {Somerville}, {Wisotzki}, \&
  {Wolf}}]{haussler07}
{H{\"a}ussler}, B., {McIntosh}, D.~H., {Barden}, M., {et~al.} 2007, \apjs, 172,
  615, \dodoi{10.1086/518836}

\bibitem[{{Hickox} \& {Alexander}(2018)}]{hickox18}
{Hickox}, R.~C., \& {Alexander}, D.~M. 2018, \araa, 56, 625,
  \dodoi{10.1146/annurev-astro-081817-051803}

\bibitem[{{Kauffmann} {et~al.}(2020){Kauffmann}, {Le F{\`e}vre}, {Ilbert},
  {Chevallard}, {Williams}, {Curtis-Lake}, {Colina}, {P{\'e}rez-Gonz{\'a}lez},
  {Pye}, \& {Caputi}}]{kauffmann20}
{Kauffmann}, O.~B., {Le F{\`e}vre}, O., {Ilbert}, O., {et~al.} 2020, \aap, 640,
  A67, \dodoi{10.1051/0004-6361/202037450}

\bibitem[{{Kennicutt}(1998)}]{kennicutt98}
{Kennicutt}, Jr., R.~C. 1998, \apj, 498, 541, \dodoi{10.1086/305588}

\bibitem[{{Kirkpatrick} {et~al.}(2015){Kirkpatrick}, {Pope}, {Sajina},
  {Roebuck}, {Yan}, {Armus}, {D{\'\i}az-Santos}, \&
  {Stierwalt}}]{kirkpatrick15}
{Kirkpatrick}, A., {Pope}, A., {Sajina}, A., {et~al.} 2015, \apj, 814, 9,
  \dodoi{10.1088/0004-637X/814/1/9}

\bibitem[{{Kirkpatrick} {et~al.}(2017){Kirkpatrick}, {Alberts}, {Pope},
  {Barro}, {Bonato}, {Kocevski}, {P{\'e}rez-Gonz{\'a}lez}, {Rieke},
  {Rodr{\'\i}guez-Mu{\~n}oz}, {Sajina}, {Grogin}, {Mantha}, {Pand ya}, {Pforr},
  {Salvato}, \& {Santini}}]{kirkpatrick17}
{Kirkpatrick}, A., {Alberts}, S., {Pope}, A., {et~al.} 2017, \apj, 849, 111,
  \dodoi{10.3847/1538-4357/aa911d}

\bibitem[{{Klaassen} {et~al.}(2020){Klaassen}, {Geers}, {Beard}, {O'Brien},
  {Cossou}, {Gastaud}, {Coulais}, {Schreiber}, {Kavanagh}, {Topinka},
  {Azzollini}, {De Meester}, {Bouwman}, {Glasse}, {Glauser}, {Law}, {Cracraft},
  {Murray}, {Sargent}, {Jones}, \& {Wright}}]{klaassen20}
{Klaassen}, P.~D., {Geers}, V.~C., {Beard}, S.~M., {et~al.} 2020, arXiv
  e-prints, arXiv:2010.15710.
\newblock \doarXiv{2010.15710}

\bibitem[{{Koekemoer} {et~al.}(2011){Koekemoer}, {Faber}, {Ferguson}, {Grogin},
  {Kocevski}, {Koo}, {Lai}, {Lotz}, {Lucas}, {McGrath}, {Ogaz}, {Rajan},
  {Riess}, {Rodney}, {Strolger}, {Casertano}, {Castellano}, {Dahlen},
  {Dickinson}, {Dolch}, {Fontana}, {Giavalisco}, {Grazian}, {Guo}, {Hathi},
  {Huang}, {van der Wel}, {Yan}, {Acquaviva}, {Alexander}, {Almaini}, {Ashby},
  {Barden}, {Bell}, {Bournaud}, {Brown}, {Caputi}, {Cassata}, {Challis},
  {Chary}, {Cheung}, {Cirasuolo}, {Conselice}, {Roshan Cooray}, {Croton},
  {Daddi}, {Dav{\'e}}, {de Mello}, {de Ravel}, {Dekel}, {Donley}, {Dunlop},
  {Dutton}, {Elbaz}, {Fazio}, {Filippenko}, {Finkelstein}, {Frazer}, {Gardner},
  {Garnavich}, {Gawiser}, {Gruetzbauch}, {Hartley}, {H{\"a}ussler},
  {Herrington}, {Hopkins}, {Huang}, {Jha}, {Johnson}, {Kartaltepe},
  {Khostovan}, {Kirshner}, {Lani}, {Lee}, {Li}, {Madau}, {McCarthy},
  {McIntosh}, {McLure}, {McPartland}, {Mobasher}, {Moreira}, {Mortlock},
  {Moustakas}, {Mozena}, {Nandra}, {Newman}, {Nielsen}, {Niemi}, {Noeske},
  {Papovich}, {Pentericci}, {Pope}, {Primack}, {Ravindranath}, {Reddy},
  {Renzini}, {Rix}, {Robaina}, {Rosario}, {Rosati}, {Salimbeni}, {Scarlata},
  {Siana}, {Simard}, {Smidt}, {Snyder}, {Somerville}, {Spinrad}, {Straughn},
  {Telford}, {Teplitz}, {Trump}, {Vargas}, {Villforth}, {Wagner}, {Wandro},
  {Wechsler}, {Weiner}, {Wiklind}, {Wild}, {Wilson}, {Wuyts}, \&
  {Yun}}]{koekemoer11}
{Koekemoer}, A.~M., {Faber}, S.~M., {Ferguson}, H.~C., {et~al.} 2011, \apjs,
  197, 36, \dodoi{10.1088/0067-0049/197/2/36}

\bibitem[{{Komatsu} {et~al.}(2011){Komatsu}, {Smith}, {Dunkley}, {Bennett},
  {Gold}, {Hinshaw}, {Jarosik}, {Larson}, {Nolta}, {Page}, {Spergel},
  {Halpern}, {Hill}, {Kogut}, {Limon}, {Meyer}, {Odegard}, {Tucker}, {Weiland},
  {Wollack}, \& {Wright}}]{komatsu11}
{Komatsu}, E., {Smith}, K.~M., {Dunkley}, J., {et~al.} 2011, \apjs, 192, 18,
  \dodoi{10.1088/0067-0049/192/2/18}

\bibitem[{{Kron}(1980)}]{kron80}
{Kron}, R.~G. 1980, \apjs, 43, 305, \dodoi{10.1086/190669}

\bibitem[{{Laidler} {et~al.}(2007){Laidler}, {Papovich}, {Grogin}, {Idzi},
  {Dickinson}, {Ferguson}, {Hilbert}, {Clubb}, \& {Ravindranath}}]{laidler07}
{Laidler}, V.~G., {Papovich}, C., {Grogin}, N.~A., {et~al.} 2007, \pasp, 119,
  1325, \dodoi{10.1086/523898}

\bibitem[{{Leitherer} {et~al.}(2002){Leitherer}, {Li}, {Calzetti}, \&
  {Heckman}}]{leitherer02}
{Leitherer}, C., {Li}, I.~H., {Calzetti}, D., \& {Heckman}, T.~M. 2002, \apjs,
  140, 303, \dodoi{10.1086/342486}

\bibitem[{{Li} {et~al.}(2019){Li}, {Xue}, {Sun}, {Liu}, {Vito}, {Brandt},
  {Hughes}, {Yang}, {Tozzi}, {Zhu}, {Zheng}, {Luo}, {Chen}, {Vignali}, {Gilli},
  \& {Shu}}]{li19}
{Li}, J., {Xue}, Y., {Sun}, M., {et~al.} 2019, \apj, 877, 5,
  \dodoi{10.3847/1538-4357/ab184b}

\bibitem[{{Li} {et~al.}(2020){Li}, {Xue}, {Sun}, {Brand t}, {Yang}, {Vito},
  {Tozzi}, {Vignali}, {Comastri}, {Shu}, {Fang}, {Fan}, {Luo}, {Chen}, \&
  {Zheng}}]{li20}
---. 2020, arXiv e-prints, arXiv:2008.05863.
\newblock \doarXiv{2008.05863}

\bibitem[{{Liu} {et~al.}(2017){Liu}, {Tozzi}, {Wang}, {Brandt}, {Vignali},
  {Xue}, {Schneider}, {Comastri}, {Yang}, {Bauer}, {Paolillo}, {Luo}, {Gilli},
  {Wang}, {Giavalisco}, {Ji}, {Alexander}, {Mainieri}, {Shemmer}, {Koekemoer},
  \& {Risaliti}}]{liu17}
{Liu}, T., {Tozzi}, P., {Wang}, J.-X., {et~al.} 2017, \apjs, 232, 8,
  \dodoi{10.3847/1538-4365/aa7847}

\bibitem[{{Luo} {et~al.}(2017){Luo}, {Brandt}, {Xue}, {Lehmer}, {Alexander},
  {Bauer}, {Vito}, {Yang}, {Basu-Zych}, {Comastri}, {Gilli}, {Gu},
  {Hornschemeier}, {Koekemoer}, {Liu}, {Mainieri}, {Paolillo}, {Ranalli},
  {Rosati}, {Schneider}, {Shemmer}, {Smail}, {Sun}, {Tozzi}, {Vignali}, \&
  {Wang}}]{luo17}
{Luo}, B., {Brandt}, W.~N., {Xue}, Y.~Q., {et~al.} 2017, \apjs, 228, 2,
  \dodoi{10.3847/1538-4365/228/1/2}

\bibitem[{{Lutz} {et~al.}(2011){Lutz}, {Poglitsch}, {Altieri}, {Andreani},
  {Aussel}, {Berta}, {Bongiovanni}, {Brisbin}, {Cava}, {Cepa}, {Cimatti},
  {Daddi}, {Dominguez-Sanchez}, {Elbaz}, {F{\"o}rster Schreiber}, {Genzel},
  {Grazian}, {Gruppioni}, {Harwit}, {Le Floc'h}, {Magdis}, {Magnelli},
  {Maiolino}, {Nordon}, {P{\'e}rez Garc{\'{\i}}a}, {Popesso}, {Pozzi},
  {Riguccini}, {Rodighiero}, {Saintonge}, {Sanchez Portal}, {Santini}, {Shao},
  {Sturm}, {Tacconi}, {Valtchanov}, {Wetzstein}, \& {Wieprecht}}]{lutz11}
{Lutz}, D., {Poglitsch}, A., {Altieri}, B., {et~al.} 2011, \aap, 532, A90,
  \dodoi{10.1051/0004-6361/201117107}

\bibitem[{{Ma{\l}ek} {et~al.}(2014){Ma{\l}ek}, {Pollo}, {Takeuchi}, {Buat},
  {Burgarella}, {Malkan}, {Giovannoli}, {Kurek}, \& {Matsuura}}]{malek14}
{Ma{\l}ek}, K., {Pollo}, A., {Takeuchi}, T.~T., {et~al.} 2014, \aap, 562, A15,
  \dodoi{10.1051/0004-6361/201321665}

\bibitem[{{Merlin} {et~al.}(2015){Merlin}, {Fontana}, {Ferguson}, {Dunlop},
  {Elbaz}, {Bourne}, {Bruce}, {Buitrago}, {Castellano}, {Schreiber}, {Grazian},
  {McLure}, {Okumura}, {Shu}, {Wang}, {Amor{\'\i}n}, {Boutsia}, {Cappelluti},
  {Comastri}, {Derriere}, {Faber}, \& {Santini}}]{merlin15}
{Merlin}, E., {Fontana}, A., {Ferguson}, H.~C., {et~al.} 2015, \aap, 582, A15,
  \dodoi{10.1051/0004-6361/201526471}

\bibitem[{{Merlin} {et~al.}(2016){Merlin}, {Bourne}, {Castellano}, {Ferguson},
  {Wang}, {Derriere}, {Dunlop}, {Elbaz}, \& {Fontana}}]{merlin16}
{Merlin}, E., {Bourne}, N., {Castellano}, M., {et~al.} 2016, \aap, 595, A97,
  \dodoi{10.1051/0004-6361/201628751}

\bibitem[{{Merloni} {et~al.}(2014){Merloni}, {Bongiorno}, {Brusa}, {Iwasawa},
  {Mainieri}, {Magnelli}, {Salvato}, {Berta}, {Cappelluti}, {Comastri},
  {Fiore}, {Gilli}, {Koekemoer}, {Le Floc'h}, {Lusso}, {Lutz}, {Miyaji},
  {Pozzi}, {Riguccini}, {Rosario}, {Silverman}, {Symeonidis}, {Treister},
  {Vignali}, \& {Zamorani}}]{merloni14}
{Merloni}, A., {Bongiorno}, A., {Brusa}, M., {et~al.} 2014, \mnras, 437, 3550,
  \dodoi{10.1093/mnras/stt2149}

\bibitem[{{Messias} {et~al.}(2012){Messias}, {Afonso}, {Salvato}, {Mobasher},
  \& {Hopkins}}]{messias12}
{Messias}, H., {Afonso}, J., {Salvato}, M., {Mobasher}, B., \& {Hopkins}, A.~M.
  2012, \apj, 754, 120, \dodoi{10.1088/0004-637X/754/2/120}

\bibitem[{{Momcheva} {et~al.}(2016){Momcheva}, {Brammer}, {van Dokkum},
  {Skelton}, {Whitaker}, {Nelson}, {Fumagalli}, {Maseda}, {Leja}, {Franx},
  {Rix}, {Bezanson}, {Da Cunha}, {Dickey}, {F{\"o}rster Schreiber},
  {Illingworth}, {Kriek}, {Labb{\'e}}, {Ulf Lange}, {Lundgren}, {Magee},
  {Marchesini}, {Oesch}, {Pacifici}, {Patel}, {Price}, {Tal}, {Wake}, {van der
  Wel}, \& {Wuyts}}]{momcheva16}
{Momcheva}, I.~G., {Brammer}, G.~B., {van Dokkum}, P.~G., {et~al.} 2016, \apjs,
  225, 27, \dodoi{10.3847/0067-0049/225/2/27}

\bibitem[{{Nandra} {et~al.}(2015){Nandra}, {Laird}, {Aird}, {Salvato},
  {Georgakakis}, {Barro}, {Perez-Gonzalez}, {Barmby}, {Chary}, {Coil},
  {Cooper}, {Davis}, {Dickinson}, {Faber}, {Fazio}, {Guhathakurta}, {Gwyn},
  {Hsu}, {Huang}, {Ivison}, {Koo}, {Newman}, {Rangel}, {Yamada}, \&
  {Willmer}}]{nandra15}
{Nandra}, K., {Laird}, E.~S., {Aird}, J.~A., {et~al.} 2015, \apjs, 220, 10,
  \dodoi{10.1088/0067-0049/220/1/10}

\bibitem[{{Netzer}(2015)}]{netzer15}
{Netzer}, H. 2015, \araa, 53, 365, \dodoi{10.1146/annurev-astro-082214-122302}

\bibitem[{{Ni} {et~al.}(2019){Ni}, {Yang}, {Brandt}, {Alexander}, {Chen},
  {Luo}, {Vito}, \& {Xue}}]{ni19}
{Ni}, Q., {Yang}, G., {Brandt}, W.~N., {et~al.} 2019, \mnras, 490, 1135,
  \dodoi{10.1093/mnras/stz2623}

\bibitem[{{Ni} {et~al.}(2020){Ni}, {Brandt}, {Yang}, {Leja}, {Chen}, {Luo},
  {Matharu}, {Sun}, {Vito}, {Xue}, \& {Zhang}}]{ni20}
{Ni}, Q., {Brandt}, W.~N., {Yang}, G., {et~al.} 2020, arXiv e-prints,
  arXiv:2007.04987.
\newblock \doarXiv{2007.04987}

\bibitem[{{Oke} \& {Gunn}(1983)}]{oke83}
{Oke}, J.~B., \& {Gunn}, J.~E. 1983, \apj, 266, 713, \dodoi{10.1086/160817}

\bibitem[{{Oliver} {et~al.}(2012){Oliver}, {Bock}, {Altieri}, {Amblard},
  {Arumugam}, {Aussel}, {Babbedge}, {Beelen}, {B{\'e}thermin}, {Blain},
  {Boselli}, {Bridge}, {Brisbin}, {Buat}, {Burgarella},
  {Castro-Rodr{\'{\i}}guez}, {Cava}, {Chanial}, {Cirasuolo}, {Clements},
  {Conley}, {Conversi}, {Cooray}, {Dowell}, {Dubois}, {Dwek}, {Dye}, {Eales},
  {Elbaz}, {Farrah}, {Feltre}, {Ferrero}, {Fiolet}, {Fox}, {Franceschini},
  {Gear}, {Giovannoli}, {Glenn}, {Gong}, {Gonz{\'a}lez Solares}, {Griffin},
  {Halpern}, {Harwit}, {Hatziminaoglou}, {Heinis}, {Hurley}, {Hwang}, {Hyde},
  {Ibar}, {Ilbert}, {Isaak}, {Ivison}, {Lagache}, {Le Floc'h}, {Levenson},
  {Faro}, {Lu}, {Madden}, {Maffei}, {Magdis}, {Mainetti}, {Marchetti},
  {Marsden}, {Marshall}, {Mortier}, {Nguyen}, {O'Halloran}, {Omont}, {Page},
  {Panuzzo}, {Papageorgiou}, {Patel}, {Pearson}, {P{\'e}rez-Fournon}, {Pohlen},
  {Rawlings}, {Raymond}, {Rigopoulou}, {Riguccini}, {Rizzo}, {Rodighiero},
  {Roseboom}, {Rowan-Robinson}, {S{\'a}nchez Portal}, {Schulz}, {Scott},
  {Seymour}, {Shupe}, {Smith}, {Stevens}, {Symeonidis}, {Trichas}, {Tugwell},
  {Vaccari}, {Valtchanov}, {Vieira}, {Viero}, {Vigroux}, {Wang}, {Ward},
  {Wardlow}, {Wright}, {Xu}, \& {Zemcov}}]{oliver12}
{Oliver}, S.~J., {Bock}, J., {Altieri}, B., {et~al.} 2012, \mnras, 424, 1614,
  \dodoi{10.1111/j.1365-2966.2012.20912.x}

\bibitem[{{Papovich} {et~al.}(2001){Papovich}, {Dickinson}, \&
  {Ferguson}}]{papovich01}
{Papovich}, C., {Dickinson}, M., \& {Ferguson}, H.~C. 2001, \apj, 559, 620,
  \dodoi{10.1086/322412}

\bibitem[{{Papovich} {et~al.}(2004){Papovich}, {Dole}, {Egami}, {Le Floc'h},
  {P{\'e}rez-Gonz{\'a}lez}, {Alonso-Herrero}, {Bai}, {Beichman}, {Blaylock},
  {Engelbracht}, {Gordon}, {Hines}, {Misselt}, {Morrison}, {Mould},
  {Muzerolle}, {Neugebauer}, {Richards}, {Rieke}, {Rieke}, {Rigby}, {Su}, \&
  {Young}}]{papovich04}
{Papovich}, C., {Dole}, H., {Egami}, E., {et~al.} 2004, \apjs, 154, 70,
  \dodoi{10.1086/422880}

\bibitem[{{Papovich} {et~al.}(2007){Papovich}, {Rudnick}, {Le Floc'h}, {van
  Dokkum}, {Rieke}, {Taylor}, {Armus}, {Gawiser}, {Huang}, {Marcillac}, \&
  {Franx}}]{papovich07}
{Papovich}, C., {Rudnick}, G., {Le Floc'h}, E., {et~al.} 2007, \apj, 668, 45,
  \dodoi{10.1086/521090}

\bibitem[{{Pirzkal} {et~al.}(2012){Pirzkal}, {Rothberg}, {Nilsson},
  {Finkelstein}, {Koekemoer}, {Malhotra}, \& {Rhoads}}]{pirzkal12}
{Pirzkal}, N., {Rothberg}, B., {Nilsson}, K.~K., {et~al.} 2012, \apj, 748, 122,
  \dodoi{10.1088/0004-637X/748/2/122}

\bibitem[{{Pouliasis} {et~al.}(2020){Pouliasis}, {Mountrichas},
  {Georgantopoulos}, {Ruiz}, {Yang}, \& {Bonanos}}]{pouliasis20}
{Pouliasis}, E., {Mountrichas}, G., {Georgantopoulos}, I., {et~al.} 2020,
  \mnras, 495, 1853, \dodoi{10.1093/mnras/staa1263}

\bibitem[{{Rieke} {et~al.}(2019){Rieke}, {Alberts}, {Shivaei}, {Colina}, \&
  {N{\o}rgaard-Nielsen}}]{rieke19}
{Rieke}, G., {Alberts}, S., {Shivaei}, I., {Colina}, L., \&
  {N{\o}rgaard-Nielsen}, H.~U. 2019, \baas, 51, 11

\bibitem[{{Salim} {et~al.}(2007){Salim}, {Rich}, {Charlot}, {Brinchmann},
  {Johnson}, {Schiminovich}, {Seibert}, {Mallery}, {Heckman}, {Forster},
  {Friedman}, {Martin}, {Morrissey}, {Neff}, {Small}, {Wyder}, {Bianchi},
  {Donas}, {Lee}, {Madore}, {Milliard}, {Szalay}, {Welsh}, \& {Yi}}]{salim07}
{Salim}, S., {Rich}, R.~M., {Charlot}, S., {et~al.} 2007, \apjs, 173, 267,
  \dodoi{10.1086/519218}

\bibitem[{{Sersic}(1968)}]{sersic68}
{Sersic}, J.~L. 1968, {Atlas de Galaxias Australes}

\bibitem[{{Shipley} {et~al.}(2016){Shipley}, {Papovich}, {Rieke}, {Brown}, \&
  {Moustakas}}]{shipley16}
{Shipley}, H.~V., {Papovich}, C., {Rieke}, G.~H., {Brown}, M. J.~I., \&
  {Moustakas}, J. 2016, \apj, 818, 60, \dodoi{10.3847/0004-637X/818/1/60}

\bibitem[{{Shivaei} {et~al.}(2017){Shivaei}, {Reddy}, {Shapley}, {Siana},
  {Kriek}, {Mobasher}, {Coil}, {Freeman}, {Sanders}, {Price}, {Azadi}, \&
  {Zick}}]{shivaei17}
{Shivaei}, I., {Reddy}, N.~A., {Shapley}, A.~E., {et~al.} 2017, \apj, 837, 157,
  \dodoi{10.3847/1538-4357/aa619c}

\bibitem[{{Spoon} {et~al.}(2007){Spoon}, {Marshall}, {Houck}, {Elitzur}, {Hao},
  {Armus}, {Brandl}, \& {Charmandaris}}]{spoon07}
{Spoon}, H.~W.~W., {Marshall}, J.~A., {Houck}, J.~R., {et~al.} 2007, \apjl,
  654, L49, \dodoi{10.1086/511268}

\bibitem[{{Stalevski} {et~al.}(2012){Stalevski}, {Fritz}, {Baes}, {Nakos}, \&
  {Popovi{\'c}}}]{stalevski12}
{Stalevski}, M., {Fritz}, J., {Baes}, M., {Nakos}, T., \& {Popovi{\'c}},
  L.~{\v{C}}. 2012, \mnras, 420, 2756, \dodoi{10.1111/j.1365-2966.2011.19775.x}

\bibitem[{{Stalevski} {et~al.}(2016){Stalevski}, {Ricci}, {Ueda}, {Lira},
  {Fritz}, \& {Baes}}]{stalevski16}
{Stalevski}, M., {Ricci}, C., {Ueda}, Y., {et~al.} 2016, \mnras, 458, 2288,
  \dodoi{10.1093/mnras/stw444}

\bibitem[{{Stalevski} {et~al.}(2019){Stalevski}, {Tristram}, \&
  {Asmus}}]{stalevski19}
{Stalevski}, M., {Tristram}, K. R.~W., \& {Asmus}, D. 2019, \mnras, 484, 3334,
  \dodoi{10.1093/mnras/stz220}

\bibitem[{{Stefanon} {et~al.}(2017){Stefanon}, {Yan}, {Mobasher}, {Barro},
  {Donley}, {Fontana}, {Hemmati}, {Koekemoer}, {Lee}, {Lee}, {Nayyeri}, {Peth},
  {Pforr}, {Salvato}, {Wiklind}, {Wuyts}, {Ashby}, {Castellano}, {Conselice},
  {Cooper}, {Cooray}, {Dolch}, {Ferguson}, {Galametz}, {Giavalisco}, {Guo},
  {Willner}, {Dickinson}, {Faber}, {Fazio}, {Gardner}, {Gawiser}, {Grazian},
  {Grogin}, {Kocevski}, {Koo}, {Lee}, {Lucas}, {McGrath}, {Nandra}, {Newman},
  \& {van der Wel}}]{stefanon17}
{Stefanon}, M., {Yan}, H., {Mobasher}, B., {et~al.} 2017, \apjs, 229, 32,
  \dodoi{10.3847/1538-4365/aa66cb}

\bibitem[{{Stern}(2015)}]{stern15}
{Stern}, D. 2015, \apj, 807, 129, \dodoi{10.1088/0004-637X/807/2/129}

\bibitem[{{Stern} {et~al.}(2005){Stern}, {Eisenhardt}, {Gorjian}, {Kochanek},
  {Caldwell}, {Eisenstein}, {Brodwin}, {Brown}, {Cool}, {Dey}, {Green},
  {Jannuzi}, {Murray}, {Pahre}, \& {Willner}}]{stern05}
{Stern}, D., {Eisenhardt}, P., {Gorjian}, V., {et~al.} 2005, \apj, 631, 163,
  \dodoi{10.1086/432523}

\bibitem[{{Stern} {et~al.}(2012){Stern}, {Assef}, {Benford}, {Blain}, {Cutri},
  {Dey}, {Eisenhardt}, {Griffith}, {Jarrett}, {Lake}, {Masci}, {Petty},
  {Stanford}, {Tsai}, {Wright}, {Yan}, {Harrison}, \& {Madsen}}]{stern12b}
{Stern}, D., {Assef}, R.~J., {Benford}, D.~J., {et~al.} 2012, \apj, 753, 30,
  \dodoi{10.1088/0004-637X/753/1/30}

\bibitem[{{Stierwalt} {et~al.}(2014){Stierwalt}, {Armus}, {Charmandaris},
  {Diaz-Santos}, {Marshall}, {Evans}, {Haan}, {Howell}, {Iwasawa}, {Kim},
  {Murphy}, {Rich}, {Spoon}, {Inami}, {Petric}, \& {U}}]{stierwalt14}
{Stierwalt}, S., {Armus}, L., {Charmandaris}, V., {et~al.} 2014, \apj, 790,
  124, \dodoi{10.1088/0004-637X/790/2/124}

\bibitem[{{Tielens}(2008)}]{tielens08}
{Tielens}, A.~G.~G.~M. 2008, \araa, 46, 289,
  \dodoi{10.1146/annurev.astro.46.060407.145211}

\bibitem[{{Ueda} {et~al.}(2014){Ueda}, {Akiyama}, {Hasinger}, {Miyaji}, \&
  {Watson}}]{ueda14}
{Ueda}, Y., {Akiyama}, M., {Hasinger}, G., {Miyaji}, T., \& {Watson}, M.~G.
  2014, \apj, 786, 104, \dodoi{10.1088/0004-637X/786/2/104}

\bibitem[{{Urry} \& {Padovani}(1995)}]{urry95}
{Urry}, C.~M., \& {Padovani}, P. 1995, \pasp, 107, 803, \dodoi{10.1086/133630}

\bibitem[{{van der Wel} {et~al.}(2012){van der Wel}, {Bell}, {H{\"a}ussler},
  {McGrath}, {Chang}, {Guo}, {McIntosh}, {Rix}, {Barden}, {Cheung}, {Faber},
  {Ferguson}, {Galametz}, {Grogin}, {Hartley}, {Kartaltepe}, {Kocevski},
  {Koekemoer}, {Lotz}, {Mozena}, {Peth}, \& {Peng}}]{van_der_wel12}
{van der Wel}, A., {Bell}, E.~F., {H{\"a}ussler}, B., {et~al.} 2012, \apjs,
  203, 24, \dodoi{10.1088/0067-0049/203/2/24}

\bibitem[{{van Dokkum}(2001)}]{van_dokkum01}
{van Dokkum}, P.~G. 2001, \pasp, 113, 1420, \dodoi{10.1086/323894}

\bibitem[{{Vito} {et~al.}(2018){Vito}, {Brandt}, {Yang}, {Gilli}, {Luo},
  {Vignali}, {Xue}, {Comastri}, {Koekemoer}, {Lehmer}, {Liu}, {Paolillo},
  {Ranalli}, {Schneider}, {Shemmer}, {Volonteri}, \& {Wang}}]{vito18}
{Vito}, F., {Brandt}, W.~N., {Yang}, G., {et~al.} 2018, \mnras, 473, 2378,
  \dodoi{10.1093/mnras/stx2486}

\bibitem[{{Xue}(2017)}]{xue17}
{Xue}, Y.~Q. 2017, \nar, 79, 59, \dodoi{10.1016/j.newar.2017.09.002}

\bibitem[{{Xue} {et~al.}(2016){Xue}, {Luo}, {Brandt}, {Alexander}, {Bauer},
  {Lehmer}, \& {Yang}}]{xue16}
{Xue}, Y.~Q., {Luo}, B., {Brandt}, W.~N., {et~al.} 2016, \apjs, 224, 15,
  \dodoi{10.3847/0067-0049/224/2/15}

\bibitem[{{Yang} {et~al.}(2019){Yang}, {Brandt}, {Alexander}, {Chen}, {Ni},
  {Vito}, \& {Zhu}}]{yang19}
{Yang}, G., {Brandt}, W.~N., {Alexander}, D.~M., {et~al.} 2019, \mnras, 485,
  3721, \dodoi{10.1093/mnras/stz611}

\bibitem[{{Yang} {et~al.}(2018{\natexlab{a}}){Yang}, {Brandt}, {Darvish},
  {Chen}, {Vito}, {Alexander}, {Bauer}, \& {Trump}}]{yang18b}
{Yang}, G., {Brandt}, W.~N., {Darvish}, B., {et~al.} 2018{\natexlab{a}},
  \mnras, 480, 1022, \dodoi{10.1093/mnras/sty1910}

\bibitem[{{Yang} {et~al.}(2014){Yang}, {Xue}, {Luo}, {Brandt}, {Alexander},
  {Bauer}, {Cui}, {Kong}, {Lehmer}, {Wang}, {Wu}, {Yuan}, {Yuan}, \&
  {Zhou}}]{yang14}
{Yang}, G., {Xue}, Y.~Q., {Luo}, B., {et~al.} 2014, \apjs, 215, 27,
  \dodoi{10.1088/0067-0049/215/2/27}

\bibitem[{{Yang} {et~al.}(2016){Yang}, {Brandt}, {Luo}, {Xue}, {Bauer}, {Sun},
  {Kim}, {Schulze}, {Zheng}, {Paolillo}, {Shemmer}, {Liu}, {Schneider},
  {Vignali}, {Vito}, \& {Wang}}]{yang16}
{Yang}, G., {Brandt}, W.~N., {Luo}, B., {et~al.} 2016, \apj, 831, 145,
  \dodoi{10.3847/0004-637X/831/2/145}

\bibitem[{{Yang} {et~al.}(2018{\natexlab{b}}){Yang}, {Brandt}, {Vito}, {Chen},
  {Trump}, {Luo}, {Sun}, {Xue}, {Koekemoer}, {Schneider}, {Vignali}, \&
  {Wang}}]{yang18}
{Yang}, G., {Brandt}, W.~N., {Vito}, F., {et~al.} 2018{\natexlab{b}}, \mnras,
  475, 1887, \dodoi{10.1093/mnras/stx2805}

\bibitem[{{Yang} {et~al.}(2020){Yang}, {Boquien}, {Buat}, {Burgarella},
  {Ciesla}, {Duras}, {Stalevski}, {Brandt}, \& {Papovich}}]{yang20}
{Yang}, G., {Boquien}, M., {Buat}, V., {et~al.} 2020, \mnras, 491, 740,
  \dodoi{10.1093/mnras/stz3001}

\end{thebibliography}
\bibliographystyle{aasjournal}



\end{CJK*}
\end{document}